\documentclass[10pt, american]{article}
\usepackage[T1]{fontenc}
\usepackage[latin1]{inputenc}
\usepackage{latexsym}
\usepackage{amsmath}
\usepackage{babel}
\usepackage{amssymb}
\usepackage{amsfonts}
\usepackage{times}
\usepackage{theorem}
\usepackage{epsfig}
\usepackage{enumerate}
\usepackage{hyperref}
\newcounter{smallarabics}
\newenvironment{arabicenumerate}
{\begin{list}{{\normalfont\textrm{(\arabic{smallarabics})}}}
  {\usecounter{smallarabics}\setlength{\itemindent}{0cm}
   \setlength{\leftmargin}{5ex}\setlength{\labelwidth}{4ex}
   \setlength{\topsep}{0.75\parsep}\setlength{\partopsep}{0ex}
   \setlength{\itemsep}{0ex}}}
{\end{list}}

\newcounter{smallroman}

\newcommand{\ben}{\begin{arabicenumerate}}  
\newcommand{\een}{\end{arabicenumerate}}

\newtheorem{theoreme}{Theorem }[section]
\newtheorem{proposition}[theoreme]{Proposition}
\newtheorem{lemma}[theoreme]{Lemma}
\newtheorem{definition}[theoreme]{Definition}

\newtheorem{corollary}[theoreme]{Corollary}
\newtheorem{remark}[theoreme]{Remark}

\def\rr{{\mathbb R}}
\def\zz{{\mathbb Z}}
\def\cc{{\mathbb C}}
\def\nn{{\mathbb N}}

\def\textsl{{}}

\def\Re{{\rm Re}\,}

\newcommand{\slim}{\mathop{\mathrm{s-lim}}\limits}

\def\c0inf{C_0^\infty}
\def\bep{\begin{proposition}}
\def\eep{\end{proposition}}

\def\proof{\noindent {\bf Proof.}\ \ }

\def\i{{\rm i}}
\newcommand{\beq}{\begin{equation}}
\newcommand{\eeq}{\end{equation}}
\newcommand{\bear}[1]{\begin{array}{#1}}
\newcommand{\ear}{\end{array}}

\def\sp{{\hat e}}

\newcommand{\e}{\mathrm{e}}
\renewcommand{\i}{\mathrm{i}}

\renewcommand{\d}{\mathrm{d}}

\def\qed{$\Box$\medskip}

\def\bel{\begin{lemma}}
\def\eel{\end{lemma}}
\def\bet{\begin{theoreme}}
\def\eet{\end{theoreme}}
\def\bed{\begin{definition}}
\def\eed{\end{definition}}
\def\bec{\begin{corollary}}
\def\eec{\end{corollary}}
\def\bar{\overline}

\def\12{\frac{1}{2}}

\def\e{{\rm e}}

\def\d{{\rm d}}
\def\Ran{{\rm Ran}\,}

\def\Dom{{\rm Dom}\,}

\def\sp{{\rm sp}}

\def\C{{\rm C}}

\def\Ent{{\rm Ent}}

\def\eq{{\rm eq}}

\newcommand{\TRI}{{\hyperref[TRIDefSect]{TRI}}}
\newcommand{\Fone}{{\hyperref[F1-def]{(F1)}}}
\newcommand{\Ftwo}{{\hyperref[F2-def]{(F2)}}}
\renewcommand{\C}{{\hyperref[C-def]{(C)}}}
\newcommand{\Eone}{{\hyperref[E1-def]{(E1)}}}
\newcommand{\Etwo}{{\hyperref[E2-def]{(E2)}}}
\newcommand{\Ethree}{{\hyperref[E3-def]{(E3)}}}
\newcommand{\EP}{{\hyperref[EP-def]{(EP)}}}
\newcommand{\Tone}{{\hyperref[T1-def]{(T1)}}}
\newcommand{\Ttwo}{{\hyperref[T2-def]{(T2)}}}
\newcommand{\Tthree}{{\hyperref[T3-def]{(T3)}}}
\newcommand{\Tfour}{{\hyperref[T4-def]{(T4)}}}
\newcommand{\Tfive}{{\hyperref[T5-def]{(T5)}}}

\newcommand{\NESStwo}{{\hyperref[NESS2-def]{(NESS2)}}}
\newcommand{\NESSthree}{{\hyperref[NESS3-def]{(NESS3)}}}
\begin{document}

\title{Entropic Fluctuations in Statistical Mechanics I.\\
Classical Dynamical Systems
\vskip 1.5truecm
}
\author{V. Jak\v{s}i\'c$^{1}$, C.-A. Pillet$^{2}$, L. Rey-Bellet$^{3}$
\\ \\ \\
$^1$Department of Mathematics and Statistics\\ 
McGill University\\
805 Sherbrooke Street West \\
Montreal,  QC,  H3A 2K6, Canada
\\ \\
$^2$Centre de Physique Th\'eorique\footnote{%
UMR 6207:
Universit\'e de Provence,
Universit\'e de la M\'editerran\'ee, 
Universit\'e de Toulon et CNRS, FRUMAM}\\
Universit\'e du Sud Toulon-Var,  B.P. 20132 \\
F-83957 La Garde Cedex, France
\\ \\
$^3$Department of Mathematics and Statistics \\ 
Lederle Graduate Research Tower, Box 34515 \\ University of 
Massachusetts\\Amherst, MA 01003-4515, USA
}
\maketitle

\vskip 2truecm
{\noindent\bf Abstract. }Within the abstract framework of dynamical system
theory we describe a general approach to the Transient (or Evans-Searles) and
Steady State (or Gallavotti-Cohen) Fluctuation Theorems of non-equilibrium
statistical mechanics. Our main objective is to display the minimal, model 
independent mathematical structure at work behind fluctuation theorems.
Besides its conceptual simplicity, another advantage
of our approach is its natural extension to quantum statistical
mechanics which will be presented in a companion paper.
We shall discuss several examples including thermostated systems, open Hamiltonian 
systems, chaotic homeomorphisms of compact metric spaces and Anosov diffeomorphisms.
\vfill\eject

\tableofcontents

\section{Introduction}
This is  the first in a series of papers  devoted to the  so-called Fluctuation Theorems of
non-equilibrium statistical mechanics. This series is a part of the research program initiated
in \cite{Pi, JP1, JP2} that concerns the development of a mathematical theory of 
non-equilibrium statistical mechanics within the framework of dynamical systems. 

The first fluctuation theorem in statistical mechanics goes back to 1905 and the celebrated 
work of Einstein on Brownian motion. The  subsequent historical developments are 
reviewed in \cite{RM} (see also the monographs \cite{GM, KTH}) and we mention here 
only the classical results of Onsager \cite{On1, On2}, Green \cite{Gr1, Gr2}, and 
Kubo \cite{Kub} which will be re-visited in this paper. Virtually all classical works on the 
subject concern the so called close to equilibrium regime in which the mechanical and 
thermodynamical forces (affinities) are weak. One of the key features of modern fluctuation
theorems, suggested by numerical experiments \cite{ECM} and established theoretically 
for the first time  by Evans and Searles \cite{ES} and  by Gallavotti and Cohen \cite{GC1, GC2},  
is that they hold for systems arbitrarily far from equilibrium and reduce to  Green-Kubo
formulas and Onsager relations in the linear regime near equilibrium. The seminal papers 
\cite{ECM, ES, GC1, GC2} were followed by a vast body of theoretical, numerical and 
experimental works which are reviewed in \cite{RM}. The Evans-Searles (ES) and 
Gallavotti-Cohen (GC) Fluctuation Theorems are the main topics of the present work.

The basic two paradigms for deterministic (dynamical system) non-equilibrium statistical 
mechanics are the so called {\em thermostated systems} and {\em open systems}. 
Thermostated systems are Hamiltonian systems (with finitely many degrees of freedom)
driven out of equilibrium by an external (non-Hamiltonian) force and constrained by a 
deterministic thermostating force to stay on a surface of constant energy. Open 
systems are Hamiltonian systems consisting of a ''small'' Hamiltonian system (with finitely 
many degrees of freedom) interacting with, say two, ''large'' reservoirs which are infinitely
extended Hamiltonian systems. The reservoirs are initially  in thermal equilibrium at distinct
temperatures and the temperature differential leads to a steady heat flux from the hotter to 
the colder reservoir across the small system. Throughout the main body of the paper we 
shall illustrate our results on an example of thermostated system and an example of open
system.

The majority of works on fluctuation theorems concern classical physics.  
In the quantum case
comparatively little  is known and there are very few mathematically  
rigorous works on the
subject (see \cite{TM, DDM, Ro,Ku2}). The  present paper, which concerns  
only the classical case,
originates in our attempts to find a proper mathematical  framework  
for the extensions of
ES and GC Fluctuation Theorems to quantum physics. One of the  
difficulties in finding
such a framework stems from the fact that it was already lacking at  
the classical level.
Indeed, even the basic examples of thermostated systems and open systems
were studied in the literature in an unrelated way and it was far from  
obvious
which aspects of the theory are model dependent and which are universal.
For example there was no clear universal rationale in the choice of  
the "entropy production" observable
(also called  "phase space contraction rate" or "dissipation  
function") which plays a central
role in the theory.

A model independent definition of  the entropy production has been  
proposed by Maes, in the context of
stochastic (Markovian or Gibbsian) dynamical systems, see e.g.  
\cite{Ma2}. We take here a different and complementary route
and  discuss non-equilibrium  statistical mechanics within
the context of deterministic dynamical systems. Our work is mostly  
of a review nature and we do not prove any new specific results.
Rather we organize the existing set of ideas and results in an axiomatic
abstract framework that unifies virtually all deterministic models  
discussed in the literature (in particular, open infinite systems and
thermostated  finite dimensional systems will be treated in a unified
manner) and clarifies the mathematical structure of the theory. The
framework has a direct extension to non-commutative dynamical
systems and in particular to quantum mechanics and this will be the
subject of the  remaining papers in the series. Our principal new results
concern the  quantum case and  we will  focus here only on those   
aspects of the classical theory  that can be extended, within the
framework of dynamical systems, to quantum statistical mechanics.

We have made an attempt to expose the results in a pedagogical way and  
the only prerequisite for the principal part of the paper is a basic knowledge  
of probability and measure theory.

The paper is organized as follows.

In Section \ref{sec-Basn} we introduce our dynamical system setup and review the
properties of relative entropies that we will need. In Section \ref{sec-finiteentropy} 
we introduce the basic objects of the theory, the entropy cocycle and the entropy 
production observable, discuss their properties, and prove  the finite time Evans-Searles
Fluctuation Theorem. The results described in this section hold under minimal regularity 
assumptions that are satisfied in virtually all models of interest. 

In Section \ref{sec-control} we start the discussion of thermodynamics by introducing
control parameters (mechanical or thermodynamical forces) to our dynamical system setup.
The finite time Evans-Searles Fluctuation Theorem is then generalized to this
setting. Following the ideas of Gallavotti \cite{Ga1} (see also \cite{LS2}) 
we use this generalization to derive  finite time Green-Kubo formulas and 
Onsager reciprocity relations. 

The results of Sections \ref{sec-finiteentropy} and \ref{sec-control} concern the system 
evolved over a finite interval of time and are very general. In particular, they do not require 
any ergodicity assumptions. Section \ref{sec-LAT} concerns the large time limit
$t\rightarrow \infty$. Under suitable ergodicity assumptions  we derive Evans-Searles 
Fluctuation Theorem on the basis of its universally valid finite time counterpart and prove 
the Green-Kubo formula and Onsager reciprocity relations. 

Section \ref{sec-gallcohen} is devoted to the Gallavotti-Cohen Fluctuation Theorem. 
After introducing the key concept of non-equilibrium steady states (NESS), 
the GC Fluctuation Theorem is stated as, essentially, an ergodic-type  hypothesis
concerning the NESS and the entropy production observable. 
The Green-Kubo formula and Onsager reciprocity relations also follow from the GC 
Fluctuation Theorem.

One  advantage of our abstract axiomatic framework is that it allows for a transparent 
comparison between the ES and GC Fluctuation Theorems. It turns out that from the 
mathematical point of view these two theorems are equivalent up to an exchange of 
limits (see Relation (\ref{exch-lim})). This exchange of limits  may fail even in some 
very simple models and its validity  can be interpreted as an ergodic property of the 
underlying  dynamical system. We raise this point to the {\em Principle of Regular 
Entropic Fluctuations} which is introduced and discussed in Section \ref{sec-pri-reg}. 
After this work was completed we have learned that related ideas have been previously
discussed in \cite{RM}.

Sections \ref{sec-examples}--\ref{sec-anosov} are devoted to  examples. In Section
\ref{sec-examples} we discuss several toy models which illustrate the optimality of our
assumptions. In Section \ref{subsection-grf} we develop the non-equilibrium statistical 
mechanics of Gaussian dynamical systems. Chaotic homeomorphisms of compact 
metric spaces are discussed in Section \ref{sec-homeo}. Finally, in Section 
\ref{sec-anosov} we discuss the non-equilibrium statistical mechanics of Anosov 
diffeomorphisms of compact manifolds. In each of these examples 
we verify the validity of the proposed principle of regular entropic fluctuations. 

For the convenience of the reader, a table of frequently used abbreviations
and symbols is provided on page \pageref{AbbTable}.

The ergodic-type hypotheses introduced in this paper typically concern existence of 
certain limits as time $t\rightarrow \infty$, the regularity (differentiability, etc) properties 
of limiting functions w.r.t. control parameters, and the validity of exchange of order 
of limits and derivatives. The introduced hypotheses are minimal ({\sl i.e.}, sufficient and 
necessary) to derive fluctuation theorems and their implications from the 
universally valid structural theory discussed in Sections 2-4. The verification of these hypotheses 
in concrete models leads to  a novel class of (analytically difficult) problems in ergodic 
theory of dynamical systems.

{\bf Acknowledgment.} The research  of V.J. was partly supported by NSERC. 
The research of C.-A.P. was partly supported by ANR (grant 09-BLAN-0098).
The research of L.R.-B. was partly supported by NSF. We wish to thank  C.~Liverani and 
D.~Ruelle for useful discussions. The results of  this paper were presented 
by its authors in mini-courses at University of Cergy-Pontoise,
Erwin Schr\"odinger Institute (Vienna), Centre de Physique Th\'eorique (Marseille and Toulon), 
University of British Columbia (Vancouver), Ecole Polytechnique (Paris), 
Institut Henri Poincar\'e (Paris) and Ecole de Physique des Houches.
The paper has gained a lot from these presentations and we wish to thank 
the  respective institutions and F.~Germinet, J.~Yngvanson, R.~Froese,
S.~Kuksin, G. Stoltz, J. Fr\"ohlich  for making these mini-courses possible.

\section{Basic notions}
\label{sec-Basn}
\subsection{Phase space, observables, states}
\label{subsecphasespace}

Let $M$ be a set and ${\cal F}$ a $\sigma$-algebra in $M$. We shall refer to the 
measure space $(M, {\cal F})$ as the phase space. If $M$ is  a topological space,  
we shall always take for ${\cal F}$ the Borel $\sigma$-algebra in $M$.

An observable is a measurable function $f: M \rightarrow  \cc$ and we denote by 
${\cal O}(M)$  the complex vector space of all  observables.  $B(M)$ denotes the subspace
of all bounded observables. Together with the  norm $\|f\|=\sup_{x\in M}|f(x)|$,  $B(M)$ 
is a Banach space. If $M$ is a topological space, $C(M)$ denotes the Banach space of all 
bounded continuous observables. The corresponding spaces of real valued observables
are denoted by ${\cal O}_\rr(M)$, $B_\rr(M)$, $C_\rr(M)$.

A state is a probability measure on $(M, {\cal F})$ and ${\cal S}$ denotes the set of
all states. The expectation value of an observable $f$ w.r.t. the state $\nu$ is denoted by 
\[
\nu(f)=\int_M f\d\nu.
\]
If ${\bf f}=(f_1, \ldots, f_N)$ is a vector-valued observable we set
$\nu({\bf f})=(\nu(f_1), \ldots, \nu(f_N))$.
We shall equip ${\cal S}$ with the weakest  topology w.r.t. which the functionals 
${\cal S}\ni \nu \mapsto \nu(f)$ are continuous for all $f \in B(M)$.
If $\theta:M\rightarrow M$ is a measurable map, we denote by 
$\nu \circ \theta^{-1}$ the measure $\mathcal F\ni A\mapsto\nu(\theta^{-1}(A))$.
Clearly $\nu\circ\theta^{-1}(f)=\nu(f\circ\theta)$ for all $f\in B(M)$. 
A map $\theta:M\rightarrow M$ is called involutive if $\theta \circ \theta(x)=x$ for all 
$x\in M$.

We  shall say that  a state $\nu$ is normal w.r.t. $\omega\in\cal S$ iff  $\nu$ is absolutely 
continuous w.r.t. $\omega$ (denoted $\nu\ll\omega$). The set of all states which are
normal w.r.t. $\omega$
 is denoted ${\cal N}_\omega$. \label{omega-N-def}
Two states $\nu$ and $\omega$ are called equivalent 
iff $\nu\ll\omega$ and $\omega\ll\nu$, {\sl i.e.,} iff $\nu$ and $\omega$ have the same sets
of measure zero. The Radon-Nikodym derivative $\d\nu/\d\omega$, 
which  will play an important role in this paper, is  defined as an element of 
$L^1(M, \d\omega)$ and is an equivalence class of functions rather
than a single function. For this reason the notion of observable is extended as follows. 
Given a state $\omega$, let 
$Z_\omega=\{ f\in {\cal O}(M)\,|\, f(x)=0 \, \, \hbox{for $\omega$-a.e. $x$}\}$ and let
${\cal O}(M)_\omega={\cal O}(M)/Z_\omega$ be the quotient vector space (the 
elements of ${\cal O}(M)_\omega$ are equivalence classes w.r.t. the relation 
$f\sim  g \Leftrightarrow f-g \in Z_\omega$). 
Similarly, $L^\infty(M, \d\omega) = B(M)/Z_\omega$. 
As usual in measure theory, dealing with equivalence classes instead of single functions is 
natural and  causes no difficulties, the classes are called functions, {\sl etc.}

In what follows, we adopt the shorthands
\[
\Delta_{\nu|\omega}=\frac{\d\nu}{\d \omega}, 
\qquad 
\ell_{\nu|\omega}=\log\Delta_{\nu|\omega}.
\label{RD-def}
\]

\subsection{Relative entropies}

The relative entropy of a state $\nu$ w.r.t. a state $\omega$ is defined by 
\begin{equation}
{\rm Ent}(\nu|\omega)=
\begin{cases}
-\infty& \text{if $\nu \not\in {\cal N}_\omega$},\\
-\nu(\ell_{\nu|\omega}) &\text{if $\nu \in {\cal N}_\omega$.}
\end{cases}
\label{rel-ent}
\end{equation}
Since $-\ell_{\nu|\omega}\le\Delta_{\nu|\omega}^{-1}-1$ and 
$\nu(\Delta_{\nu|\omega}^{-1})=1$, relative entropy is well defined as a map from 
$\mathcal S\times\mathcal S$ to $[-\infty,0]$. Its basic properties are  (see e.g. \cite{OP}):
\bet\label{gen-ent}
\begin{enumerate}[{\rm (1)}]
\item For $\omega,\nu\in\cal S$,
${\rm Ent}(\nu|\omega)=\inf _{f \in B_{\mathbb R}(M)}
\left[\log \omega(\e^f)- \nu(f)\right]$.
\item  For $\omega \in {\cal S}$ and $f \in B_\rr(M)$,  
$\log \omega(\e^f)=\sup_{\nu \in {\cal S}}\left[ 
{\rm Ent}(\nu|\omega)+ \nu(f)\right]$.
\item Concavity:  For $\omega_1,\omega_2,\nu_1,\nu_2\in\cal S$
and $\lambda \in [0,1]$ ,
\[
\Ent(\lambda\nu_1+(1-\lambda)\nu_2|\lambda\omega_1+(1-\lambda)\omega_2)\geq
\lambda\Ent(\nu_1|\omega_1)+(1-\lambda)\Ent(\nu_2|\omega_2).
\]
\item $\Ent(\nu|\omega)\leq 0$  for all $\omega,\nu\in\cal S$ and 
$\Ent(\nu|\omega)=0$ if and only if $\nu=\omega$.
\item If $\theta: M\rightarrow  M$ is a measurable bijection, then 
${\rm Ent}(\nu\circ \theta^{-1} | \omega\circ \theta^{-1})={\rm Ent}(\nu|\omega)$.
\item The relative entropy is an upper semicontinuous map from 
${\cal S}\times{\cal S}$ to $[-\infty,0]$, that is
$$
\Ent(\nu|\omega)\ge\limsup_\alpha\Ent(\nu_\alpha|\omega_\alpha),
$$
for all convergent nets $\nu_\alpha\rightarrow\nu$ and
$\omega_\alpha\rightarrow\omega$ in ${\cal S}$. 
\item For any $\omega \in {\cal S}$ and any finite constant  $C$, the set
$\left\{ \nu \in {\cal S}\,|\,{\rm Ent}(\nu|\omega)\geq C\right\}$
is compact in ${\cal S}$.
\end{enumerate}
\eet

The  R\'enyi relative entropy of order $\alpha\in\rr$, \cite{Re}, is defined by 
\begin{equation}
{\rm Ent}_\alpha(\nu|\omega)
=
\begin{cases} -\infty& \text{if $\nu \not\in {\cal N}_\omega$}
\\
\log \omega(\Delta_{\nu|\omega}^\alpha) &\text{if $\nu \in {\cal N}_\omega$.}
\end{cases}
\label{renyi-rel-ent}
\end{equation}
This generalization of relative entropy has found numerous applications (see  \cite{BS, OP}
for references and additional information). We list below several  properties of the R\'enyi 
relative entropy that are relevant for our purposes:
\bep Suppose that $\nu \in {\cal N}_\omega$. 
\begin{enumerate}[{\rm (1)}]
\item If $\theta: M\rightarrow  M$ is a measurable bijection, then 
${\rm Ent}_\alpha(\nu\circ \theta^{-1} |\omega\circ \theta^{-1})
={\rm Ent}_\alpha(\nu|\omega)$.
\item $\rr \ni \alpha \mapsto {\rm Ent}_\alpha(\nu|\omega)\in]-\infty, \infty]$
is a convex function. It is real analytic and non-positive on $]0,1[$.
It is positive for $\alpha \not\in [0,1]$.
\item
${\displaystyle
\lim_{\alpha \uparrow 1}\frac{1}{1-\alpha}{\rm Ent}_\alpha(\nu|\omega)
={\rm Ent}(\nu|\omega).
}$

\bigskip
In the remaining statements we assume that $\nu$ and $\omega$ are equivalent.

\item ${\rm Ent}_0(\nu|\omega)={\rm Ent}_1(\nu|\omega)=0$.
\item ${\rm Ent}_\alpha(\nu|\omega)={\rm \Ent}_{1-\alpha}(\omega|\nu)$.
\item ${\rm Ent}_\alpha(\nu|\omega)\geq \alpha {\rm Ent}(\omega|\nu)$.
\end{enumerate}
\label{prop-renyi}
\eep
 
\subsection{Dynamics}
\label{subsec-flows}

Let ${\cal I}$ be an index set whose elements are interpreted as instances of time. We 
shall always assume that ${\cal I}=\zz$ (the discrete time  case) or ${\cal I}=\rr$ (the
continuous time case).  A dynamics $\phi=\{\phi^t\,|\,t\in{\cal I}\}$ on $M$ is a  group
of invertible measurable transformations $\phi^t:M\rightarrow M$ describing the evolution
of the system. More precisely, we shall assume:
\begin{quote}
\label{F1-def}
{\bf (F1)} $\phi^0$ is the identity map and $\phi^{t+s}=\phi^t \circ \phi^s$ for all 
$s, t \in {\cal I}$. In particular, for all $t\in\cal I$, $\phi^t$ is an automorphism of the 
measurable space $(M,\cal F)$. 

\label{F2-def}
{\bf (F2)} The map $(t, x)\mapsto \phi^t(x)$ is measurable.
\end{quote}
The assumption (F2) is relevant only in the case ${\cal I}=\rr$ (in this case, the
dynamics $\phi$ is a {\sl flow} on $M$). In the discrete time case ${\cal I}=\zz$,
the dynamics is obtained by iterating the time $1$ map $\phi=\phi^1$ and
its inverse $\phi^{-1}$. We will sometimes write $\phi^n$ instead of $\phi^t$. 

A dynamics $\phi$ on $M$ induces transformation groups on ${\cal O}(M)$  and 
${\cal S}$  by $f_t=f\circ \phi^t$, $\nu_t=\nu \circ \phi^{-t}$.
They are clearly related by $\nu_t(f)=\nu(f_t)$.  A state $\nu$ is called steady 
(or stationary) if $\nu_t= \nu$ for all $t$. We denote by ${\cal S}_I$ the set of all 
steady states.\label{SI-def}

\subsection{Reference state} 
\label{sec-cds}

The starting point of our discussion is a classical dynamical system 
$(M, \phi, \omega)$, where $\phi$ is a  given dynamics on $M$ and $\omega$ a given
{\sl reference state} satisfying the following regularity assumption: 
\begin{quote}
\label{C-def}
{\bf  (C)} $\omega_t$ and $ \omega$  are equivalent for all $t\in {\cal I}$.
\end{quote}
In non-trivial models that arise in non-equilibrium statistical mechanics  
${\cal N}_\omega\cap{\cal S}_{I}=\emptyset$. In particular, $\omega\not\in {\cal S}_I$. 
In this  important aspect our starting point  differs from the usual one in the ergodic theory of 
classical dynamical systems where the reference state $\omega$ is assumed to be 
invariant under the dynamics. 

Assumption \C{} ensures that $\phi$ preserves $Z_\omega$ and hence naturally induces
a group of transformations of ${\cal O}(M)_\omega$ and $L^\infty(M, \d\omega)$.

Assumptions \Fone{}, \Ftwo{} and \C{} are our fundamental working hypothesis and will be 
assumed in the following without further notice.

\subsection{Time reversal invariance}
\label{TRIDefSect}

A time reversal of the dynamics $\phi$ on $M$ is an involutive measurable transformation 
$\vartheta: M\rightarrow M$ such that $\vartheta \circ\phi^t=\phi^{-t}\circ \vartheta$
for all $t\in\cal I$. 

A state $\omega\in{\cal S}$ is  called time reversal invariant (TRI) if 
$\omega\circ \vartheta=\omega$. In this case $\vartheta$ preserves $Z_\omega$
and induces an involution on $\cal O(M)_\omega$ and $L^\infty(M,\d\omega)$.
Note that if $\omega$ is TRI, then $\omega_t\circ \vartheta=\omega_{-t}$.

\label{TRI-def}
The dynamical system $(M, \phi, \omega)$ is called TRI if $M$ is equipped with a time 
reversal $\vartheta$ of $\phi$ such that $\omega$ is TRI. 

Time-reversal invariance will play a central role in our discussion. Other symmetries can have
important consequences on statistical properties of the dynamics, e.g. the conformally symplectic 
structure of some systems leads to symmetries in their Lyapunov
spectrum (see \cite{WL,MD}). Such symmetries, however, will not play a role in our work.

If the system $(M, \phi, \omega)$ is not TRI, for the purpose of model 
building the following construction is useful. Set 
\[
\widetilde M=M\times M, \qquad \widetilde \phi^t(x, y)=(\phi^t(x), \phi^{-t}(y)), \qquad 
\d\widetilde \omega = \d\omega \otimes \d\omega.
\]
Then $(\widetilde M, \widetilde \phi, \widetilde \omega)$ is TRI with the  time reversal 
$\vartheta(x, y)=(y, x)$. 

\section{Finite time entropy production}
\label{sec-finiteentropy}
\subsection{Entropy cocycle}
\label{subsec-entcocyc}

Since $\omega_t$ and $\omega$ are equivalent measures, 
\[
c^t=\ell_{\omega_t|\omega}\circ\phi^t\in{\cal O}(M)_\omega,
\label{Ecocycle-def}
\]
is well-defined. It satisfies the following additive cocycle property:

\bep For all $t, s \in {\cal I}$ one has 
\[c^{t+s}=c^s+c^t\circ\phi^s.\] In particular
$c^0=0$ and $c^{-t}=-c^t\circ\phi^{-t}$.
\label{cocy1}
\eep
\proof 
We adopt the shorthand $\Delta^t= \Delta_{\omega_t|\omega}$.
For $f\in B(M)$ and $s,t\in {\cal I}$ one has $\omega_{t+s}(f)=\omega(\Delta^{t+s}f)$
and 
\[
\omega_{t+s}(f)=\omega_s(f_t)=
\omega(\Delta^s\,f_t)
=\omega((\Delta^s\circ\phi^{-t}\,f)\circ\phi^t)=\omega_t(\Delta^s\circ\phi^{-t}\,f)
=\omega(\Delta^s\circ\phi^{-t}\,\Delta^t\,f). 
\]
Hence,
\begin{equation}
\Delta^{t+s} = \Delta^s\circ \phi^{-t}\,\Delta^t,
\label{JCocycle}
\end{equation}
where the equality is in ${\cal O}(M)_\omega$. Taking the logarithm we derive 
$$
\ell_{\omega_{t+s}|\omega}=\ell_{\omega_{s}|\omega}\circ\phi^{-t}
+\ell_{\omega_{t}|\omega}.
$$
Our first identity follows immediately. The second one follows from the
substitution $s=0$ and the third one is obtained by setting $s=-t$.
\qed

We shall call $c^t$ the {\sl entropy cocycle} of the dynamical system $(M,\phi,\omega)$.
The entropy cocycle of a \TRI{} dynamical system enjoys the following additional property.

\bep\label{TriEll}If the system $(M,\phi,\omega)$ is  {\rm\TRI{}}
with a time reversal $\vartheta$, then
\begin{equation}
c^t\circ \vartheta=c^{-t},
\label{second}
\end{equation}
holds for all $t\in\mathcal I$.
\eep
\proof
Setting again $\Delta^t= \Delta_{\omega_t|\omega}$ we have, for any $f\in B(M)$ and
$t\in\cal I$,
$$
\omega(\Delta^t\circ\vartheta\, f)=\omega\circ\vartheta(\Delta^t\, f\circ\vartheta)=
\omega(\Delta^t f\circ\vartheta)=\omega_t(f\circ\vartheta)=\omega(f\circ\vartheta\circ\phi^t)
=\omega(f\circ\phi^{-t}\circ\vartheta)=\omega(f\circ\phi^{-t})=\omega(\Delta^{-t}f).
$$
The resulting identity $\Delta^{-t}=\Delta^t\circ\vartheta$ further leads to
$\Delta^{-t}\circ\phi^{-t}=\Delta^t\circ\vartheta\circ\phi^{-t}=\Delta^t\circ\phi^t\circ\vartheta$.
Taking the logarithm gives the result.
\qed

\subsection{Entropy balance equation}
\label{subsec-entbal}

By Definition (\ref{rel-ent}) of the relative entropy one has
\[
\Ent(\omega_t|\omega)= -\omega_t(\ell_{\omega_t|\omega})
=-\omega(\ell_{\omega_t|\omega}\circ\phi^t)=-\omega(c^t).
\label{ent-balance}
\]
Since $\Ent(\omega|\omega)=0$ this identity can be rewritten as 
$$
\omega\left(\Sigma^t\right)=-\frac{1}{t}\left(
\Ent(\omega_t|\omega)-\Ent(\omega|\omega)
\right),
$$
where
\begin{equation}
\Sigma^t=\frac{c^t}t.
\label{SigmaDef}
\end{equation}
Thus, we can interpret $\Sigma^t$ as the observable of {\sl mean entropy production 
rate over the time interval $[0, t]$.} We shall call the relation
\begin{equation}
\Ent(\omega_t|\omega)=-t\,\omega(\Sigma^t),
\label{entbal}
\end{equation}
the {\sl entropy balance equation}. Its immediate consequence is the important
inequality
\begin{equation}
\omega(\Sigma^t)\in[0,\infty],
\label{ent-ba-equ}
\end{equation}
which holds for all $t>0$. 

The cocycle property yields
$$
\Sigma^t=-\frac{c^{-t}\circ\phi^t}t=\Sigma^{-t}\circ\phi^t.
$$
We note for later reference that if $(M, \phi, \omega)$ is \TRI{}, then Proposition
\ref{TriEll} further leads to
\begin{equation}
\Sigma^t\circ\vartheta=\frac{c^{-t}}t=-\Sigma^{-t}
=-\Sigma^{t}\circ\phi^{-t}.
\label{TriSigma}
\end{equation}

\subsection{Finite time Evans-Searles symmetry}
\label{FiniES}

Let $P_t$ be the law of the real-valued random variable $\Sigma_t$,
{\sl i.e.,} the Borel probability  measure on $\rr$ such that  for any 
$f\in B(\rr)$,  
\[
P_{t}(f)=\omega (f(\Sigma^t)).
\]
Let ${\mathfrak r}: \rr\rightarrow \rr$ be the reflection 
${\mathfrak r}(s)=-s$ and 
define the reflected measure $\bar P_t=P_t\circ{\mathfrak r}$.

\bep If $(M, \phi, \omega)$ is {\rm\TRI{}} then, for any $t\in\mathcal I$, 
the  measures $P_{t}$ and $\bar P_{t}$  are equivalent and 
\begin{equation}
\frac{\d\bar P_{t}}{\d P_{t}}(s)=\e^{-t s}.
\label{ES-rel}
\end{equation}
\label{es-thm}
\eep
\proof For  $f\in B(\rr)$, Equ. (\ref{TriSigma}) and the fact that 
$\omega_{t}\circ\vartheta=\omega_{-t}$ yield
\[
\bar P_{t}(f)= \omega(f(-\Sigma^t)) = \omega_t(f(-\Sigma^t\circ\phi^{-t}))=
\omega_t(f(\Sigma^t\circ\vartheta))=\omega_{-t}(f(\Sigma^t))
=\omega(\e^{-t\Sigma^t}f(\Sigma^t)),
\]
and the statement follows. \qed

To our knowledge, the relation (\ref{ES-rel}) was first obtained by Evans and Searles in
\cite{ES} and is sometimes called the {\sl transient fluctuation theorem.} We shall call it
{\em the finite time ES-identity}. We stress its universal character: besides the
\TRI{}\ assumption it only relies on the minimal hypothesis \Fone{}, \Ftwo{} and \C{}.
In a loose sense, it can be understood as a dynamical
form of the second law of thermodynamics: on the finite time interval $[0,t]$, the 
probability to observe a negative mean entropy production rate $-s$ is exponentially
small compared to the probability to observe the positive value $s$.

The ES-identity can be re-formulated in  terms of  R\'enyi entropy. For $\alpha\in\rr$, 
we adopt the shorthand
\begin{equation}
e_t(\alpha)=\Ent_\alpha(\omega_t|\omega)
=\log \omega (\e^{\alpha \ell_{\omega_t|\omega}})
=\log\omega(\e^{\alpha t\Sigma^{-t}}).
\label{renyi-e}
\end{equation}
By Theorem \ref{gen-ent} (2), if $\ell_{\omega_t|\omega}\in L^\infty(M, \d\omega)$, then 
\begin{equation}
 e_t(\alpha) =\sup_{\nu \in {\cal N}_\omega}[ \Ent (\nu|\omega) + \alpha \nu(\ell_{\omega_t|\omega})].
\label{cergy-var-1}
\end{equation}
This  variational characterization will play an important role  in the extension of the theory 
of entropic fluctuations to non-commutative dynamical systems. 

The basic properties of the functional (\ref{renyi-e}) follow directly from Proposition 
\ref{prop-renyi}. We list them for later reference:

\bep
\begin{enumerate}[{\rm(1)}]
\item For all $t\in{\cal I}$ the function
\[
\rr\ni\alpha\mapsto e_t(\alpha)\in]-\infty,\infty],
\]
is convex, satisfies $e_t(0)=e_t(1)=0$ and
$$
\left\{
\begin{array}{ll}
e_t(\alpha)\in]-\infty,0]&\text{if } \alpha \in [0,1],\\[3.5pt]
e_t(\alpha)\in[0,\infty]&\text{otherwise}.
\end{array}
\right.
$$
\item It satisfies the lower bound
\begin{equation}
e_t(\alpha)\geq\min\left(
\alpha\,\Ent(\omega|\omega_t),(1-\alpha)\Ent(\omega_t|\omega)
\right).
\label{not-possible1}
\end{equation}
\item It is real analytic on the interval $]0,1[$.
\end{enumerate}
\label{hang-over}
\eep
{\bf Remark.} The relation $e_t(1)=0$ is sometimes called the 
{\em non-equilibrium partition identity} or {\sl Kawasaki identity}, see 
\cite{CWW}.

Note that if the system is \TRI{}, Equ. (\ref{TriSigma}) implies
$\omega\bigl(\e^{\alpha t\Sigma^{-t}}\bigr)
=\omega\bigl(\e^{-\alpha t\Sigma^{t}\circ\vartheta}\bigr)
=\omega\bigl(\e^{-\alpha t\Sigma^{t}}\bigr)
$
so that
$$
e_t(\alpha)=\log \omega\left(\e^{-\alpha t\Sigma^t}\right).
$$

\bep\label{es-sym1}
\begin{enumerate}[{\rm (1)}]
\item For any $t\in{\cal I}$ and $\alpha \in \rr$ one has $e_t(\alpha)=e_{-t}(1-\alpha)$.
\item If $(M, \phi, \omega)$ is {\rm\TRI{}} then $e_{-t}(\alpha)=e_t(\alpha)$ and hence
\begin{equation}
 e_t(\alpha)=e_t(1-\alpha).
\label{es-sym}
\end{equation}
\end{enumerate}
\eep

\proof  Parts (1) and (5) of  Proposition \ref{prop-renyi} imply 
\[
e_t(\alpha)={\rm Ent}_{\alpha}(\omega_t|\omega)=
{\rm Ent}_{1-\alpha}(\omega|\omega_t)=
{\rm Ent}_{1-\alpha}(\omega_{-t}|\omega)=e_{-t}(1-\alpha).
\]
Since \TRI{} implies $\omega_{t}\circ \vartheta=\omega_{-t}$, Part (1) of  Proposition
\ref{prop-renyi} allows us to conclude
$$
e_{-t}(\alpha)=
{\rm Ent}_{\alpha}(\omega_{-t}|\omega)=
{\rm Ent}_{\alpha}(\omega_{t}\circ \vartheta|\omega\circ \vartheta)=
{\rm Ent}_{\alpha}(\omega_{t}|\omega)=e_t(\alpha).
$$
\qed

We shall call Relation (\ref{es-sym}) the {\sl finite time ES-symmetry}.
We finish this section with the observation that the finite
time ES-symmetry is an equivalent formulation of the finite time ES-identity.

\bep For each $t\in{\cal I}$, the following statements are equivalent:
\begin{enumerate}[{\rm (1)}]
\item The  measures $P_{t}$ and $\bar P_{t}$ are equivalent and satisfy the
ES-identity (\ref{ES-rel}).
\item For all $\alpha \in \rr$, $e_{-t}(\alpha)=e_{-t}(1-\alpha)$.
\end{enumerate}
\label{ESEquiv}
\eep
\proof It suffices to notice the relation between the functional $e_{-t}(\alpha)$
and the Laplace transform of the measure $P_t$. One has
$$
e_{-t}(\alpha)=\log\omega\left(\e^{-\alpha t\Sigma^t}\right)
=\log\int\e^{\alpha ts}\d\bar P_t(s),
$$
and hence
$$
e_{-t}(1-\alpha)=\log\int\e^{-(1-\alpha) ts}\d P_t(s)=
\log\int\e^{\alpha ts}\,\e^{-ts}\d P_t(s).
$$
\qed

\subsection{Entropy production observable}
\label{subsec-entobs}

For a discrete time dynamical system the cocycle property  
$c^{t+1}=c^t+c^1\circ\phi^t$ implies
\begin{equation}
c^t={\sum_{s=0}^{t-1}\sigma_{s}},
\label{logJFormuladiscrete}
\end{equation}
where 
\begin{equation}
\sigma=c^1=\ell_{\omega_1|\omega}\circ\phi.
\label{sigmadefdiscrete}
\end{equation}
In particular, we can express the mean entropy production rate observable as
$$
\Sigma^t=\frac{c^t}t=
\frac1t\sum_{s=0}^{t-1}\sigma_{s}.
$$
Consequently, the entropy balance equation (\ref{entbal}) becomes
$$
\Ent(\omega_t|\omega)=-\sum_{s=0}^{t-1}\omega(\sigma_{s}).
$$
We shall call $\sigma$ the {\em entropy production} observable. For obvious reasons 
the  entropy production observable is also  often called {\em phase space contraction rate.}

Basic properties of the entropy production observable are:

\bep $\omega(\sigma)\geq 0$, $\omega_{-1}(\sigma)\leq 0$, and 
if $(M, \phi, \omega)$ is {\rm\TRI{}} then $\sigma\circ \vartheta=-\sigma_{-1}$.
\label{bas-ep}
\eep
\proof $-\omega(\sigma)=-\omega_1(\ell_{\omega_1|\omega})
={\rm Ent}(\omega_1|\omega)\leq 0$ implies 
$\omega(\sigma)\geq 0$. 
Jensen's inequality 
\[\e^{\omega_{-1}(\sigma)}\leq\omega_{-1}(\e^{\sigma})=
\omega(\Delta_{\omega_1|\omega})=1,\] 
implies $\omega_{-1}(\sigma)\leq 0$.
The last statement follows from (\ref{second}) and the cocycle property
$c^{-1}=-c^1\circ\phi^{-1}$.
\qed

It is not possible to define the entropy production observable of a continuous time
dynamical system at the current level of generality. We shall make some
minimal regularity assumptions to ensure
that the entropy cocycle has a generator $\sigma$,
{\sl i.e.,} that the continuous time analog of Equ. (\ref{logJFormuladiscrete}) holds.
\begin{quote}
\noindent{\bf (E1)} 1. The  function 
$\rr\ni t\mapsto\Delta_{\omega_t|\omega}\in L^1(M,\d\omega)$
is strongly $C^1$.

\noindent\phantom{{\bf (E1)} }2. The entropy production observable 
\[
\sigma=\left.\frac{\d\ }{\d t}\Delta_{\omega_t|\omega}\,\right|_{t=0},
\label{r-sigma-def}
\]
\phantom{{\bf (E1)} 2. }is such that the function $\rr\ni t\mapsto\sigma_{t}\in L^1(M,\d\omega)$ is
strongly continuous.
\label{E1-def}
\end{quote}
{\bf Remark.} If $M$ is a complete, separable and metrizable space, then
the Koopman operators $T^t:f\mapsto\Delta_{\omega_t|\omega}f_{-t}$
form a strongly continuous group of isometries of $L^1(M,\d\omega)$.
Denote by $L$ its generator, {\sl i.e.,} $T^t=\e^{tL}$. Since
$\Delta_{\omega_t|\omega}=T^t1$, Part 1 of Assumption \Eone{} is equivalent to
$1\in{\rm Dom}(L)$, the domain of $L$, and then $\sigma=L1$.

\bep Suppose that  {\rm (E1)} holds. Then:
\begin{enumerate}[{\rm (1)}]
\item For all $x\in M$ the function $t\mapsto\Delta_{\omega_t|\omega}(x)$ is
absolutely continuous and
$$
\frac{\d\ }{\d t}\Delta_{\omega_t|\omega}(x)
=\Delta_{\omega_t|\omega}(x)\sigma_{-t}(x),
$$
holds for $\omega$-almost all $x\in M$ and Lebesgue almost all $t\in\rr$.
\item For all $t\in\rr$ the identities
$$
\Delta_{\omega_t|\omega}=\e^{\int_0^t\sigma_{-s}\,\d s},
$$
and
\begin{equation}
c^t=\int_0^t\sigma_{s}\,\d s,
\label{logJFomulacontin}
\end{equation}
hold in $\mathcal O(M)_\omega$.
\item $\omega(\sigma)=0$.
\item If $(M, \phi, \omega)$ is {\rm\TRI{}}, then $\sigma \circ \vartheta=-\sigma$. 
\end{enumerate}
\label{bas-ep-cont}
\eep
\proof We again set $\Delta^t=\Delta_{\omega_t|\omega}$. 

(1) The cocycle property (\ref{JCocycle})
and Part (1) of Assumption \Eone{} yield
$$
\omega\left(\left|\Delta^{t+s}-\Delta^t-s\Delta^t\sigma_{-t}\right|\right)=
\omega\left(\Delta^t\left|\Delta^s\circ\phi^{-t}-1-s\sigma_{-t}\right|\right)
=\omega\left(\left|\Delta^s-1-s\sigma\right|\right)=o(s),
$$
as $s\to0$, from which we conclude that
$$
\frac{\d\ }{\d t}\Delta^t
=\Delta^t\sigma_{-t},
$$
holds strongly in $L^1(M,\d\omega)$. By Part (1) of Assumption \Eone{}
we can assume that for any $x\in M$ the function $t\mapsto \Delta^t(x)$ is absolutely 
continuous and that for $\omega$-almost all $x\in M$
$$
\frac{\d\ }{\d t}\Delta^t(x)=\Delta^t(x)\sigma_{-t}(x),
$$
holds for Lebesgue almost all $t\in\rr$ (see e.g. Theorem 3.4.2 in \cite{HP}).

(2) By Part (2) of Assumption \Eone{}, the Riemann integral
$$
\ell_t=\int_0^s\sigma_{-s}\,\d s,
$$
defines a strongly $C^1$ function $t\mapsto\ell_t\in L^1(M,\d\omega)$. As before,
we can assume that $t\mapsto\ell_t(x)$ is absolutely continuous for all $x\in M$
and that for $\omega$-almost all $x\in M$
$$
\frac{\d\ }{\d t}\ell_t(x)=\sigma_{-t}(x),
$$
holds for Lebesgue almost all $t\in\rr$. Consequently, for all $x\in M$ the function
$t\mapsto F_t(x)=\Delta^t(x)\e^{-\ell_t(x)}$ is absolutely continuous and for $\omega$-almost
all $x\in M$
$$
\frac{\d\ }{\d t}F_t(x)=0,
$$
holds for Lebesgue almost all $t\in\rr$. We conclude that for all $t\in\rr$, 
$F_t(x)=F_0(x)=1$, {\sl i.e.,} $\Delta^t(x)=\e^{\ell_t(x)}$
for $\omega$-almost all $x\in M$. Equ. (\ref{logJFomulacontin}) follows immediately.

(3) 
Differentiating the identity $e_t(1)=0$  w.r.t. $t$ at $t=0$ we derive $\omega(\sigma)=0$. 

(4) Follows from the identity (\ref{second}).
\qed

{\bf Remark.} Under the strong continuity condition of Assumption \Eone{}, the identity 
(\ref{logJFomulacontin}) holds in $\mathcal O(M)_\omega$ with a Lebesgue integral.
 It also holds in $L^1(M,\d\omega)$ with a strong Riemann integral.

\bigskip
The relation (\ref{logJFomulacontin}) yields
$$
\Sigma_t=\frac1t\int_0^t\sigma_{s}\,\d s,
$$
and  so the entropy balance equation and the ES-functional can be written as
$$
\Ent(\omega_t|\omega)=-\int_0^t\omega(\sigma_{s})\,\d s,
$$
$$
e_t(\alpha)=\log \omega\left(\e^{\alpha\int_0^t\sigma_{-s}\,\d s}\right).
$$
For \TRI{} systems one has $\sigma_{-s}\circ\vartheta=-\sigma_s$ and in this case
$$
e_t(\alpha)=\log \omega\left(\e^{-\alpha\int_0^t\sigma_{s}\,\d s}\right).
$$

Unless otherwise stated, in the sequel we will only consider continuous time dynamical
systems. The discrete time case is very similar, time-integrals being replaced by
appropriate sums.

\subsection{$L^p$-Liouvilleans}
\label{lp-liou}

In this section, in order to avoid unessential technicalities,
we shall assume in addition to \Eone{}, 
\begin{quote}
\label{E2-def}
{\bf (E2)} $\sigma\in L^\infty(M,\d\omega)$.
\end{quote}
Let $p\in ]-\infty, \infty]$, $p\not=0$,  and  $f \in L^\infty(M, \d\omega)$. We shall consider
the following special class of Ruelle transfer operators 
\begin{equation}
 U_p(t) f=\Delta_{\omega_t|\omega}^{\frac{1}{p}}f_{-t}
=\e^{\frac{1}{p}\int_0^t \sigma_{-s}\d s}f_{-t}.
\label{ruelle1}
\end{equation}
One easily shows:
\bep\label{lp-li}Under Assumptions {\rm \Eone{}-\Etwo{}}, Equ. (\ref{ruelle1})  defines a 
family of bounded linear operators on $L^\infty(M, \d\omega)$ which satisfies:
\begin{enumerate}[{\rm (1)}]
\item $U_p(0)= I$ and $U_p(t+s)=U_p(t)U_p(s)$.
\item If $p^{-1} + q^{-1}=1$, then $\omega([U_p(t)f] [U_q(t)g])=\omega(fg)$.
\item $\omega(|U_p(t)f |^p)=\omega( |f|^p)$. For $p\in[1,\infty]$, $U_p(t)$ extends to a 
group of isometries of $L^p(M, \d\omega)$.
\item $U_p(t)$ extends to a group of bounded operators on $L^2(M, \d\omega)$ such 
that $U_p^\ast(t)=U_q(-t)$ and 
\[
\|U_p(t)\|\leq \e^{|t|m_p},\qquad m_p=\frac{|2-p|}{|p|} \,\|\sigma\|_\infty.
\]
\item Suppose that $U_p(t)$ is strongly continuous on $L^2(M, \d\omega)$ and let  
$L_p$ be its  generator, $U_p(t)=\e^{t L_p}$. Then $L_p^\ast = -L_q$, 
${\rm sp}(L_p)\subset\{ z\,|\, |{\rm Re} z|\leq m_p\}$, 
${\rm Dom}(L_p)= {\rm Dom}(L_\infty)$, and for $f\in \Dom (L_p)$
\[
L_p f = L_\infty f + \frac{\sigma}{p} f.
\]
\end{enumerate}
\eep

We shall call the operator $L_p$ {\em the $L^p$-Liouvillean.} In Part (5), 
${\rm Dom}(A)$ denotes the domain of the operator $A$ and $\sp(A)$ its spectrum.

If $\alpha=1/p$, then
\begin{equation}
e_t(\alpha) = \log 
(1, \e^{  t L_p} 1) =\log \int_M \e^{tL_p}1\, \d\omega.
\label{cergy-var-2}
\end{equation}
Similarly to (\ref{cergy-var-1}), this operator characterization  of 
$e_t(\alpha)$ will play an important role  in the extension of the theory of entropic
fluctuations to the non-commutative setting.

If $\sigma$ is unbounded, the operator $U_p(t)$ could  be unbounded and 
Proposition \ref{lp-li} may not hold as formulated. The technical aspects of its extension 
are then best carried out in the context of concrete models.

In the discrete time case the concept of $L^p$-Liouvillean is not very natural and
instead one deals directly with the transfer operator 
\begin{equation}
U_p f = \e^{\frac{1}{p}\sigma}f_{-1}.
\label{disc-transf}
\end{equation}
If $\sigma\in L^\infty(M,\d\omega)$, then Parts
(1)--(4) of Proposition \ref{lp-li} obviously hold, $U_p(n)=U_p^n$ and 
$e_n(\alpha)=\log (1, U_p^n 1)$.

\subsection{Examples: Differentiable dynamics and thermostated systems}
\label{Differentiable-dynamics}

Let $U \subset \rr^n$ be an open  connected set and let $\phi:U \rightarrow U$ be a 
$C^1$-diffeomorphism. Denote by $D\phi$ its derivative.
Let $M\subset U$ be compact  and suppose
that  $\phi(M)\subset M$. Finally denote by $\omega$ the normalized Lebesgue measure
on $M$. The entropy production observable of the discrete time dynamical system 
$(M, \phi, \omega)$ is given by
\[
\sigma=-\log |\det D\phi|\,\bigr|_M.
\]

To describe the continuous time case let $X$ be a $C^1$-vector field on $U$.
Assume that the flow $\phi$ generated by the differential equation
\[
\frac{\d\ }{\d t} x_t = X(x_t),
\]
satisfies $\phi^t(M)\subset M$.  Then $\phi$ is $C^1$ on $\rr\times M$. The entropy
cocycle of  $(M, \phi, \omega)$ is $c^t=-\log\left|\det D\phi^t\right|\bigr|_M$ and the 
entropy production observable  is given by
\[
 \sigma=-\left.\frac{\d\ }{\d t}\log |\det D\phi^t|\right|_{M,t=0}
=-\mathrm{div}\,X\bigr|_M.
\]
Assumptions \Eone{}-\Etwo{} are clearly satisfied.

A special class  of differentiable dynamics is provided by the so called 
Gaussian thermostated systems. Consider a Hamiltonian system with $n$ degrees of 
freedom. The  phase space of the system is $\rr^{n}\oplus\rr^n$
(or more generally the cotangent bundle of a smooth manifold).
For simplicity, we shall assume that its Hamiltonian $H$ is $C^2$
and that the finite energy subsets $\{(p,q)\,|\,H(q, p)\le E\}$ are compact.
These assumptions ensure that the equations of motion
\[
\dot p_t=-\nabla_q H(p_t, q_t)\,,\qquad 
\dot  q_t=\nabla_p H(p_t, q_t),
\label{hamiltonian}
\]
define a global $C^1$ Hamiltonian flow $\phi_H^t(p, q)= (p_t, q_t)$
which preserves the energy, $H\circ \phi_H^t= H$,  and 
Lebesgue measure on $\rr^{n}\oplus\rr^n$ (Liouville's theorem).

To drive this system out of equilibrium, an external non-Hamiltonian force $F(q)$ is
applied. To prevent it from heating up, the energy supplied by this force is removed by 
a thermostat (the so-called Gaussian thermostat). This leads to the modified
equations of motion
\[
\dot p_t=-\nabla_q H(p_t, q_t)+F(q_t)-\Theta(q_t,p_t),\qquad
\dot  q_t=\nabla_p H(p_t, q_t),
\label{thermostat}
\]
where the thermostating force is given by
\[
-\Theta(p,q)=
-\frac{ F(q)\cdot{\nabla_p}H(p,q)}{|{\nabla_p}H(p,q)|^2}{\nabla_p}H(p,q)\,.
\]
One easily checks that the flow $\phi$ generated by this system satisfies $H\circ \phi^t= H$
and therefore  preserves the total energy. This flow, however,  does not preserve Lebesgue 
measure and the entropy production observable
\[
\sigma =\nabla_p\cdot\Theta(q,p),
\]
measures the local rate of phase space contraction.

Fixing $E\in\Ran H$ we see that  Gaussian thermostated systems are 
special cases of differentiable dynamics with 
\[
M=\{(p,q)\in\rr^{n}\oplus\rr^n\,|\,H(p, q)=E\}.
\]

Two other well-known thermostating mechanisms are the isokinetic
and Nos\'e-Hoover thermostats. Models using these thermostats have
been constructed to describe various phenomena like shear flows
\cite{ECM,CL}, heat conduction \cite{HHP,PH},  and turbulent fluids
\cite{Ga3,GRS}.  They all have in common that the dynamics is
described by a deterministic finite-dimensional dynamical system on a 
compact manifold and are very convenient for numerical  studies \cite{Ho,EM,Do}.

A well-known  model in this class is  a Sinai billiard with an external
electric field \cite{CELS1,CELS2,Ch1,Ch2,Yo,RY}. 
General mathematical results concerning thermostated Hamiltonian models 
can be found in \cite{GC1,GC2,Ga1,Ga2,Ru2,Ru3,Ru4,Ru5}.

\subsubsection{A micro-canonical ideal gas out of equilibrium.}
\label{mucangasfinitet}

In this section we consider an exactly solvable thermostated system -- a gas of 
$N>1$ identical, non-interacting particles moving on a circle. The phase space is 
$\rr^N\times\mathbb T^N$ equipped with Lebesgue measure and the Hamiltonian is
$$
H(L,\theta)=\frac12|L|^2.
$$
The flow $\phi$ is generated
by the system
\begin{equation}
\left.
\begin{array}{l}
\dot L_j=F-\lambda(L) L_j,\\[10pt]
\dot\theta_j=L_j,
\end{array}
\right\}\qquad (j=1,\ldots,N).
\label{mucanflow}
\end{equation}
Here, $F\in\rr$ denotes the constant strength of the external forcing and $-\lambda(L) L_j$ 
is the thermostating force,
\begin{equation}
\lambda(L)=F\frac{\ell}{u},\qquad \ell=\frac1N\sum_{k=1}^NL_k,
\quad u=\frac1N\sum_{k=1}^NL_k^2.
\label{lambdaEqu}
\end{equation}
The mean kinetic energy per particle $u$ is constant under the flow $\phi$ and we consider
the dynamical system $(M,\phi,\omega)$ where
$$
M=\{(L,\theta)\,|\, u=\epsilon\}\simeq S^{N-1}\times\mathbb T^N,
$$
for some $\epsilon >0$ and $\omega$ is the normalized micro-canonical measure
$$
\omega(f)=\frac1Z\int_{\rr^N\times\mathbb T^N}f(L,\theta)
\delta\left(u-\epsilon\right)\prod_{j=1}^N\d L_j\d\theta_j.
$$

The Cauchy-Schwarz inequality ensures that $\ell^2\le u$ on $M$ and the observable
$$
\xi=-\frac12\log\left(\frac{\sqrt u-\ell}{\sqrt u+\ell}\right),
$$
is well defined.
One derives $\dot\xi=\mu$ where $\mu=F\epsilon^{-1/2}$ is a constant. It follows that
$\ell=\sqrt\epsilon\,\mathrm{th}\,\xi$ which, once inserted in (\ref{lambdaEqu}), allows
to integrate the equations of motion (\ref{mucanflow}) to obtain
\begin{align}
\theta_{jt}&=\theta_{j0}+\frac{\sqrt{\epsilon}}{F}
(L_{j0}\,\mathrm{ch}\,\xi_0-\sqrt{\epsilon}\,\mathrm{sh}\,\xi_0)
(\mathrm{arctg}(\mathrm{sh}\,\xi_t)-\mathrm{arctg}(\mathrm{sh}\,\xi_0))
+\frac \epsilon F\log\frac{\mathrm{ch}\,\xi_0}{\mathrm{ch}\,\xi_t},
\label{gasdynone}\\
L_{jt}&=
\frac{L_{j0}\,\mathrm{ch}\,\xi_0-\sqrt{\epsilon}\,\mathrm{sh}\,\xi_0}{\mathrm{ch}\,\xi_t}
+\sqrt{\epsilon}\,\mathrm{th}\,\xi_t,\label{gasdyntwo}
\end{align}
where $\xi_t=\xi_0+\mu t$. 
The entropy production observable is
\begin{equation}
\sigma=(N-1)\frac F{\sqrt{\epsilon}}\,\mathrm{th}\,\xi.
\label{sigmagas}
\end{equation}
Assumptions \Eone{}-\Etwo{} are satisfied in this model. Moreover,
the map $\vartheta(L,\theta)=(-L,\theta)$ is a time reversal for the flow $\phi$ and 
the measure $\omega$ is \TRI{}. 

One easily computes
$$
\exp\left(-\alpha\int_0^t\sigma_s\,\d s\right)
=\left(\frac{\mathrm{ch}\,\xi_0}{\mathrm{ch}\,\xi_t}\right)^{(N-1)\alpha}.
$$
Since the distribution of $\xi$ induced by the measure $\omega$ is
$$
\omega(f(\xi))=\frac{\Gamma(N/2)}{\sqrt\pi\Gamma((N-1)/2)}
\int_{-\infty}^\infty f(\xi)(\mathrm{ch}\,\xi)^{-(N-1)}\,\d\xi,
$$
we conclude that
\begin{equation}
e_t(\alpha)=\log\left(\frac{\Gamma(N/2)}{\sqrt\pi\Gamma((N-1)/2)}
\int_{-\infty}^\infty (\mathrm{ch}\,\xi)^{-(N-1)(1-\alpha)}
(\mathrm{ch}(\xi+\mu t))^{-(N-1)\alpha}\,\d\xi
\right).
\label{gasESfunction}
\end{equation}
The validity of the finite time ES-symmetry, ensured by Proposition \ref{es-sym1},
can be explicitly checked by noticing that
\begin{align*}
\int_{-\infty}^\infty (\mathrm{ch}\,\xi)^{-(N-1)(1-\alpha)}
(\mathrm{ch}(\xi+\mu t))^{-(N-1)\alpha}\,\d\xi
&=\int_{-\infty}^\infty (\mathrm{ch}(\xi-\mu t))^{-(N-1)(1-\alpha)}
(\mathrm{ch}\,\xi)^{-(N-1)\alpha}\,\d\xi\\
&=\int_{-\infty}^\infty (\mathrm{ch}(-\xi-\mu t))^{-(N-1)(1-\alpha)}
(\mathrm{ch}(-\xi))^{-(N-1)\alpha}\,\d\xi\\
&=\int_{-\infty}^\infty (\mathrm{ch}(\xi+\mu t))^{-(N-1)(1-\alpha)}
(\mathrm{ch}\,\xi)^{-(N-1)\alpha}\,\d\xi.
\end{align*}
This example continues in Sections \ref{gasprom} and \ref{gasness}.

\section{Thermodynamics}
\label{sec-control}

\subsection{Basic notions}
\label{sec-cont-basic}

Suppose that our dynamical system $(M, \phi_X, \omega_X)$ depends on some 
control parameters $X=(X_1, \cdots, X_N)\in\rr^N$. One can think 
of the $X_j$'s as mechanical or thermodynamical forces (affinities in the language of
non-equilibrium thermodynamics) acting on the system. 
When dealing with such families of systems,  we shall always  assume that 
\Fone{}-\Ftwo{}, \C{} and \Eone{} hold for each system $(M, \phi_X, \omega_X)$. 
The  entropy production observable of $(M, \phi_X, \omega_X)$ is denoted 
$\sigma_X$. We shall also assume:
\begin{quote}
\label{T1-def}
{\bf (T1)}   $\omega_0$ is $\phi_0^t$ invariant. 
\end{quote}
We will  write $\phi^t=\phi_0^t$, $\omega=\omega_0$, $\omega_{0t}=\omega_t$, etc, 
and  refer to the value $X=0$ as {\em equilibrium}. Under assumption \Tone{} the entropy 
cocycle satisfies $c^t=c_0^t=0$ for all $t\in\rr$ and 
consequently $\sigma=\sigma_0=0$.

\begin{definition} We call a family of vector-valued real observables 
${\bf \Phi}_X= (\Phi_X^{(1)}, \cdots, \Phi_X^{(N)})$, $X\in \rr^N$,
a flux relation if, for  all $X$,
\begin{equation}
\sigma_X=X\cdot{\bf\Phi}_X
=\sum_{j=1}^N X_j \Phi_X^{(j)}.
\label{ent-flux}
\end{equation}
\end{definition}
In what follows our discussion of thermodynamics concerns a family of quadruples  
$(M, \phi_X, \omega_X, {\bf \Phi}_X)$, where  ${\bf \Phi}_X$ is  a given flux relation. 
In concrete models arising in physics, physical requirements typically select  a unique 
flux relation ${\bf \Phi}_X$ 
(see Section \ref{sec-open-sys} for  an example). We will refer  to $\Phi_X^{(j)}$ as 
the flux (or current) observable associated  to the force $X_j$.   
Since  $\sigma_0=0$, if the map $X \mapsto \sigma_X$ is smooth we can always 
pick the  fluxes as 
$$
{\bf \Phi}_X=\int_0^1 \nabla \sigma_Y|_{Y=uX}\d u.
$$

{\bf Remark.} To simplify the notation,  unless otherwise stated we shall always assume that
$(M, \phi_X, \omega_X, {\bf \Phi}_X)$ is defined for all $X\in\rr^N$.
In concrete situations (e.g. like in the class of examples introduced in Section 
\ref{sec-open-sys}), the systems may only be defined on a restricted range of
the physical parameters $X_1,\ldots,X_N$. This causes no difficulties--one can either trivially
extend the range of parameters to all of $\rr^N$ or indicate in each statement the range of 
parameters to which they apply. 

\bigskip
Our second general assumption concerns time reversal.
\begin{quote}
\label{T2-def}
{\bf (T2)} The dynamical systems  $(M, \phi_X, \omega_X)$ are time-reversal 
invariant and 
\[
{\bf\Phi}_X\circ \vartheta_X=- {\bf\Phi}_X.
\] 
\end{quote}
This assumption implies that  $\omega_X({\bf\Phi}_X)=0$ for all $X$. 

\subsection{Finite  time Generalized Evans-Searles symmetry}
Let
\[{\bf\Sigma}_{X}^t =\frac{1}{t}\int_0^t {\bf \Phi}_{Xs}\d s
=\left(\frac{1}{t}\int_0^t \Phi_{Xs}^{(1)}\d s, \ldots,
\frac{1}{t}\int_0^t \Phi_{Xs}^{(N)}\d s\right), 
\label{BSigma-def}
\]
where ${\bf \Phi}_{Xs}={\bf \Phi}_X \circ \phi_X^s$,  
$\Phi_{Xs}^{(j)}=\Phi_X^{(j)}\circ \phi_X^s$.
 The entropy cocycle can be written as
\[
c_X^t=t\,X\cdot{\bf\Sigma}_{X}^t.
\]

Let  $P^t_X$ be the law of  ${\bf\Sigma}_{X}^t$, {\sl i.e.,}  
the Borel probability measure on $\rr^N$ such that
$P_{X}^t(f)=\omega_X(f({\bf\Sigma}_{X}^t))$
for any $f\in B(\rr^N)$.
Let ${\mathfrak r}:\rr^N \rightarrow \rr^N$ be the reflection
$\mathfrak r(s)=-s$ and $\bar P_{X}^t= P_{X}^t \circ {\mathfrak r}$. 

\bep If Assumptions {\rm \Tone{}-\Ttwo{}} hold, then
for any $t\in\rr$ the measures $P_{X}^t$ and $\bar P_{X}^t$ are equivalent and 
\begin{equation}
\frac{\d\bar P_{X}^t}{\d P_{X}^t}(s)=\e^{-t X\cdot s}.
\label{ges-univ}
\end{equation}
\label{GESprop}
\eep
The proof of this proposition is very similar to the proof of Proposition \ref{es-thm} and 
we will omit it.

We shall call the universal relation (\ref{ges-univ}) the {\sl finite time generalized 
Evans-Searles (GES) identity.} As for the finite time ES identity, one can reformulate
(\ref{ges-univ})  in terms of the Laplace transform of $P_{X}^t$. To this end, consider
the functional
\begin{equation}
g_{t}(X, Y) = \log \omega_X\left(\e^{-Y\cdot \int_0^t {\bf \Phi}_{Xs}\d s}\right)
=\log\int\e^{-tY\cdot s}\,\d P_{X}^t(s).
\label{GESfunc}
\end{equation}
One easily sees that it inherits many properties of the R\'enyi entropy
$e_t(\alpha)$.  For fixed $X$ it is a convex function of $Y\in\rr^N$ which satisfies
$g_t(X,0)=g_t(X,X)=0$. The lower bounds
$$
g_t(X,Y)\ge\left\{
\begin{array}{l}
- t\, Y\cdot\omega_X({\bf\Sigma}_{X}^t),\\[3.5pt]
- t\, (X-Y)\cdot\omega_X({\bf\Sigma}_{X}^t),
\end{array}
\right.
$$
hold, and in particular $g_t(X,Y)>-\infty$.
Most importantly, Proposition \ref{GESprop} is equivalent to the {\sl finite time 
GES-symmetry} expressed by the following proposition.

\bep Under Assumptions {\rm \Tone{}-\Ttwo{}} one has
\begin{equation}
g_t(X, Y)=g_t(X, X-Y),
\label{ges-symmetry}
\end{equation}
for any $X, Y\in \rr^N$ and any $t\in\rr$.
\label{GESEquiv}
\eep
We again omit the proof which follows the same lines as the proof of
Proposition \ref{ESEquiv}.

For later applications we recall the following elementary result (explicit in \cite{LS2} and 
implicit in \cite{Ga1}) which we shall  call the {\sl symmetry lemma.}  We say that a 
function $a(X,Y)$ is $C^{1,2}$ in an open set $O\subset \rr^N \times \rr^N$ if all the 
partial derivatives $\partial_{X_i} a$,  $\partial_{Y_i} a$,  $\partial_{Y_i}\partial_{Y_j}a$,
$\partial_{X_j}\partial_{Y_i}a$ and $\partial_{Y_i}\partial_{X_j}a$ exist and are continuous in  $O$. 

\bel  
Let the function $a(X,Y)$ be $C^{1,2}$  in a neighborhood of
$(0,0)\in\rr^N\times\rr^N$ and such that  
\[
a(X, Y)=a(X, X-Y).
\]
Then 
\[
\partial_{X_k}\partial_{Y_j} a(X, Y)|_{X=Y=0}
=-\frac{1}{2}\partial_{Y_j}\partial_{Y_k} a(X, Y)|_{X=Y=0}.
\]
\label{sym-lemma}
\eel
\proof The identity
\[
\partial_{X_k}a(X,Y)|_{X=0}=\partial_{X_k}a(X,X-Y)|_{X=0}
=(\partial_{X_k}a)(0,-Y)+(\partial_{Y_k}a)(0,-Y), 
\]
leads to
\[
\partial_{Y_j}\partial_{X_k}a(X,Y)|_{X=Y=0}=
-\partial_{Y_j }\partial_{X_k}a(X,Y)|_{X=Y=0}
-\partial_{Y_j }\partial_{Y_k}a(X,Y)|_{X=Y=0}.
\]
The equality of mixed partial derivatives $\partial_{Y_j }\partial_{X_k}a
=\partial_{X_k}\partial_{Y_j}a$ implies the statement.
\qed

\subsection{Finite time linear response theory}
\label{sec-finite-time}

For any real or vector valued observable $f$ we set
\begin{equation*}
\langle f\rangle_t=\frac1t\int_0^t\omega(f_s)\,\d s.
\label{T-Av-def}
\end{equation*}
Finite time linear response theory is concerned with the first order perturbation 
theory w.r.t.  $X$ of  $\langle{\bf\Phi}_X\rangle_t$.
Hence, in addition to \Tone{}-\Ttwo{} we assume: 
\begin{quote}
\label{T3-def}
{\bf (T3)} The function $X\mapsto \langle{\bf\Phi}_X\rangle_t$ is differentiable at 
$X=0$ for all $t$.
\end{quote}
The finite time kinetic transport coefficients are defined by 
\[
L_{jkt} =\partial_{X_k}\langle \Phi_X^{(j)}\rangle_t\big|_{X=0}.
\label{FTL-def}
\]
Since 
\begin{equation}
\langle \sigma_X\rangle_t =X\cdot\langle{\bf\Phi}_X\rangle_t =
\sum_{j,k}X_jX_kL_{jkt} + o(|X|^2)\geq 0,
\label{ent-kin}
\end{equation}
the {\em real} quadratic form determined by $[L_{jkt}]$ is positive semi-definite. 
This fact does not depend on the TRI assumption \Ttwo{} and  {\em does not} imply 
that $L_{jkt}=L_{kjt}$.   We shall call the relations 
\[
L_{jkt}=L_{kjt},
\]
the {\sl finite time Onsager reciprocity relations} (ORR).
As general structural relations, they can hold only for \TRI{} systems.

Another immediate consequence of Equ. (\ref{ent-kin}) is:
\bep Suppose that {\rm\Tone{}} holds and let ${\bf\Phi}_X, \widetilde {\bf\Phi}_X$ be 
two flux relations satisfying {\rm\Tthree{}}. Then the corresponding finite time transport 
coefficients satisfy
\[
L_{jkt} + L_{kjt} = \widetilde L_{jkt} +\widetilde L_{kjt}.
\]
If the finite time ORR hold, then $L_{jkt}=\widetilde L_{jkt}$.
\eep

In the next proposition we  shall show that the finite time ORR follow from the finite 
time  GES-symmetry establishing along the way the finite time Green-Kubo formula.

\bep  Suppose that {\rm\Tone{}-\Ttwo{}} hold and that the function  $g_t(X, Y)$ 
is $C^{1,2}$ in a neighborhood of $(0, 0)$. Then {\rm\Tthree{}} holds and:
\begin{enumerate}[{\rm (1)}]
\item The finite time Green-Kubo formula holds, 
\begin{equation}
L_{jkt}= \frac{1}{2}\int_{-t}^t \omega(\Phi^{(k)}\Phi_s^{(j)})
\left(1 - \frac{|s|}{t}\right)\d s.
\label{finite-gk}
\end{equation}
\item The finite time Onsager reciprocity relations hold,
\begin{equation}
L_{jkt}=L_{kjt}.
\label{finite-orr}
\end{equation}
\end{enumerate}
\label{fin-lin-resp}
\eep

\proof  From the definition (\ref{GESfunc}) we derive
$\partial_{Y_j}g_t(X, Y)\big|_{Y=0}=-t\langle \Phi_X^{(j)}\rangle_t$,
and hence
\[
L_{jkt}=\partial_{X_k}\langle \Phi_X^{(j)}\rangle_t\big|_{X=0}=-\frac{1}{t}
\partial_{X_k}\partial_{Y_j}g_t(X, Y)\big|_{X=Y=0}.
\]
The  finite time GES-symmetry and the symmetry lemma yield 
\[
L_{jkt}=\frac{1}{2t}\partial_{Y_j}\partial_{Y_k}g_t(X, Y)|_{X=Y=0}
=\frac{1}{2t}\int_0^t\int_0^t \omega(\Phi^{(k)}_{s_1}\Phi^{(j)}_{s_2})\d s_1\d s_2.
\]
Assertion (2) follows from the equality of mixed derivatives 
$\partial_{Y_j}\partial_{Y_k}g_t=\partial_{Y_k}\partial_{Y_j}g_t$.
Since $\omega$ is invariant, we further get
$$
L_{jkt}
=\frac{1}{2t}\int_0^t\int_0^t \omega(\Phi^{(k)}\Phi^{(j)}_{s_2-s_1})
    \d s_1\d s_2
=\frac{1}{2}\int_{-t}^t \omega(\Phi^{(k)}\Phi_s^{(j)})\left(1 - \frac{|s|}{t}\right)\d s,
$$
which proves  Assertion (1).\qed

We finish this section with two remarks. 

{\bf Remark 1.} The identity 
$(\Phi^{(k)}\Phi^{(j)}_{s}) \circ \vartheta = \Phi^{(k)}\Phi^{(j)}_{-s}$ implies that 
\[ 
L_{jkt} = \int_{0}^t \omega(\Phi^{(k)}\Phi_s^{(j)})\left(1 - \frac{s}{t}\right)\d s. 
\]

{\bf Remark 2.} The covariance matrix  ${\bf D}_t=[D_{jkt}]$ of the vector-valued random
variable 
\[
\frac{1}{\sqrt t}\int_0^t{\bf \Phi}_s\d s,
\]
with respect to  $\omega$ is  
\begin{equation}
D_{jkt}=\int_{-t}^t  \omega(\Phi^{(k)}\Phi_s^{(j)})\left(1 - \frac{|s|}{t}\right)\d s.
\label{fin-clt}
\end{equation}
The time-reversal plays no role in (\ref{fin-clt}). However, if the assumptions of Proposition 
\ref{fin-lin-resp} hold and ${\bf L}_t=[L_{jkt}]$, then obviously 
\[
{\bf D}_t = 2{\bf L}_t.
\]
These are the {\sl finite time Einstein relations} which link the finite time covariance of
fluxes in equilibrium to the finite time kinetic transport coefficients. Together with 
Proposition \ref{fin-lin-resp} they constitute the {\em finite time Fluctuation-Dissipation 
Theorem.} We shall return to this topic at the end of  Section \ref{sec-no-end}. 
%
%

\subsection{Example: Thermally driven open systems}
\label{sec-open-sys}

We consider a system $S$, with phase space 
$M_{S}=\rr^{n_{S}}\oplus\rr^{n_{S}}$ and Hamiltonian 
$H_{S}(p_{S},q_{S})$, coupled to $N$ heat reservoirs 
$R_1,\ldots,R_N$. 
The phase space and the Hamiltonian of the $j$-th reservoir are 
$M_{j}=\rr^{n_j}\oplus\rr^{n_j}$ and $H_j(p_j,q_j)$.
The phase space and the Hamiltonian of the composite system are
$$
M=M_{S}\oplus M_1\oplus\cdots\oplus M_N,\qquad
H_0(p,q)=H_{S}(p_{S},q_{S})+H_1(p_1,q_1)+\cdots
+H_N(p_N,q_N),
$$
and we denote by $m$ the Lebesgue measure on $M$. 

The coupling between the system $S$ and the 
$j$-th reservoir is described by the Hamiltonian 
$V_j(p_{S},p_j,q_{S},q_j)$. 
The full Hamiltonian is
$$
H(p,q)=H_0(p,q)+ V(p,q)
=H_0(p,q)+\sum_{j=1}^NV_j(p_{S},p_j,q_{S},q_j).
$$
We assume that  $H$ is $C^2$
and that the finite energy subsets $\{(p,q)\,|\,H(q, p)\le E\}$ are compact.
These assumptions ensure that $H$ generates a  global Hamiltonian flow $\phi^t$ of 
class $C^1$ on $M$. For any $C^1$ observable  $F$,
\[ 
\frac{\d F_t}{\d t}=\{H, F\}_t,
\]
where $\{\,\cdot\,, \,\cdot\,\}$ denotes the Poisson bracket, 
$\{F,G\}=\nabla_q G\cdot\nabla_p F -\nabla_p G\cdot\nabla_q F$.

The state of the combined system in which each reservoir is at thermal equilibrium 
at inverse temperature $\beta_j$ and the system $S$ at inverse temperature
$\beta$ is the product measure
$$
\frac{1}{Z}\e^{-\beta H_{S}-\sum_{j=1}^N\beta_j H_j}\,m.
$$
Introducing the control parameters $X_j =\beta-\beta_j$, we can rewrite it as
\begin{equation}
\omega_X=\frac{1}{Z}\e^{-\beta H_0+\sum_{j=1}^N X_j H_j}\,m.
\label{wrongchoice}
\end{equation}
The dynamics does not depend on $X$ and we set $\phi_X^t=\phi^t$. Note  that 
$\omega_0$ is not invariant under the flow $\phi^t$. In order to satisfy hypothesis \Tone{} 
one modifies (\ref{wrongchoice}) as
\begin{equation}
\omega_X=\frac{1}{ Z}\e^{-\beta(H_{S}+V)-\sum_{j=1}^N\beta_j H_j}\,m=
\frac{1}{Z}\e^{-\beta H + \sum_{j=1}^N X_j H_j}\,m.
\label{goodchoice}
\end{equation}
With this definition, $\omega_0$ is the Gibbs canonical ensemble at
inverse temperature $\beta$ and is invariant under $\phi^t$.
Moreover, if the reservoirs have a large spatial extension and
the coupling Hamiltonians $V_j$ are well localized, the states
(\ref{wrongchoice}) and (\ref{goodchoice}) describe the same
thermodynamics. 

For the reference state (\ref{goodchoice}) the entropy cocycle is given by
\[
c_X^t=-\sum_{j=1}^N X_j(H_{jt} - H_j)
=-\sum_{j=1}^N X_j\int_0^t\frac{\d\ }{\d s}H_{js}\,\d s,
\]
and we have the flux relation
\[
\sigma_X=-\sum_{j=1}^N X_j\{H, H_j\}=\sum_{j=1}^N X_j\{H_j, V\}.
\]
The flux observables do not depend on $X$ and are given  by
$\Phi^{(j)}=\{H_j,  V\}$.  Since 
\[
H_{jt}- H_j =- \int_0^t \Phi_s^{(j)}\d s,
\]
the flux observable $\Phi^{(j)}$ describes the flow of energy out of  the $j$-th reservoir.
The time reversal in physical systems is usually given by the map $\vartheta(p, q)=(-p,q)$
and the system $(M,\phi_X,\omega_X)$ is then \TRI{} provided $H\circ\vartheta=H$.

We shall investigate a simple example of thermally driven open system in the remaining 
part of this section, to be continued in Sections \ref{harmonicbig} and
\ref{chainNESS}.

\subsubsection{The harmonic chain}
\label{subsection-chains}
Quadratic Hamiltonians  provide instructive examples of open systems  whose 
non-equilibrium characteristics can be computed in a closed from \cite{LS1}. 
From the mathematical point of view they are special cases of the Gaussian 
dynamical systems discussed in Section \ref{subsection-grf}.
Since the entropic fluctuations of such models are studied in detail in the 
forthcoming paper \cite{JLTP}, for reasons of space we shall be brief.

For a finite subset $\Gamma=\{n,n+1,\ldots,n+k-1\}\subset\zz$ we set
$M^\Gamma= \rr^{\Gamma}\oplus \rr^{\Gamma}$ and we define
$$
M^\Gamma\ni(p,q)\mapsto
H^\Gamma(p, q)=\sum_{x\in\zz} \frac{p_x^2 +q_x^2}{2}
+\frac{(q_x-q_{x-1})^2}{2},
$$
where we set $p_x=q_x=0$ for $x\not\in\Gamma$ (Dirichlet boundary conditions).

For some integer $m>0$, let $\Gamma_{S}=\{-m,\ldots,m\}$ and set
$H_{S}=H^{\Gamma_{S}}$. This Hamiltonian describes
a finite harmonic chain. We shall couple it to two large heat reservoirs,
$R_L$ and $R_R$, at its two ends. For this purpose,
let $n\gg m$ and set $\Gamma_L=\{-n,\ldots,-m-1\}$, 
$\Gamma_R=\{m+1,\ldots,n\}$. The Hamiltonians of the two reservoirs are
$$
H_L=H^{\Gamma_L},\qquad H_R=H^{\Gamma_R}.
$$
The Hamiltonian of the composite (but still decoupled) system is
$$
H_0=H^{\Gamma_L}+H^{\Gamma_{S}}+H^{\Gamma_R}
=H_L+H_{S}+H_R.
$$
Finally, define the Hamiltonian of the coupled system by
$$
H=H^{\Gamma_L\cup\Gamma_{S}\cup\Gamma_R}.
$$
The coupling Hamiltonian is given by
$$
V=H-H_0=V_L+V_R=-q_{-m-1}q_{-m}-q_{m}q_{m+1},
$$
and is independent of $n$. Since the equations of motion induced by $H_0$ and $H$
are linear, the associated Hamiltonian flows are linear group which 
we write as  $\e^{t\mathcal L_0}$ and $\e^{t\mathcal L}$ respectively.

Let us denote by $h$, $h_L$, $h_R$ the real symmetric  
matrices corresponding to the quadratic forms $2H$, $2H_L$, 
$2H_R$. The reference state $\omega_X$ is the centered Gaussian measure 
of covariance
$$
D_X=(\beta h-k(X))^{-1},
$$
where
$$
k(X)=X_Lh_L\oplus X_Rh_R.
$$
For $\beta>0$ the set $O_\beta=\{X\in\rr^2\,|\,\beta h-k(X)>0\}$ is
an open neighborhood of $0$. The dynamical system thus obtained 
is well defined for $X\in O_\beta$, is \TRI{}, and clearly satisfies Assumptions \Tone{} and \Ttwo{}.

It is a simple exercise in Gaussian integrals to show that Assumption \Eone{} is satisfied
for $X$ small enough. However, note that the flux observables
\begin{align*}
\Phi^{(L)}&=\{H_L,V\}=-p_{-m-1}q_{-m},\\
\Phi^{(R)}&=\{H_R,V\}=-p_{m+1}q_{m},
\end{align*}
as well as entropy production
$\sigma_X=X_L\Phi^{(L)}+X_R\Phi^{(R)}$
are unbounded. Thus, Hypothesis \Etwo{} is not satisfied. 

Propositions \ref{GESprop} and \ref{GESEquiv} apply. Moreover, the functional 
$g_t(X,Y)$ reduces to a Gaussian integral that can be computed explicitly
$$
g_{t}(X,Y)=-\frac12\log\det\left(
I-D_X(\e^{t\mathcal L^\ast}k(Y)\e^{t\mathcal L}-k(Y))
\right),
$$
with the convention that $\log x=-\infty$ for $x\le0$. Since the groups $\e^{t\mathcal L}$
is uniformly bounded, the function $(X,Y)\mapsto g_t(X,Y)$ is real analytic on an open 
neighborhood of $(0,0)$ in $O_\beta\times\rr^2$ which is independent of $t\in\rr$ and
Proposition \ref{fin-lin-resp} applies too.

The validity of the finite time GES-symmetry, ensured by Proposition \ref{GESEquiv},
can be explicitly checked as follows.
Energy conservation, $\e^{t\mathcal L^\ast}h\e^{t\mathcal L}=h$, yields
\begin{align*}
D_X^{-1}-(\e^{t\mathcal L^\ast}k(Y)\e^{t\mathcal L}-k(Y))
&=\beta h-k(X-Y)-\e^{t\mathcal L^\ast}k(Y)\e^{t\mathcal L}\\
&=\e^{t\mathcal L^\ast}\left(
\beta h-k(Y)-\e^{-t\mathcal L^\ast}k(X-Y)\e^{-t\mathcal L}\right)\e^{t\mathcal L}\\
&=\e^{t\mathcal L^\ast}\left(
D_X^{-1}-(\e^{-t\mathcal L^\ast}k(X-Y)\e^{-t\mathcal L}-k(X-Y))\right)\e^{t\mathcal L}.
\end{align*}
This relation and Liouville's theorem, $\det\e^{t\mathcal L}=\det\e^{t\mathcal L^\ast}=1$,
imply that $g_t(X,Y)=g_{-t}(X,X-Y)$. Finally, \TRI{} yields $g_{-t}(X,X-Y)=g_t(X, X-Y)$ and
the finite time GES-symmetry follows. 

\section{The large time limit}
\label{sec-LAT}
\subsection{Entropy production}
\label{entropyprod}

For any observable $f$ we set
$$
\langle f\rangle_+=\lim_{t\to\infty}\langle f\rangle_t,
\label{T-As-def}
$$
whenever this limit exists.

In this section in addition to \Eone{}-\Etwo{} we assume: 
\begin{quote}
\label{E3-def}
{\bf (E3)} The limit $\langle \sigma\rangle_+$ exists and is finite.
\end{quote}
The entropy balance equation yields the basic result:
\begin{proposition}\label{ep-positivity}
$\langle \sigma\rangle_+\geq 0$.
\end{proposition}

We shall say that the dynamical system  $(M, \phi, \omega)$ is
{\it entropy producing}  if it satisfies
\begin{quote}
\label{EP-def}
{\bf (EP)} $\langle \sigma\rangle_+>0$. 
\end{quote}

The validity of \Ethree{}  and \EP{}
are dynamical problems that can only be answered in the 
context of concrete models. In this section we shall discuss several structural results 
which shed some light on these central issues. 

\bep  Suppose that 
\begin{equation}
 \langle \sigma\rangle_t = \langle \sigma\rangle_+  + O(t^{-1}),
\label{hypolite}
\end{equation}
as $t \rightarrow \infty$. Then $\langle \sigma\rangle_+ =0$  implies that there exists
$\nu \in {\cal S}_I\cap {\cal N}_\omega$ satisfying 
${\rm Ent}(\nu|\omega)>-\infty$.
\eep

{\bf Proof:} We shall use the properties of relative entropy listed in Theorem \ref{gen-ent}. 
Suppose that $\langle \sigma \rangle_+=0$. The entropy balance equation and (\ref{hypolite})
yield
\[ 
|\Ent(\omega_t|\omega)|=\left|\int_0^t (\omega(\sigma_s)-\langle \sigma\rangle_+)\d s\right|= O(1).
\]
Hence, there is $C$ such that $\Ent(\omega_t|\omega)\geq C$ for all $t\geq 0$. Set 
$\tilde \omega_t(\cdot)=t^{-1}\int_0^t \omega_s(\cdot)\d s$. The concavity and the 
upper-semicontinuity of the relative entropy yield $\Ent(\tilde \omega_t|\omega)\geq C$. 
By compactness, there exists $\nu\in\cal S$ and a net
$t_\alpha \rightarrow \infty$ such that $\tilde\omega_{t_\alpha}\rightarrow \nu$.
It follows easily that $\nu \in {\cal S}_I$ and the upper-semicontinuity implies
${\rm Ent}(\nu|\omega)>-\infty$. \qed

\bep 
Let $\nu \in {\cal S}_I \cap {\cal N}_\omega$. Then $\nu(\sigma)=0$. 
\label{long-mik}
\eep

Before proving Proposition \ref{long-mik}, we need a preliminary result which is of 
independent interest. In what follows we shall say that a sequence $t_n\uparrow \infty$ 
is {\em regular} if $\sum_n \e^{ -a t_n}<\infty$ for all $a>0$. 

\bel \label{epp1} Let  $t_n$ be a regular sequence. Then,  for
$\omega$-a.e. $x$,
\begin{equation}
\liminf_{n \rightarrow \infty}\frac{1}{t_n}\int_0^{t_n} \sigma_s(x)\d s 
\geq 0,\qquad
\limsup_{n \rightarrow \infty}\frac{1}{t_n}\int_0^{t_n} \sigma_{-s}(x)\d s
\leq 0.
\label{ok11}
\end{equation}
\eel

\proof We will prove the first relation in  (\ref{ok11}), a similar argument yields
the second.  

Let $X_n= t_n^{-1}\int_0^{t_n} \sigma_s\d s $ and 
$A=\{x\in M\,|\,\liminf_{n\rightarrow\infty}X_n(x)<0\}$.
We need to show that $\omega(A)=0$. Since $A=\cup_{k\ge 1}A_k$ with
$A_k=\{x\in M\,|\,\liminf_{n\rightarrow\infty}X_n(x)\le-1/k\}$,
it suffices to show that $\omega(A_k)=0$ for all integers $k\ge 1$. 
Set $\Delta^t=\Delta_{\omega_t|\omega}$ and note that $\omega(\Delta^{-t})=\omega_{-t}(1)=1$,
for all $t$. The Markov inequality gives
$$
\omega(\{x\in M\,|\,\Delta^{-t}(x)\ge\lambda\})\le\lambda^{-1},
$$
for $\lambda>0$. Since $\Delta^{-t_n}=\e^{-t_n\,X_n}$, we have 
$$
\omega(\{x\in M\,|\,X_n(x)\le-a\})
=\omega(\{x\in M\,|\,\Delta^{-t_n}(x)\geq \e^{a t_n}\})\le\e^{-a t_n}.
$$
Hence, 
\[
\sum_{n}\omega(\{x\in M\,|\,X_n(x)\le 1/k\}) <\infty,
\]
and the Borel-Cantelli lemma yields that $\omega(A_k)=0$. \qed

\noindent{\bf Proof of Proposition \ref{long-mik}.}
By the ergodic theorem, there is $\bar \sigma\in L^1(M, \d\nu)$ such that 
 \[
 \lim_{n\rightarrow \infty}\frac{1}{n}\int_{0}^n \sigma_{\pm s}(x)\d s=\bar \sigma(x),
 \]
for $\nu$-a.e. $x$. Since $\nu$ is invariant one has $\nu(\sigma)=\nu(\bar \sigma)$.
Since $\nu$ is normal w.r.t. $\omega$, Lemma \ref{epp1} implies that 
 $\nu(\bar \sigma)=0$ and the statement follows. \qed

 \begin{corollary} Suppose that $ {\cal S}_I \cap {\cal N}_\omega\not=\emptyset$ and
that there exists a sequence  $t_n \uparrow \infty$  such that   
\[
\lim_{n \rightarrow \infty} \frac{1}{t_n}\int_0^{t_n} \sigma_s(x) \d s
= \langle \sigma \rangle_+,
\]
for $\omega$-a.e. $x$. Then  $\langle \sigma\rangle_+=0$.
\label{no-no-end}
\end{corollary}
\proof  Let $\nu \in {\cal S}_I \cap {\cal N}_\omega$. Then 
\[\langle \sigma \rangle_+=\nu\left(\lim_{n \rightarrow \infty} 
\frac{1}{t_n}\int_0^{t_n} \sigma_s(x) \d s\right)=\nu(\sigma)=0.
\]
\qed

The results of this section establish that, under very general 
conditions, the dynamical system $(M, \phi, \omega)$ is entropy producing
iff ${\cal S}_I \cap {\cal N}_\omega=\emptyset$.

\subsection{Linear response theory}
\label{sec-no-end}
Consider a family  $(M, \phi_X, \omega_X, {\bf \Phi}_X)$ satisfying \Tone{}, \Ttwo{} and \Tthree{}.
In this section we are interested in the large time limit of the time averaged expectation
values of individual fluxes and validity of the linear response theory.  In addition to 
(T1)-(T3) we assume: 
\begin{quote}
\label{T4-def}
{\bf (T4)} The limit $\langle {\bf\Phi}_X\rangle_+$
exists for $X$  small enough and is differentiable at $X=0$. 

\bigskip
\label{T5-def}
{\bf (T5)} $\omega(\Phi^{(k)}\Phi_t^{(j)})=O(t^{-1})$ as $t\to\infty$. 
\end{quote}
The verification of (T4) and (T5)  is  a dynamical problem that can be answered only in
the context of concrete models. \Ttwo{} and \Tfive{} imply that 
$\omega(\Phi^{(k)}\Phi_t^{(j)})=O(t^{-1})$ as $t\to-\infty$.

The kinetic transport coefficients are defined by 
\begin{equation}
 L_{jk} =\partial_{X_k}\langle \Phi_X^{(j)}\rangle_+\big|_{X=0}.
\label{lin2}
\end{equation}
Since $\langle \sigma_X\rangle_+ =\sum_j X_j \langle \Phi_X^{(j)}\rangle_+ \geq 0$, 
the real quadratic form determined by $[L_{jk}]$ is positive semi-definite. 

The kinetic transport coefficients satisfy the Onsager reciprocity relations (ORR) if 
\begin{equation}
 L_{jk}=L_{kj},
\label{full-orr}
\end{equation}
and the Green-Kubo formula holds if 
\begin{equation}
L_{jk}=\frac{1}{2}\int_{-\infty}^\infty \omega(\Phi^{(k)}\Phi_{s}^{(j)})\d s,
\label{gk-full}
\end{equation}
where, unless otherwise specified, 
$\int_{-\infty}^\infty=\lim_{t\rightarrow \infty}\int_{-t}^t$.  
Note that (\ref{gk-full}) $\Rightarrow$ (\ref{full-orr}).

The finite time linear response theory leads to a  natural axiomatic program  for the 
verification of (\ref{full-orr}) and (\ref{gk-full}) based on the following

\bep Suppose that {\rm (T1)-(T5)} hold. Then the following statements are
equivalent: 
\begin{enumerate}[{\rm (1)}]
\item The Green-Kubo formulas {\rm (\ref{gk-full})} hold. 
\item $\lim_{t\rightarrow \infty}L_{jkt}=L_{jk}$.
\end{enumerate}
\label{neverland}
\eep

\proof Set
$$
F(t)=\frac12\int_{-t}^t\omega(\Phi^{(k)}\Phi_s^{(j)})\d s,
$$
and notice that 
$$
L_{jkt}=\frac12\int_{-t}^t\omega(\Phi^{(k)}\Phi_s^{(j)})\left(1-\frac{|s|}{t}\right)\d s
=\frac1t\int_{0}^tF(s)\d s.
$$
By the fundamental property of Ces\`aro's mean,
(1) $\Rightarrow$ $\lim_{t\to\infty}F(t)=L_{jk}$ $\Rightarrow$ (2).
On the other hand, Hypothesis \Tfive{} and  Hardy-Littlewood's Tauberian 
theorem (see e.g. \cite{Ko}) yield
(2) $\Rightarrow$ $\lim_{t\to\infty}F(t)=L_{jk}$ $\Rightarrow$ (1).
\qed

We would like to add several remarks regarding Proposition \ref{neverland}.

{\bf Remark 1.} Finite time linear response theory requires the minimal regularity 
assumptions \Tone{}-\Tthree{}, and in particular  no  ergodicity assumption. It is valid in 
practically all models of interest. Assumption \Tfour{}  states that the basic objects 
of  linear response theory are well-defined (existence of the $L_{jk}$'s)
and is of course necessary to have a meaningful theory. Condition (2) of Proposition 
\ref{neverland} can be reformulated
as 
\begin{equation}
\partial_{X_k}\left.\left(\lim_{t \rightarrow \infty}
\langle \Phi_X^{(j)}\rangle_t\right)\right|_{X=0}=
\lim_{t \rightarrow \infty}\left(\partial_{X_k}\left.
\langle \Phi_X^{(j)}\rangle_t\right|_{X=0}\right),
\label{lin3}
\end{equation}
{\sl i.e.,} as an exchange of the two limits $t\to\infty$ and $X\to0$.
Even though the existence of the improper integral in (\ref{gk-full})
does not require any decay of the correlation function 
$t\mapsto\omega(\Phi^{(k)}\Phi_t^{(j)})$,
Assumption  \Tfive{} provides the minimal decay assumption which ensures that
$$
\lim_{t\to\infty}L_{jkt}
=\lim_{t\to\infty}\frac{1}{2}\int_{-t}^t \omega(\Phi^{(k)}\Phi_{s}^{(j)})\d s.
$$ 
Note however that Assumption \Tfive{} is needed only for the Green-Kubo formula and that 
Condition (2) automatically implies ORR. 
Assumptions \Tfour{} and \Tfive{} are ergodic in nature and are typically difficult to 
verify in physically interesting models. A particularly delicate aspect is differentiability of 
the function $X\mapsto \langle{\bf \Phi}_X \rangle_+$.

{\bf Remark 2.}  The proposed program for the derivation of linear response theory is in 
a certain sense minimal.  On physical grounds one would like  to have an additional 
estimate 
\begin{equation}
\langle {\bf \Phi}_X\rangle_t= {\bf L}X + {\rm Er}(X, t),
\label{kampen1}
\end{equation}
where the error term satisfies 
\[
\lim_{X\rightarrow 0}\sup_{t>0}\frac{| {\rm Er}(X, t)|}{|X|}=0,
\]
with the rate of convergence/range of parameters that allow to draw physical/numerical 
conclusion from (\ref{kampen1}). This point is  related to van Kampen's objections 
against linear response theory \cite{Ka, KTH}, see \cite{CELS2} for a discussion.

{\bf Remark 3.} 
In some models the  following well-known result (the multivariable Vitali theorem) can be
effectively used to verify (\ref{lin3}) (see \cite{JPP}). 
Let $I_\epsilon=\{X\in \rr^N\,|\, |X|<\epsilon\}$ and
$D_\epsilon=\{ X\in \cc^N\,|\, |X|<\epsilon\}$.
\bep For all $t>0$ let $F_t: D_\epsilon \rightarrow {\mathbb C}$
be an analytic function such that 
\[
\sup_{X\in D_\epsilon, t>0}|F_t(X)|<\infty,
\]
and assume that
\begin{equation}
\lim_{t \rightarrow \infty}F_t(X)=F(X),
\label{vitali}
\end{equation}
exists for $X\in I_\epsilon$. Then the limit (\ref{vitali}) exist for all $X\in D_\epsilon$ and 
is an analytic function on $D_\epsilon$. Moreover, as 
$t\rightarrow \infty$,  all derivatives of $F_t$ converge
uniformly on compact subsets of $D_\epsilon$ to the corresponding derivatives of $F$.
\label{vitali-van}
\eep

We shall  point out some  mathematical intricacies regarding the  interchange of the
limit and derivative in (\ref{lin3}) on a simple  example in Section \ref{sec-half-line}. 

Our final topic in this section is the Fluctuation-Dissipation Theorem (recall its finite 
time counterpart discussed in Remark 2 of  Section \ref{sec-finite-time}). 

\begin{definition} Suppose that {\rm (T1)-(T4)} hold.  We shall say that the
Fluctuation-Dissipation Theorem  holds  for $(M, \phi_X, \omega_X, {\bf \Phi}_X))$ if: 
\begin{enumerate}[{\rm (1)}]
\item The Green-Kubo formulas (and hence the Onsager reciprocity relations) hold for the
kinetic transport coefficients ${\bf L}=[L_{jk}]$.

\item The Central Limit Theorem holds for ${\bf \Phi}=(\Phi^{(1)}, \cdots, \Phi^{(N)})$
w.r.t. $(M, \phi, \omega)$ with covariance matrix 
\begin{equation}
{\bf D}=2{\bf L},
\label{einstein1}
\end{equation}
i.e., for any Borel set $B\subset\rr^N$, 
\[
\lim_{t \rightarrow \infty} \omega\left(\left\{ x\in M\,\,\bigg|\,\, 
\frac{1}{\sqrt t}\int_0^t{\bf \Phi}_s\d s \in B\right\}\right)
=\mu_{\bf D}(B),
\]
where $\mu_{\bf D}$ is the centered Gaussian measure of covariance $\bf D$ on $\rr^n$.
\end{enumerate}
\end{definition}

{\bf Remark.} The celebrated Einstein's relations (\ref{einstein1})  link  equilibrium  
fluctuations to kinetic transport coefficients. 

\bigskip
Just like Proposition \ref{neverland}, the Fluctuation-Dissipation Theorem is "forced" by 
its universally valid finite time counterpart. With regard to the proof of the Central Limit 
Theorem, we mention the following result of Bryc \cite{Bry}. 
$D_\epsilon$ and $I_\epsilon$ are as in Proposition \ref{vitali-van}.

\bep Assume that {\rm\Tone{}} holds. Suppose that for some $\epsilon>0$ the function
$$
g_t(0, Y)= \log \omega(\e^{-Y \cdot \int_0^t {\bf \Phi}_{s}\d s}),
$$
is analytic in $D_\epsilon$, satisfies 
\[
\sup_{Y\in D_\epsilon,t>1}\frac{1}{t}|g_t(0, Y)|<\infty,
\]
and that 
\begin{equation}
\lim_{t\rightarrow \infty}\frac{1}{t}g_t(0, Y),
\label{bryc1}
\end{equation}
exists for all $Y\in I_\epsilon$. 
Then the Central Limit Theorem holds  for ${\bf \Phi}=(\Phi^{(1)}, \cdots, \Phi^{(N)})$ 
w.r.t. $(M, \phi, \omega)$ with covariance matrix 
\begin{equation}
D_{jk}=\lim_{t\rightarrow \infty}\int_{-t}^t \omega(\Phi^{(k)}\Phi_s^{(j)})\left(1 - \frac{|s|}{t}\right)\d s.
\label{lim-bryc}
\end{equation}
 \label{bryc-prop}
\eep
{\bf Remark.} If $\Phi^{(j)}\in L^\infty(M,\d\omega)$, then the function
$Y\mapsto g_t(0, Y)$ is real analytic. The location of the complex zeros
of the entire analytic function 
$Y\mapsto  \omega(\e^{-Y \cdot \int_0^t {\bf \Phi}_{s}\d s})$ determines the region 
of complex plane to which $g_t(0,Y)$ extends analytically. 
If in addition \Ttwo{}, \Tthree{} and \Tfive{} hold,  the existence of  the limit (\ref{lim-bryc}) implies 
that 
\[
\lim_{t\rightarrow \infty}L_{jkt}
= \frac{1}{2}\int_{-\infty}^\infty \omega(\Phi^{(k)}\Phi_s^{(j)})\d s
= \frac{1}{2}D_{jk}.
\]

The Fluctuation-Dissipation Theorem is the pillar of non-equilibrium statistical mechanics in 
the regime where  the thermodynamic 
forces are weak.  The far from equilibrium case is discussed in the next section.

\subsection{The Evans-Searles  Fluctuation Theorem}
\label{subsection-evans-searls}

We start by recalling some basic facts of the
Large Deviation Theory (see e.g. \cite{DZ, El}).   

\begin{definition}\label{LDPdef} 
A vector-valued observable ${\bf f}=(f^{(1)}, \cdots, f^{(N)})$ satisfies a 
Large Deviation Principle  w.r.t. to  $(M, \phi, \omega)$ if there exists an
upper-semicontinuous function  $I : \rr^N \to [-\infty,0]$ with compact level sets  
such that for all Borel sets  $G \subset \rr^N$ we have

\begin{equation}
\begin{array}{rcccl}
\displaystyle
\sup_{Z\in \mathring{G}} I(Z) &\leq&\displaystyle \liminf_{t \rightarrow \infty} 
\frac{1}{t} \log \omega\left(
\left\{ x\in M\,\, \bigg|\,\, \frac{1}{t}\int_0^t{\bf f}_{s}(x) \d s\in G\right\}\right)&& \\[12pt]
&\leq &
\displaystyle\limsup_{t \rightarrow \infty} 
\frac{1}{t} \log \omega\left(
\left\{ x\in M\,\, \bigg|\,\, \frac{1}{t}\int_0^t{\bf f}_{s}(x)\d s\in G\right\}\right)& \leq &
\displaystyle\sup_{Z\in \bar{G}} I(Z).
\end{array}
 \label{largedeviationbounds}
\end{equation}
where $ \mathring{G}$ denotes the interior of $G$ and $\bar{G}$ its closure.
\end{definition}

The following standard result  goes under the name of 
G\"artner-Ellis Theorem and will be used repeatedly, see e.g. \cite{El,DZ} for a proof.  

\bep\label{gartnerellis}
Assume that  the limit
\[ 
h(Y) = \lim_{t \rightarrow \infty}\frac{1}{t}\log\omega 
( e^{- \int_0^t  Y \cdot {\bf f_s} \d s}),
\]
exists in $[-\infty,+\infty]$ for all $ Y\in \rr^N$ and is finite for $Y$
in some open neighborhood of $0\in\rr^N$.  
\begin{enumerate}[{\rm (1)}]
\item Suppose that  $h(Y)$ is differentiable at  $Y=0$. Then, the limit 
\[
\langle {\bf f} \rangle_+
=\lim_{t \rightarrow \infty}\frac{1}{t}\int_0^{t} \omega({\bf f}_s) \d s,
\]
exists and $\langle {\bf f} \rangle_+=-\nabla h(0)$. Moreover,
for any regular sequence $t_n$ one has
\[
\lim_{n \rightarrow \infty}\frac{1}{t_n}\int_0^{t_n} {\bf f}_s(x) \d s 
= \langle {\bf f} \rangle_+,
\]
for $\omega$-a.e. $x$.  
\item Suppose that $h(Y)$ is a lower semicontinuous function on $\rr^N$ which is
differentiable on the interior of the set $\mathcal D=\{Y\in\rr^N\,|\,h(Y)<\infty\}$
and satisfies
$$
\lim_{\mathring{\mathcal D}\ni Y\to Y_0}|\nabla h(Y)|=\infty,
$$
for all $Y_0\in\partial\mathcal D$. Then the  Large Deviation Principle holds for ${\bf f}$
w.r.t. $(M, \phi, \omega)$ with  the rate function  
\[
I( Z) = \inf_{  Y \in \rr^N }(  Y \cdot  Z + h( Y)),
\]
i.e., $-I(Z)$ is the Legendre transform of $h(-Y)$. In particular, $I(Z)$ is concave. 
\end{enumerate}
\eep

\begin{remark}\label{WeakGE}{\rm
The conclusion of Part (2) holds in particular if $h(Y)$ is differentiable on $\rr^N$.
There are other (local) versions of G\"artner-Ellis theorems that  are 
useful in applications.  Suppose, for example,  that the function $h(Y)$ is finite, strictly convex and continuously  
differentiable in some open neighborhood $B \subset \rr^N$ of the origin.
Then Part (1) holds and a weaker version of part (2) also holds:  the large deviation bounds
(\ref{largedeviationbounds}) hold provided the set $G$ is contained in a sufficiently
small neighborhood of the mean $\langle {\bf f} \rangle_+$ (see Lemma XIII.2 of   \cite{HH} and Section 4.5 of \cite{DZ}). 
}
\end{remark}

Let $(M, \phi, \omega)$ be a \TRI{} system. Recall that  
\[
e_t(\alpha)= \log \omega\left(\e^{-\alpha \int_0^t \sigma_s \d s}\right).
\]
We suppose: 
\begin{quote}
\label{ES-functional-def}
{\bf (ES)} The  {\em  Evans-Searles  functional} (ES-functional for short)
\[
\rr\ni\alpha\mapsto
e(\alpha)=\lim_{t \rightarrow \infty} \frac{1}{t}e_t(\alpha)\in[-\infty,\infty],
\]
exists.
\end{quote}

Propositions \ref{hang-over} and \ref{es-sym1} yield the  basic properties of the 
ES-functional :
\bep
\begin{enumerate}[{\rm (1)}]
\item $e(\alpha)$ is a convex function of $\alpha$.
\item It satisfies the {\sl ES-symmetry}
\begin{equation}e(\alpha)=e(1-\alpha).
\label{hypolite2}
\end{equation}
\item $$e(0)=e(1)=0,$$
$$
\left\{
\begin{array}{ll}
e(\alpha)\leq 0&\text{if } \alpha \in [0,1],\\[2.5pt]
e(\alpha)\geq 0&\text{otherwise}.
\end{array}
\right.
$$
\item It satisfies the lower bound
\begin{equation}
e(\alpha)\geq\left(\left|\alpha-\frac12\right|-\frac12\right)\Sigma^+
\label{not-possible}
\end{equation}
with
\[
\Sigma^+=\limsup_{t\rightarrow \infty}\omega(\Sigma^t).
\]
In particular, if {\rm \Ethree{}} holds then $\Sigma^+=\langle\sigma\rangle_+$.
\end{enumerate}
\eep
We emphasize that the ES-symmetry (\ref{hypolite2}) is an immediate consequence 
of the finite time ES-symmetry.

Using Proposition \ref{gartnerellis} we obtain 

\bep 
\begin{enumerate}[{\rm (1)}]
\item Suppose that  $e(\alpha)$ is differentiable at  $\alpha=0$. 
Then {\rm \Ethree{}} holds
and $\langle\sigma\rangle_+=-e'(0)$.
\item Suppose that $e(\alpha)$  is differentiable for all $\alpha\in \rr$. Then the  Large 
Deviation Principle hold for the entropy production observable $\sigma$ w.r.t. 
$(M, \phi, \omega)$ with the concave rate function 
$I(s) = \inf_{\alpha \in \rr}( s \alpha + e(\alpha))$.  Moreover
\begin{equation}\label{es-symm} 
I(s)= s + I(-s).
\end{equation}
\end{enumerate}
\label{evans-searles1}
\eep
\proof  We only need to prove (\ref{es-symm}). Using (\ref{hypolite2}) we have
$$
I(s)=\inf_{\alpha} ( s \alpha + e(\alpha))
     = \inf_{\alpha} (s(1-\alpha) + e(1-\alpha))
     =s+\inf_{\alpha} (-s\alpha + e(\alpha))
     =s+I(-s).  
$$
\qed

The relation (\ref{es-symm}) is called the {\em ES-symmetry} for the rate function $I(s)$.

Consider now a  family  $(M, \phi_X, \omega_X, {\bf \Phi}_X)$ indexed by $X\in \rr^N$ and
satisfying Assumptions \Tone{}, \Ttwo{} and \Tthree{}. We assume:

\begin{quote}
\label{GES-functional-def}
{\bf (GES)} The  {\em Generalized Evans-Searles functional} (GES-functional)  
\[ 
g(X, Y)=   \lim_{t \rightarrow \infty}\frac{1}{t}g_t(X, Y),
\]
exists for all $X, Y$. 
\end{quote}

$g(X,Y)$ is a convex function of $Y$ and the finite time GES-symmetry implies  that
\[
g(X, Y)=g(X, X-Y).
\] 
We shall refer to this relation as the {\em GES-symmetry}. 

\bep
\begin{enumerate}[{\rm (1)}]
\item Suppose that $Y\mapsto g(X,Y)$ is differentiable at $0$. Then {\rm\Tfour{}} holds,
$$
\langle {\bf \Phi}_X\rangle_{+}=-\nabla_Yg(X, Y)|_{Y=0},
$$
and if  $t_n$ is  a regular sequence, then 
\[
\lim_{n\rightarrow \infty}\frac{1}{t_n}\int_0^{t_n}{\bf \Phi}_{Xs}(x)\d s
= \langle {\bf \Phi}_X\rangle_+,
\]
for $\omega_X$-a.e. $x$.
\item Suppose that $Y\mapsto g(X, Y)$ is differentiable for all $Y$. Then the Large Deviation
Principle holds for the flux observables w.r.t. $(M, \phi_X, \omega_X)$ with the concave rate
function $I_{X}(s)= \inf_{Y \in {\mathbb R^N}}(Y\cdot s + g(X, Y))$. Moreover  
 \[
I_{X}(s)= X \cdot s + I_{X}(-s).
\]
\item Suppose that $g(X, Y)$ is $C^{1,2}$ in a neighborhood of $(0,0)$. Then the kinetic
transport coefficients are defined and satisfy the Onsager reciprocity relations. 
\item In addition to the assumption of {\rm (3)} suppose  that {\rm\Tfive{}} holds  and that 
for some  $\epsilon >0$,
\begin{equation}
\sup_{Y\in D_\epsilon, t>1}\frac{1}{t}|g_t(0, Y)|<\infty.
\label{vit-g11}
\end{equation}
Then the Fluctuation-Dissipation Theorem holds for $(M, \phi_X, \omega_X, {\bf \Phi}_X)$.
\end{enumerate}
\label{whistler-2}
\eep
\proof (1) and (2) are immediate from Proposition \ref{gartnerellis} and the symmetry
of $I_X(s)$ is proved as in Proposition \ref{evans-searles1}.  

(3) By Assertion (1), $\langle \Phi_X^{(j)}\rangle_{+}=\partial_{Y_j}g(X,Y)\big|_{Y=0}$,
hence the GES-symmetry and the symmetry lemma yield
\begin{equation}
L_{jk}=\partial_{X_k}\partial_{Y_j}g(X,Y)\big|_{X=Y=0}
=-\frac{1}{2}\partial_{Y_j}\partial_{Y_k}g(X, Y)\big|_{X=Y=0}.
\label{cigale}
\end{equation}
Since the partial derivatives on the right hand side are symmetric in $j,k$, we have
$L_{jk}=L_{kj}$.

(4) From (\ref{cigale}) and the fact that
\[
\partial_{Y_j}\partial_{Y_k}\frac{1}{t}\log g_{t}(0, Y)\big|_{Y=0}=-
\int_{-t}^t \omega(\Phi^{(k)}\Phi_s^{(j)}) \left( 1 - \frac{|s|}{t}\right)\d s,
\]
we see that the Green-Kubo formula holds iff  the limit and the derivative in the expression
\begin{equation}
\partial_{Y_j}\partial_{ Y_k}g(0, Y)|_{Y=0}
=\partial_{Y_j}\partial_{Y_k}\lim_{t\rightarrow \infty} \frac{1}{t}g_{t}(0, Y)\big|_{Y=0},
\label{gc-inter}
\end{equation}
can be interchanged. This is ensured by the assumption (\ref{vit-g11}) and Proposition 
\ref{vitali-van}. Similarly, Proposition \ref{bryc-prop} yields the  CLT. \qed

We shall say that a given \TRI{} model satisfies  the {\em Evans-Searles Fluctuation Theorem} 
if the respective functionals $e(\alpha)$/$g(X,Y)$ exist and are differentiable/$C^{1,2}$.
It follows from Propositions \ref{evans-searles1} and \ref{whistler-2}   that the the 
Evans-Searles  Fluctuation Theorem can be  interpreted as an extension of the 
fluctuation-dissipation to the far from equilibrium region.

{\bf Remark 1.} The sufficient and necessary condition for the validity of the Green-Kubo 
formula is that the limit and the derivative in the formula  (\ref{gc-inter})  can be 
interchanged.  Assumption  (\ref{vit-g11}) provides a convenient  criterion for validity 
of this exchange which will be satisfied in several examples that we will consider.  In general, 
however, there may exist other mechanisms that will lead to the justification of 
(\ref{gc-inter}), see for example the proof of the Fluctuation-Dissipation theorem 
for the Sinai billiard with small  external forces in \cite{CELS1,CELS2,Ch1,Ch2}.

{\bf Remark 2.} It is instructive to compare (3) and (4) to the finite time based derivation of the linear response theory presented in Section \ref{sec-no-end}. 

{\bf Remark 3.} In some  models where the entropy production observable is unbounded
the ES-functional $e_t(\alpha)$  is finite only on an open interval  containing $[0,1]$. 
In this case one can still formulate a meaningful Evans-Searles Fluctuation Theorem, 
see Section \ref{subsection-grf} for an example. The same remark applies to the 
GES-functional (see Section \ref{harmonicbig}).

\subsection{The resonance interpretation of $e(\alpha)$}

Under suitable regularity conditions, the identity 
\begin{equation}
 e_t(\alpha)=\log (1, \e^{ t L_p} 1),
\label{cergy-var-3}
\end{equation}
for $p=1/\alpha$ leads to identification  of $e(\alpha)$ with  a resonance of 
$L^p$-Liouvillean $L_p$.  In this section we  state  two general results regarding 
this identification. 

Since $\sp (L_p) \subset \{ z\,|\, |\Re z|\leq m_p\}$, the resolvent $(z-L_p)^{-1}$ is a 
well-defined operator valued function analytic in the half-plane $\Re z>m_p$. 
Note that $|e(\alpha)|\leq m_p$ and that
$$
(1, (z-L_p)^{-1} 1)=\int_0^\infty\e^{e_t(p^{-1})-tz}\,\d t,
$$
for $\Re z>m_p$. 

In the next two propositions $\alpha \in \rr$ is fixed and $p=1/\alpha$.

\bep  Suppose that for some $\gamma>0$ and $c\in\rr$,
\[
e_t(\alpha)= t e(\alpha)+c+ O(\e^{-\gamma t}),
\]
as $t\to\infty$. Then  the function $z\mapsto (1, (z-L_p)^{-1} 1)$
has a meromorphic continuation from the half-plane $\Re z>m_p$ to the the half-plane 
$\Re z>e(\alpha)-\gamma$ and its only singularity there is a simple pole at $z=e(\alpha)$
with residue $\e^c$. Moreover, for any $\epsilon>0$ and $j\in\{0,1\}$,
\[ 
\sup _{x >e(\alpha)-\gamma +\epsilon}
\int_{|y|>\epsilon}|(1, (x+\i y-L_p)^{j-2} 1)|^{j+1} \d y <\infty.
\]
\label{res-prop1}
\eep

This result has the following converse:

\bep
Suppose that  the function $z\mapsto (1, (z-L_p)^{-1} 1)$
has a meromorphic continuation from the half-plane $\Re z>m_p$ to the  half-plane 
$\Re z>e(\alpha)-\gamma$  for some $\gamma>0$ and that its only singularity there
is a simple pole  at $z=e(\alpha)$. Suppose also that for some $\epsilon >0$
and any $j\in\{0,1\}$,
\[ 
\sup _{x >e(\alpha)-\gamma}\int_{|y|>\epsilon}|(1, (x+\i y-L_p)^{j-2} 1)|^{j+1} \d y
<\infty.
\]
Then 
\[
e_t(\alpha)= t e(\alpha)+c+ O(\e^{-\gamma t}),
\]
as $t\rightarrow \infty$.
\label{res-prop2}\eep

The proofs of Propositions \ref{res-prop1} and \ref{res-prop2} are standard 
(see \cite{JP3}) and for reasons of space we will omit them.

After introduction of a suitable transfer operator the resonance interpretation of  
$g(X, Y)$ is very similar.

In the discrete time case (recall (\ref{disc-transf})) instead of the resolvent 
$(z- L_p)^{-1}$ one considers
\[
{\cal R}(z)=\sum_{n=0}^\infty \e^{-n z}(1, U_p^n 1).
\]
Propositions \ref {res-prop1} and \ref{res-prop2} hold in the discrete case after
obvious modifications.

The reader familiar with classical results in spectral theory of Ruelle transfer operators 
\cite{Bo2, Ru1, Ba1, BKL, GL1,GL2,Ba2,Ba3} might be surprised at our insistence on the 
Hilbert space framework. It is however precisely   in this framework and through the link 
with Tomita-Takesaki theory \cite{BR} that  Ruelle transfer operators ($L^p$-Liouvilleans) 
naturally extend to the non-commutative setting. The Banach space framework, which is 
dominant in the classical presentations, emerges through the complex spectral deformation 
technique which is a natural tool to study resonances of the $L^p$-Liouvilleans  in 
the non-commutative setting (see \cite{JP3} for  the case  $p=\infty$).

\subsection{Examples}

\subsubsection{The microcanonical ideal gas}
\label{gasprom}
In this section we investigate the large time limit in the example of Section \ref{mucangasfinitet}.
For $F\not=0$ it follows from Equ. (\ref{sigmagas}) that
$$
\lim_{t\to\infty}\sigma_t=(N-1)\frac{|F|}{\sqrt\epsilon},
$$
holds for $\omega$-a.e. $(L,\theta)\in M$, hence
\begin{equation}
\langle\sigma\rangle_+=(N-1)\frac{|F|}{\sqrt\epsilon}>0.
\label{gasEP}
\end{equation}
The generating function (\ref{gasESfunction}) can be expressed in terms of the 
associated Legendre function $P_n^m$ as
$$
e_t(\alpha)=\log\left(
\Gamma(N/2)\left(\frac2{\mathrm{sh}\mu t}\right)^{(N-2)/2}
P_{(N-1)|\alpha-1/2|-1/2}^{-(N-2)/2}(\mathrm{ch}\mu t)
\right).
$$
From the asymptotic behavior for $z\to+\infty$ (see e.g. Equ. (8.766) in \cite{GR})
$$
P_n^m(z)=\frac1{\sqrt\pi}\left(
\frac{\Gamma(n+1/2)}{\Gamma(1+n-m)}(2z)^n
+\frac{\Gamma(-n-1/2)}{\Gamma(-n-m)}(2z)^{- n-1}
\right)\left(1+O(z^{-2})\right),
$$
we obtain the ES-functional
$$
e(\alpha)=\lim_{t\to+\infty}\frac1t e_t(\alpha)
=-\langle\sigma\rangle_+
\left(\frac12-\left|\alpha-\frac12\right|\right).
$$
This function is not differentiable at $\alpha=1/2$. However, it is differentiable
near $\alpha=0$, we conclude that the entropy production observable satisfies
a (local) Large Deviation Principle with rate function
$$
I(s)=\left\{
\begin{array}{ll}
-\infty&\text{if }|s|>\langle\sigma\rangle_+,\\[10pt]
\displaystyle\frac12(s-\langle\sigma\rangle_+)&\text{if }|s|\le\langle\sigma\rangle_+,
\end{array}
\right.
$$
near its mean value $\langle\sigma\rangle_+$.

With $F$ as control parameter, we have the flux relation
$\sigma=F\,(N-1)\epsilon^{-1/2}\mathrm{th}\xi$,
and hence the flux observable
$$
\Phi=\frac{N-1}{\sqrt\epsilon}\mathrm{th}\xi.
$$
From Equ. (\ref{gasEP}) we conclude that
$\langle\Phi\rangle_+=(N-1)\epsilon^{-1/2}\mathrm{sign}\,F$ is not differentiable
at $F=0$ and linear response theory fails for this model.  We remark that the finite time
Green-Kubo formula reads
$$
L_t=\partial_F\left.\left(\frac1t\int_0^t\omega(\Phi_s)\,\d s\right)\right|_{F=0}=
\frac12\int_{-t}^t\omega(\Phi\Phi_s)\left(1-\frac{|s|}t\right)\,\d s=
\frac t{2}\omega(\Phi^2)
=\frac{(N-1)^2}{N}\frac{t}{2\epsilon},
$$
and that $L_t$ diverges as $t\to\infty$.

\subsubsection{The harmonic chain}
\label{harmonicbig}

We continue with the example of Section \ref{subsection-chains}. We again
omit the details of the calculations which the interested reader may find in
\cite{JLTP}.

Since the  reservoirs in Section \ref{subsection-chains} are all finite,
one has
$$
\langle\Phi^{(L/R)}\rangle_+=
\lim_{t\to\infty}\frac1t\int_0^t\omega_X(\Phi^{(L/R)}_s)\,\d s=
\lim_{t\to\infty}\frac1{2t}\mathrm{tr}\left(D_X
\left(h_{L/R}-\e^{t\mathcal L^\ast}h_{L/R}\e^{t\mathcal L}\right)\right)
=0.
$$
In particular, $\langle\sigma_X\rangle_+
=X_L\langle\Phi^{(L)}\rangle_++X_R\langle\Phi^{(R)}\rangle_+=0$.

To get a non-vanishing entropy production, we must perform the thermodynamic limit 
of the reservoirs ({\sl i.e.,} take $n\to\infty$ keeping $m$ fixed) before taking $t\to\infty$.
We shall not be concerned here with the existence of a limiting dynamical system.
However, it should be clear from our discussion that the limiting dynamical system 
exists as a special instance of the Gaussian dynamical systems of Section 
\ref{subsection-grf}. In the following, we denote the dependence on $n$ of various 
objects of interest by the superscript ${}^{(n)}$.
 
The phase space of the composite system has a natural embedding in the real
Hilbert space $\mathcal H=\ell_\rr^2(\zz)\oplus\ell_\rr^2(\zz)$. We denote by
$\mathcal H=\mathcal H_L\oplus\mathcal H_S\oplus\mathcal H_R$
the decomposition of this space corresponding to the partition
$$
\zz=\{x\in\zz\,|\,x<-m\}\cup\{x\in\zz\,|\,-m\le x\le m\}\cup\{x\in\zz\,|\,x>m\},
$$
and by $p_L$, $p_\mathcal{S}$, $p_R$ the corresponding orthogonal projections.
The operators $h^{(n)}$, $h_L^{(n)}$, $h_R^{(n)}$ and $k^{(n)}(X)$ have strong limits
in this Hilbert space as $n\to\infty$. We shall denote these bounded self-adjoint
limits by $h$, $h_L$, $h_R$ and $k(X)$. For example
$$
\slim_{n\to\infty}h^{(n)}=h=\left(\begin{array}{cc}
I&0\\
0&I-\Delta
\end{array}\right),
$$
where $\Delta$ is the finite difference Laplacian on $\ell^2(\zz)$
$$
(\Delta u)_x=u_{x-1}-2u_x+u_{x+1}.
$$
In the same way, the generators $\mathcal L_0^{(n)}$, $\mathcal L^{(n)}$
as well as their adjoints $\mathcal L_0^{(n)\ast}$, $\mathcal L^{(n)\ast}$
have bounded strong limits $\mathcal L_0$, $\mathcal L$ and  $\mathcal L_0^\ast$, 
$\mathcal L^\ast$. It follows that
$$
\slim_{n\to\infty}\e^{t\mathcal L_0^{(n)}}=\e^{t\mathcal L_0},\qquad
\slim_{n\to\infty}\e^{t\mathcal L^{(n)}}=\e^{t\mathcal L},
$$
and similar relations for the adjoint groups hold uniformly on compact time intervals.

Denote by $\phi_{L/R}$ the finite rank self-adjoint operator associated
to the quadratic form $2\Phi^{(L/R)}$ (recall that it does not depend on $n$).
For $Y=(Y_L,Y_R)\in\rr^2$ set $\phi(Y)=Y_L\phi_L+Y_R\phi_R$. It follows from
$$
\e^{t\mathcal L^{(n)\ast}}k^{(n)}(Y)\e^{t\mathcal L^{(n)}}-k^{(n)}(Y)
=-\int_0^t \e^{s\mathcal L^{(n)\ast}}\phi(Y)\e^{s\mathcal L^{(n)}}\,\d s,
$$
that
$$
\lim_{n\to\infty}\e^{t\mathcal L^{(n)\ast}}k^{(n)}(Y)\e^{t\mathcal L^{(n)}}
-k^{(n)}(Y)=
\e^{t\mathcal L^{\ast}}k(Y)\e^{t\mathcal L}-k(Y)
=-\int_0^t \e^{s\mathcal L^{\ast}}\phi(Y)\e^{s\mathcal L}\,\d s,
$$
holds in the trace norm for any finite $t\in\rr$. Since $(\beta h^{(n)}-k^{(n)}(X))^{-1}$
is uniformly bounded and
$$
\slim_{n\to\infty}(\beta h^{(n)}-k^{(n)}(X))^{-1}=(\beta h-k(X))^{-1},
$$
we conclude that
$$
\lim_{n\to\infty}(\beta h^{(n)}-k^{(n)}(X))^{-1}
\left(\e^{t\mathcal L^{(n)\ast}}k^{(n)}(Y)\e^{t\mathcal L^{(n)}}-k^{(n)}(Y)\right)
=-\int_0^t(\beta h-k(X))^{-1}
 \e^{s\mathcal L^{\ast}}\phi(Y)\e^{s\mathcal L}\,\d s,
$$
holds in trace norm. This finally yields
\begin{equation}
g_t(X,Y)=
\lim_{n\to\infty}g_t^{(n)}(X,Y)=-\frac12\log\det\left(
I+\int_0^t(\beta h-k(X))^{-1}
 \e^{s\mathcal L^{\ast}}\phi(Y)\e^{s\mathcal L}\,\d s
\right),
\label{sohot}
\end{equation}
with the convention that $\log x =-\infty$ for $x\leq 0$. 

We are now in position to perform the $t\to\infty$ limit. The wave operators
\begin{equation}
W_\pm
=\slim_{t\to\pm\infty}h^{1/2}\e^{-t\mathcal L}e^{t\mathcal L_0}h_0^{-1/2}(p_L+p_R)
=\slim_{t\to\pm\infty}h^{-1/2}\e^{t\mathcal L^\ast}e^{-t\mathcal L_0^\ast}h_0^{1/2}
(p_L+p_R),
\label{wavedef}
\end{equation}
exists and are partial isometries from $\mathcal H_L\oplus\mathcal H_R$ 
to $\mathcal H$. The scattering
matrix $S=W_+^\ast W_-$ is unitary on $\mathcal H_L\oplus\mathcal H_R$ 
and commutes with the
self-adjoint operator $L_0=h_0^{1/2}\mathcal L_0 h_0^{-1/2}$. Denoting by 
$\mathcal H_L\oplus\mathcal H_R=\int^\oplus\mathfrak h(\lambda)\,\d\lambda$ the
spectral decomposition induced by $L_0$ and by $S(\lambda)$ the fiber of the scattering
matrix $S$ acting on $\mathfrak h(\lambda)$, one has
$$
g(X,Y)=
\lim_{t\to\infty} \frac1t g_t(X,Y)
=-\frac1{4\pi}\int\log\mathrm{det}_{\mathfrak h(\lambda)}\left(
I-(\beta I-\widehat X)^{-1}(S(\lambda)^\ast\widehat YS(\lambda)-\widehat Y)
\right)\,\d\lambda,
$$
where $\widehat X=X_Lp_L+X_Rp_R$. We note that this formula remains valid for
arbitrary {\sl finite} harmonic system coupled to a finite number of
infinite harmonic reservoirs, as long as the coupling
$v=h-h_0$ is trace class (see \cite{JLTP}). The fluxes are given by
$$
\langle\Phi^{(j)}\rangle_+=-\partial_{Y_j}g(X,Y)\bigl|_{Y=0}=
\frac1{4\pi}\int\mathrm{tr}_{\mathfrak h(\lambda)}\left(
(\beta I-\widehat X)^{-1}(p_j-S(\lambda)^\ast p_j S(\lambda))
\right)\,\d\lambda,
$$
which is a classical version of the B\"uttiker-Landauer formula (see \cite{AJPP}).

Explicit calculation of the scattering matrix yields the result
$$
g(X,Y)=-\kappa\log\left(
\frac{((\beta-X_L)-(Y_R-Y_L))((\beta-X_R)+(Y_R-Y_L))}{(\beta-X_L)(\beta-X_R)}
\right),\qquad\kappa=\frac{\sqrt5-1}{2\pi}.
$$
For fixed $X\in\rr^2$ such that $\max(X_L,X_R)<\beta$,  the function $Y\mapsto g(X,Y)$ 
is a real analytic in the open strip $\{Y\in\rr^2\,|\,-(\beta-X_R)<Y_R-Y_L<\beta-X_L\}$.

By Proposition \ref{gartnerellis}
\begin{equation}
\langle\Phi^{(L)}\rangle_+=-\langle\Phi^{(R)}\rangle_+
=\kappa\left(\frac1{\beta-X_L}-\frac1{\beta-X_R}\right)=\kappa(T_L-T_R),
\label{phiharmony}
\end{equation}
where $T_{L/R}=(\beta-X_{L/R})^{-1}$ denotes the temperature of the $L/R$ reservoir.
In particular, entropy production
$$
\langle\sigma_X\rangle_+=X_L\langle\Phi^{(L)}\rangle_++X_R\langle\Phi^{(R)}\rangle_+
=\kappa\frac{(X_L-X_R)^2}{(\beta-X_L)(\beta-X_R)}=
\kappa\frac{(T_L-T_R)^2}{T_LT_R},
$$
is strictly positive provided $T_L\not=T_R$. By Proposition 
\ref{gartnerellis}, the flux observables $(\Phi^{(L)},\Phi^{(R)})$ satisfy a 
large  deviation principle w.r.t. $\omega_X$, with rate function
$$
I_X(s_L,s_R)=\left\{
\begin{array}{ll}
-\infty&\text{if } s_L+s_R\not=0,\\[10pt]
\displaystyle F(\theta)&\text{if }s_L=-s_R=\displaystyle\frac\kappa{\beta_0}\mathrm{sh}\theta,
\end{array}
\right.
$$
where
$$
F(\theta)=-\kappa\left[
2\,\mathrm{sh}^2\frac\theta2-\frac{\delta}{\beta_0}\mathrm{sh}\theta
-\log\left(\left(1-\frac{\delta^2}{\beta_0^2}\right)\mathrm{ch}^2\frac\theta2\right)
\right ],
$$
$\beta_0=\beta-(X_L+X_R)/2$ and $\delta=(X_L-X_R)/2$.

Writing Equ. (\ref{sohot}) as $g_t(X,Y)=-\frac12\log\det(I+A_t)$, one easily shows that
the trace norm of $A_t$ is bounded by
$$
\|A_t \|_1\le C \,|Y|\,|t|,
$$
while its operator norm satisfies
$$
\|A_t\|\le C\, |Y|,
$$
for any $Y\in\cc^2$ and $t\in\rr$ with a constant $C$ depending only on $\beta$ and $X$.
It follows that the bound (\ref{vit-g11}) is satisfied for sufficiently small $\epsilon>0$.
Finally, as a consequence of the local decay estimate for the discrete Klein-Gordon
equation
$$
|(\delta_x,\e^{-\i t\sqrt{-\Delta+1}}\delta_y)|\le C_{x,y}|t|^{-1/2},
$$
Hypothesis \Tfive{} is satisfied. It follows from assertions (3) and (4) of Proposition 
\ref{whistler-2} that the fluctuation dissipation theorem holds.

The equality $\langle\Phi^{(L)}\rangle_+=-\langle\Phi^{(R)}\rangle_+$ in 
(\ref{phiharmony}) is a consequence of energy conservation. More generally,
for an open system as described in Section \ref{sec-open-sys}, energy conservation
implies $\sum_j\langle\Phi^{(j)}\rangle_+=0$. For the same reason, the rate function 
$I_X(s)$ takes the value $-\infty$ outside of the subspace $\sum_js_j=0$ and the 
covariance matrix $\bf D$ in the central limit theorem is singular on this subspace. 
One can easily avoid all these singularities by reducing the number of parameters. 
In fact, one observes that the large time characteristics of the system do not depend 
on the initial inverse temperature $\beta$ of the small subsystem $S$. 
Indeed, the GES-functional $g(X,Y)$ only depends on the inverse temperatures of the 
reservoirs $\beta-X_j$ (this is a general feature of open systems). This suggests to fix 
the parameter $\beta$ at the mean inverse temperature of the reservoirs by restricting 
the parameters $X$ to the hyperplane $\sum_jX_j=0$ of $\rr^N$. This reduces the number
of parameters and consequently the number of associated fluxes by one. 
In our simple example
with two reservoirs, this amounts to set $X_L=-X_R$.

Harmonic systems are very special in that  the study of their dynamics can be 
effectively reduced to an application of the trace class scattering theory.  The
more difficult case of a finite anharmonic chain coupled to infinitely extended 
harmonic reservoirs has been analyzed in \cite{EPR1, EPR2, EH1, EH2, RT1, RT2}.

\section{Non-equilibrium steady  states} 
\label{sec-gallcohen}
\subsection{Basic notions}
\label{subsec-bn-steady}

To discuss Non-Equilibrium Steady States (NESS) we need several additional  
assumptions on $(M, \phi, \omega)$. The first  is:
\begin{quote}
\label{NESS1-def}
{\bf (NESS1)} $M$ is a complete separable metric space.
\end{quote}
In this case it is natural to equip ${\cal S}$ with the topology of weak convergence,
{\sl i.e.,} the minimal topology w.r.t. which all  the functionals 
\[
{\cal S}\ni \nu \mapsto \nu(f), \qquad f \in C(M),
\]
are continuous. This topology is metrizable and ${\cal S}$ is a complete separable metric 
space.  A sequence $\nu_n\in {\cal S}$ converges to $\nu$ iff $\nu_n(f)\rightarrow \nu(f)$
for all $f \in C(M)$.

With regard to Theorem \ref{gen-ent}, in (1) $B_\rr(M)$ could be replaced by $C_\rr(M)$.
(6)  and (7) are valid as formulated except that in (6) the convergent nets can be replaced
with convergent sequences. 

${\cal S}$ is compact iff $M$ is compact. More generally, a set 
${\mathfrak S}\subset {\cal S}$ is precompact (its closure is compact) iff 
${\mathfrak S}$ is tight, {\sl i.e.,} for any $\epsilon >0$ there is a compact set
$K_\epsilon \subset M$ such that $\nu(K_\epsilon)>1-\epsilon$ for all 
$\nu \in{\mathfrak S}$. 

The remaining additional assumptions are: 
\begin{quote}
\label{NESS2-def}
{\bf (NESS2)} $\phi^t$ is a  group of  homeomorphisms of $M$ and the map 
$(t, x)\mapsto \phi^t(x)$ is continuous.

\label{NESS3-def}
{\bf (NESS3)} $\sigma \in C(M)$.

\label{NESS4-def}
{\bf (NESS4)} The set of states 
\[ {\mathfrak S}(\omega)=\left\{ \frac{1}{t}\int_0^t \omega_s\d s\,\bigg|\, t\geq 1 \right\},
\]
is precompact in ${\cal S}$. 
\end{quote}
We denote by ${\cal S}_+(\omega)$ the set of limit points of ${\mathfrak S}(\omega)$
as $t\rightarrow \infty$. ${\cal S}_+(\omega)$ is non-empty and 
$\omega_+\in {\cal S}_+(\omega)$ iff there exists a sequence 
$t_n\rightarrow \infty$ such that, for all $f \in C(M)$,
\begin{equation}
\lim_{n\rightarrow \infty}\frac{1}{t_n}\int_0^{t_n}\omega_s(f)\d s =\omega_+(f).
\label{socket-1}
\end{equation}
\begin{definition}
\label{NESS-def}
We shall call the elements of ${\cal S}_+(\omega)$ the NESS of $(M, \phi, \omega)$.
\end{definition}
Two basic properties of a NESS $\omega_+$ are: 
\bep
\begin{enumerate}[{\rm (1)}]
\item $\omega_+ \in {\cal S}_I$. 
\item $\omega_+(\sigma)\geq 0$.
\end{enumerate}
\eep
\proof The statements follow  from  the relations (\ref{socket-1})  and (\ref{ent-ba-equ}).  \qed

The assumptions  (NESS1)-(NESS4) naturally apply to a family
$(M, \phi_X, \omega_X, {\bf \Phi}_X)$ except that in this case in (NESS3) one also
requires that $\Phi_X^{(j)}\in C(M)$.

{\bf Remark 1.} In the study of specific models it is an important and often very difficult
problem to prove that  ${\cal S}_+(\omega)$ is a singleton, namely that there exists 
$\omega_+\in {\cal S}$ such that for all $f\in C(M)$, 
\[
\lim_{t\rightarrow \infty}\frac{1}{t}\int_0^{t}\omega_s(f)\d s =\omega_+(f).
\]

{\bf Remark 2.} The  regularity assumptions (NESS2) and (NESS3)  are made for simplicity
of presentations and can be relaxed, see Sections \ref{chainNESS} and  
\ref{subsection-grf} for examples.

{\bf Remark 3.}  The NESS property is related to the SRB property in dynamical
systems, see \cite{Ru2} for more details.

\subsection{The Gallavotti-Cohen Fluctuation Theorem}

Let  $(M, \phi, \omega)$ be a \TRI{} system and let $\omega_+\in {\cal S}_+(\omega)$
be given.  Let 
\[
e_{t+}(\alpha)=\log \omega_+(\e^{-\alpha \int_0^t \sigma_s \d s}).
\]
Note that $e_{t+}(\alpha)$ is a convex function of the parameter $\alpha$ and that 
$e_{t+}(\alpha)\geq -\alpha t \omega_+(\sigma)$.

We suppose:
\begin{quote}
\label{GC-functional-def}
{\bf (GC)} The {\em Gallavotti-Cohen functional} (GC-functional)
\[ e_+(\alpha) = \lim_{t \rightarrow \infty} 
\frac{1}{t}e_{t+}(\alpha),
\]
exists for all $\alpha \in \rr$.
\end{quote}
If the GC-functional satisfies 
\[
e_+(\alpha)=e_+(1-\alpha),
\]
for all $\alpha$, we shall say that the {\em GC-symmetry} holds.  
$e_+(\alpha)$ is a convex function and the GC-symmetry implies that 
$e_+(0)=e_+(1)=0$, $e_+(\alpha)\leq 0$ for $\alpha\in [0,1]$, and 
$e_+(\alpha)\geq 0$ for $\alpha \not\in [0,1]$.

In comparison with the ES-symmetry, we remark that in general the relation 
$e_{t+}(\alpha)=e_{t+}(1-\alpha)$ does not hold for finite $t$ and that GC-symmetry 
may fail even  in some very  simple models  (see Subsection \ref{sec-half-line}). In 
contrast, whenever $e(\alpha)$ exists,  the universally  valid  finite time ES-symmetry 
$e_{t}(\alpha)=e_{t}(1-\alpha)$ implies that $e(\alpha)=e(1-\alpha)$.

\bep 
\begin{enumerate}[{\rm (1)}]
\item Suppose that $e_+(\alpha)$ is differentiable at $\alpha=0$. Then {\rm \Ethree{}} 
holds and 
\[
\langle \sigma \rangle_+=\omega_+(\sigma)
=\lim_{t \rightarrow \infty}\frac{1}{t}\int_0^{t} \omega(\sigma_s) \d s
=-e_+^\prime(0).
\]
If  in addition the GC-symmetry holds, then $\langle \sigma \rangle_+=0$ iff 
$e_+(\alpha)=0$ for $\alpha \in [0,1]$.  For any regular sequence $t_n$, 
\[
\lim_{n\rightarrow \infty} \frac{1}{t_n}\int_0^{t_n}\sigma_s(x) \d s
=\langle \sigma \rangle_+,
\]
for $ \omega_+$-a.e. $x\in M$. If  $\langle \sigma \rangle_+>0$, then 
$\omega$ and $\omega_+$ are mutually singular.

\item  Suppose that $e_+(\alpha)$ is differentiable for all $\alpha$. Then the  
Large Deviation Principle holds for  $\sigma$ w.r.t. $(M, \phi, \omega_+)$  with the 
concave rate function $I_+(s)= \inf_{\alpha \in {\mathbb R}}(\alpha s+ e_+(\alpha))$. 
If the GC-symmetry holds, then  
\[I_+(s)= s + I_+(-s).\]
The last relation is called the  GC-symmetry for the rate function $I_+$.
\end{enumerate}
\label{spencer}
\eep

\proof The only part that requires a proof  is the last statement in Part (1).  Suppose that 
$\langle \sigma \rangle_+>0$ and let $\omega_+=\nu_1 +\nu_2$, 
$\nu_1 \ll \omega$, $\nu_2\perp \omega$, be the Radon-Nikodym decomposition of 
$\omega_+$ w.r.t. $\omega$. Since $\omega_+\in {\cal S}_I$,  Assumption \C{} and the 
uniqueness of the Radon-Nikodym decomposition imply that $\nu_1, \nu_2\in {\cal S}_I$. 
If $\nu_1$ is non-trivial, then Corollary \ref{no-no-end} implies that 
$\langle \sigma \rangle_+=0$, a contradiction.
\qed

Consider  a family $(M, \phi_X, \omega_X, {\bf \Phi}_X)$, $X\in \rr^N$, satisfying \Tone{}-\Ttwo{}
and let $\omega_{X+}\in {\cal S}_+(\omega_X)$ be given. Let
\[
g_{t+}(X, Y) = \log \omega_{X+}\left(\e^{-Y \cdot  \int_0^t {\bf \Phi}_{Xs}\d s}\right).
\]

We suppose:
\begin{quote}
\label{GGC-functional-def}
{\bf (GGC)} The {\em Generalized Gallavotti-Cohen functional} (GGC-functional)
\[
g_{+}(X, Y)= \lim_{t\rightarrow \infty}\frac{1}{t}g_{t+}(X, Y),
\]
exists for all $X, Y$. 
\end{quote}
If  the GGC-functional satisfies 
\[
g_+(X,Y)=g_+(X,X-Y),
\]
for all $X,Y$, we shall say that {\em GGC-symmetry} holds. 

\bep
\begin{enumerate}[{\rm (1)}]
\item Suppose that $Y\mapsto g_+(X, Y)$ is differentiable at $Y=0$.
Then 
\[
\langle {\bf \Phi}_X\rangle_+
=\lim_{t \rightarrow \infty}\frac{1}{t}\int_0^{t}\omega_X( {\bf \Phi}_{Xs}) \d s,
\]
exists and
$\langle {\bf \Phi}_X\rangle_+ =\omega_{X+}({\bf \Phi}_X)=-\nabla_Y g_+(X,Y)|_{Y=0}$.
Moreover, for any regular sequence $t_n$, 
\[
\lim_{n\rightarrow \infty}\frac{1}{t_n}\int_0^{t_n}{\bf \Phi}_{Xs}(x)\d s
= \omega_{X+}({\bf \Phi}_X),
\]
for $\omega_{X+}$-a.e. $x\in M$.

\item Suppose that $Y\mapsto g_+(X, Y)$ is differentiable for all $Y$. Then the Large
Deviation Principle holds  for ${\bf \Phi}_X$ w.r.t. $(M,\phi_X, \omega_{X+})$ with the 
concave rate function $I_{X+}(s)= \inf_{Y \in {\mathbb R^N}}(Y\cdot s + g_+(X, Y))$.
If the  GGC-symmetry  holds, then
\[
I_{X+}(s)= X \cdot s + I_{X+}(-s).
\]

\item Suppose that $g_+(X, Y)$ is $C^{1,2}$ in a neighborhood of $(0,0)$ and that 
the GGC-symmetry holds. Then the kinetic transport coefficients
are defined and satisfy the Onsager reciprocity relations $L_{jk}=L_{kj}$.

\item In addition to the assumptions of {\rm (3)}   suppose  that for some 
$\epsilon >0$,
\begin{equation}
\sup_{Y\in D_\epsilon, t>1 }\frac{1}{t}|g_{t+}(0, Y)|<\infty.
\label{vit-g}
\end{equation}
Then the Fluctuation-Dissipation Theorem holds. 
\end{enumerate}
\label{whistler-3}
\eep
The proof of this proposition is the same as the proof of Proposition \ref{whistler-3}. 
Apart from the last statement in (1) and 
(2), the conclusions of these two propositions are also identical.
Note that $g_{t+}(0,Y)=g_{t}(0, Y)$ and so the parts (4) of the two propositions are 
in fact identical (we included the statement for completeness).  

We shall say that a given \TRI{} system satisfies the {\em Gallavotti-Cohen Fluctuation Theorem}
if the respective functionals $e_+(\alpha)$/$g_+(X,Y)$ exist and are differentiable/$C^{1,2}$
and satisfy the GC/GGC-symmetry.
It follows from Propositions \ref{spencer} and \ref{whistler-3}  that the Gallavotti-Cohen 
Fluctuation Theorem  is also an extension of the fluctuation-dissipation theorem to  the  
far from equilibrium region. 
\label{sec-cla-ness}

\subsection{The resonance interpretation of $e_+(\alpha)$}

Let $\omega_+$ be a NESS of $(M, \phi, \omega)$ (in particular, we assume that 
(NESS1)-(NESS4) hold). For  $p\in (-\infty, \infty]$, $p\not=0$ and  $f \in C(M)$  let 
\[
U_p(t)f =\e^{\frac{1}{p} \int_0^t \sigma_{-s}}f_{-t}.
\]

One easily shows that:

\bep 
\begin{enumerate}[{\rm (1)}]
\item $\omega_+([U_{p}(t)f] [U_{-p}(t)g])=\omega_+(fg)$.

\item $U_p(t)$ extends to a strongly continuous group of bounded operators on 
$L^2(M, \d\omega_+)$ such that  $U_p^\ast(t)=U_{-p}(-t)$ and 
\[
\|U_p(t)\|\leq \e^{|t|m_{p+}},
\]
where $m_{p+}=\sup_{x\in M}|\sigma(x)| /|p|$. 
Let  $L_{p+}$ be the generator of $U_p(t)$, $U_p(t)=\e^{t L_{p+}}$. Then 
$L_{p+}^\ast =- L_{-p+}$, ${\rm sp}(L_{p+})\subset\{ z\,|\, |\Re  z|\leq m_{p+}\}$, 
${\rm Dom}(L_{p+})= {\rm Dom}(L_{\infty +})$, and for $f\in \Dom L_{p+}$
\[
L_{p+}f= L_{\infty +}f + \frac{\ \sigma}{p}f.
\]
\end{enumerate}
\eep

We shall call the operator $L_{p+}$ the NESS $L^p$-Liouvillean. 
If $\alpha=-1/p$, then 
\[
e_{t+}(\alpha) = 
\log (1, \e^{ t L_{p+}} 1)_+ =\log \int_M\e^{t L_{p+}}1\, \d\omega_+.
\]

Under suitable regularity condition this relation leads to the identification of   $e_+(\alpha)$
with a complex resonance of    $L_{p+}$.  With the  obvious notational changes
Propositions \ref{res-prop1} and \ref{res-prop2} apply to $e_+(\alpha)$ and $L_{p+}$.

\subsection{Examples}

\subsubsection{The microcanonical ideal gas}
\label{gasness}

In the example of Section \ref{mucangasfinitet}, it follows from
(\ref{gasdynone}), (\ref{gasdyntwo}) that the unique NESS of the system is
$$
\d\omega_+=\prod_{j=1}^N\delta(L_j-\sqrt{\epsilon})\frac{\d L_j\d\theta_j}{2\pi}.
$$
We note that it is singular w.r.t. the reference measure $\omega$.
One immediately computes
$$
e_{+t}(\alpha)=-\alpha t\langle\sigma\rangle_+,
$$
and observes that the GC-functional
$$
e_+(\alpha)=-\alpha\langle\sigma\rangle_+,
$$
does {\sl not} satisfy the GC-symmetry. Entropy production does not fluctuate in
the NESS $\omega_+$. Accordingly, the rate function for its large deviation
is
$$
I_+(s)=\left\{\begin{array}{ll}
0&\text{if } s=\langle\sigma\rangle_+,\\[3.5pt]
-\infty&\text{otherwise}.
\end{array}
\right.
$$
The GC-fluctuation theorem fails in this model.

\subsubsection{The harmonic chain}
\label{chainNESS}

To compute the NESS in the example of Section \ref{harmonicbig}
(in the thermodynamic limit) we note that $\omega_{X t}$ is the centered Gaussian 
measure of covariance
$$
D_t=
(\e^{-t\mathcal L^\ast}(\beta h-k(X))\e^{-t\mathcal L})^{-1}
=h^{-1/2}(\beta-h^{-1/2}\e^{-t\mathcal L^\ast}h_0^{1/2}\widehat X
h_0^{1/2}\e^{-t\mathcal L}h^{-1/2})^{-1}h^{-1/2}.
$$
By Equ. (\ref{wavedef}), this covariance converges strongly to the limit
$$
D_+=
h^{-1/2}W_-(\beta-\widehat X)^{-1}W_-^\ast h^{-1/2}.
$$
It follows that the system has a unique NESS $\omega_{X+}$ which is Gaussian 
with covariance $D_+$. In particular, one has
$$
g_{t+}(X,Y)=-\frac12\log\det\left(
I+\int_0^t D_+ \e^{s\mathcal L^{\ast}}\phi(Y)\e^{s\mathcal L}
\,\d s
\right).
$$
As for the functional $g_t(X,Y)$, one can compute the infinite time limit
and get (see \cite{JLTP})
\[
g_+(X,Y)=\lim_{t\to\infty}\frac1tg_{t+}(X,Y)=g(X,Y),
\]
{\sl i.e.,} the GGC-functional coincide with the GES-functional. In particular,
it satisfies the GGC-symmetry and the Gallavotti-Cohen fluctuation theorem holds.
Note that in this example Assumptions \NESStwo{} and \NESSthree{} do not hold.
 
\section{The principle of regular entropic fluctuations}
\label{sec-pri-reg}

The  mathematical  similarity between Propositions \ref{evans-searles1}, \ref{whistler-2} 
on one side and Propositions \ref{spencer}, \ref{whistler-3} on the other side is not accidental.
The principal distinction is that the Evans-Searles symmetries are universal---they hold for any TRI
dynamical systems for which the objects in question are defined. The mechanism 
behind Gallavotti-Cohen symmetries {\em a priori} could be model dependent and in general
they may fail. Perhaps surprisingly, a  careful look at all principal classes of models for which
the symmetries have been rigorously established  reveals that the respective functionals satisfy
\begin{equation}
 e(\alpha)=e_+(\alpha), \qquad g(X, Y)=g_+(X, Y).
 \label{sym-eq}
 \end{equation}
This is true for the toy models (see Section \ref{sec-examples}), for Hamiltonian open systems 
systems treated in \cite{RT1} (see Section \ref{sec-open-sys})  and for Anosov diffeomorphisms
of compact manifolds  \cite{GC1, GC2} (see Section \ref{sec-anosov}). In each single case  the 
strong ergodic properties of the  model force the relations (\ref{sym-eq}).  Note that the first
relation in (\ref{sym-eq}) holds iff the limits in the expression
\begin{equation}
e_+(\alpha)=\lim_{t \rightarrow \infty}\lim_{u\rightarrow \infty}\frac{1}{t}\log \omega_u(\e^{-\alpha \int_0^t \sigma_s \d s}),
\label{exch-lim}
\end{equation}
can be interchanged and a similar remark applies to the second relation.
This leads to a  transparent mechanism 
for validity of GC-symmetries: The strong ergodicity (chaoticity) of the model forces the identities
 (\ref{sym-eq}) and  the universal ES-symmetries imply  that  the GC-symmetries hold.  This
leads  to the {\em principle of regular entropic fluctuations}. 

\begin{definition} We shall say that $(M, \phi, \omega, \omega_+)$ satisfies the principle of 
regular entropic fluctuations if $e(\alpha)$ and $e_+(\alpha)$ exist, are differentiable
in a neighborhood of $\alpha=0$, and satisfy
\[
e(\alpha)=e_+(\alpha),
\]
for all $\alpha$. In the presence of control parameters, we say that family 
$(M, \phi_X, \omega_X, \omega_{X+}, {\bf \Phi}_X)$ satisfy  the principle of 
regular fluctuations if $g(X,Y)$ and $g_+(X, Y)$ exist, are $C^{1,2}$ in a neighborhood
of $(0,0)$ and satisfy
\[
g(X, Y)=g_+(X, Y).
\]
\end{definition}

Combining Propositions \ref{evans-searles1}, \ref{whistler-2}, \ref{spencer}, \ref{whistler-3}
one derives the implications of the proposed principle. The principle will naturally extend to
quantum statistical mechanics.  

We emphasize that the principle of regular entropic fluctuations is ergodic in nature. It would
be very interesting to exhibit examples of non-trivial physically relevant models for which the
principle fails in the sense that $e(\alpha)$ and $e_+(\alpha)$ exist and are differentiable, 
the GC-symmetry holds, but $e(\alpha)\not=e_+(\alpha)$.

If the  principle of regular entropic fluctuations holds in a given model, then one may 
consider the Evans-Searles and Gallavotti-Cohen symmetries as mathematically equivalent. 
This {\em does not} mean that implications of  Evans-Searles and Gallavotti-Cohen 
Fluctuation Theorems are identical. 
If $(M, \phi, \omega)$ has non vanishing entropy production, then $\omega$ and 
$\omega_+$ are mutually singular,  and the corresponding Large Deviation Principles (LDPs)
are very different statements. The principle only asserts that these LDPs hold with the same 
rate function and explains the origin of the Gallavotti-Cohen symmetry.  

Note that the Evans-Searles fluctuation theorem requires less regularity  than 
the Gallavotti-Cohen fluctuation theorem and in particular does not require 
the existence of a NESS. This  does not make the notion  of NESS redundant---the fine 
studies of systems far from equilibrium are critically  centered around the NESS. The 
situation is somewhat analogous to studies of phase transitions in spin systems---the 
pressure functional  provides some information about the phase transitions but a much 
finer information is encoded in the set of equilibrium states at the critical temperature.

In Sections \ref{sec-homeo} and \ref{sec-anosov} we shall examine the ergodic mechanism 
behind the principle of regular entropic fluctuations  in the  case of chaotic homeomorphisms 
of compact metric spaces and in particular Anosov diffeomorphisms.  Starting with the works
\cite{GC1, GC2} these models have been a basic  paradigm for any theory of entropic
fluctuations.

\section{Toy models}
\label{sec-examples}

In this section we discuss a few additional examples for which  the Evans-Searles
and Gallavotti-Cohen  functionals can be computed explicitly. 

\subsection{Bernoulli shift}
Let $M=\{0,1\}^{\mathbb Z}$. The elements of $M$ are sequences $x=(x_j)_{j\in \zz}$,
$x_j\in\{0,1\}$. Let $\nu_p$ be the usual Bernoulli measure on $\{0,1\}$, 
$\nu_p(\{1\})=p$, $\nu_p(\{0\})=1-p$. Let 
\[
\omega =\left(\bigotimes_{n=-\infty}^{0}\nu_p\right)\otimes
\left(\bigotimes_{n=1}^\infty\nu_q\right)
\]
where $p, q\in]0,1[$ are given. The dynamics $\phi$ is  the left shift
\[
\phi(x)_j=x_{j+1}.
\]
The Radon-Nikodym derivative of $\omega_1=\omega \circ \phi^{-1}$ w.r.t. $\omega$ is 
\[
\Delta_{\omega_1|\omega}(x)=\frac{
\nu_p(\{x_0\})}{\nu_q(\{x_0\})},
\]
and the entropy production observable is 
\[\sigma(x)=\log \Delta_{\omega_1|\omega}(x) =\log\frac{\d\nu_q}{\d\nu_p}(x_0)=
\begin{cases}
\log\frac{q}{p} &\mbox{if  } x_0=1, \\[3mm]
 \log \frac{1-q}{1-p} &\mbox{if  } x_0=0.
 \end{cases}
 \]
Note that for all $n\geq1$, 
\[
\omega(\sigma_{n})=q\log\frac{q}{p} + (1-q)\log\frac{1-q}{1-p}=
-{\rm Ent}(\nu_q|\nu_p)\geq 0,
\]
so that the system is entropy producing,
\[
\langle \sigma \rangle_+=-{\rm Ent}(\nu_q|\nu_p)>0,
\]
iff $p\not=q$. Similarly, for all $n\geq 1$, 
\[
\frac{1}{n}\log\omega\left({\rm e}^{\alpha\sum_{j=1}^{n}\sigma_{-j}}\right)
=\log\left[p^{1-\alpha}q^\alpha + (1-p)^{1-\alpha}(1-q)^{\alpha}\right]
={\rm Ent}_\alpha(\nu_q|\nu_p),
\]
and so the ES-functional exists and is given by
\[
e(\alpha)= \log\left[p^{1-\alpha}q^\alpha + (1-p)^{1-\alpha}(1-q)^{\alpha}\right].
\]
Assuming $p\not=q$, the ES-symmetry $e(\alpha)=e(1-\alpha)$ holds iff $q=1-p$. 
In particular, if $q\not=1-p$, the model is not time reversal invariant. If $q=1-p$, then  the  
time reversal is $\vartheta(x)_j=1-x_{-j}$.

Let 
\[
\omega_+ = \bigotimes_{n\in {\mathbb Z}}\nu_q.
\]
$M$ is a compact metric space and for any $f\in C(M)$, 
\[
\lim_{n\rightarrow \infty}\omega_n(f)=\omega_+(f).
\]
Hence, $\omega_+$ is the NESS of $(M, \phi, \omega)$. Moreover, for any $n\geq 1$, 
\[
\frac{1}{n}
\log\omega_+\left({\rm e}^{-\alpha\sum_{j=0}^{n-1}\sigma_{j}}\right)=
\log\left[q^{1-\alpha}p^\alpha + (1-q)^{1-\alpha}(1-p)^{\alpha}\right],
\]
and so the GC-functional is
\[
e_+(\alpha)=e(1-\alpha).
\]
The GC-symmetry holds iff the ES-symmetry does, and in this case the GC-functional
coincide with the ES-functional.

If $p=1-q$,  then one can consider $X=p-1/2$ as a control parameter. The respective 
Fluctuation-Dissipation Theorem follows easily \cite{Sh}.

\subsection{Baker transformation}
 Let $ M=[0,1[\times [0,1[$,  $\d\omega=\d x \d y$ and 
\[
 \phi(x, y)=
\begin{cases} (x/p, y q)& \text{if $x\in [0,p[$,}
\\
((x-p)/q, py + q)&\text{if $x\in [p, 1[$,}
\end{cases}
\]
where $p, q\in]0,1[$.
This  model has been studied in \cite{DGT}, see also \cite{Do, TG, Sh}. Define  a map 
$T:\{0, 1\}^\zz\to M$ by $T \{a_j\}_{j\in \zz}=(x,y)$ where 
\[ 
x =p\left( a_0 + \sum_{n=1}^\infty a_np^{n-\sum_{i=0}^{n-1}a_i} q^{\sum_{i=0}^{n-1}a_i}\right), \qquad 
y = q\left(a_{-1} + \sum_{n=1}^\infty a_{-n-1}q^{n -\sum_{i=0}^{n-1} a_{-i-1}}q^{\sum_{i=0}^{n-1}a_{-i-1}}\right). 
\]
The map $T$ is a mod $0$ isomorphism between the Bernoulli shift considered in 
the previous section and the Baker transformation. Hence, the ES-functional
$e(\alpha)$ of these two models are the same. The non-equilibrium steady states of the two
models are also related by the map $T$ and their GC-functionals are equal 
\cite{Sh}. 

\subsection{Dilation on a half-line}
\label{sec-half-line}
Let $M=[0, \infty]$ (the compactified positive half-line), 
$\phi_\gamma^t(x)=\e^{\gamma t}x$ where $\gamma \in \rr$, and
\[
\d\omega =\frac{2}{\pi}\frac{\d x}{1+x^2}.
\]
The map $\vartheta(x)=x^{-1}$ is  a time-reversal of $(M, \phi_\gamma, \omega)$.
The reference state $\omega$ is invariant under $\phi_0$ and for $\gamma\not=0$ the
system has a unique NESS
$$
\omega_{\gamma+}=\left\{
\begin{array}{ll}
\delta_0&\text{if }\gamma<0,\\
\delta_\infty&\text{if }\gamma>0.
\end{array}
\right.
$$
One has
\[
\Delta_{\omega_t|\omega}(x)= \e^{-\gamma t}
\frac{1+x^2}{1 + \e^{-2\gamma t}x^2}, \qquad 
\sigma_\gamma(x)  =-\gamma \frac{1-x^2}{1+x^2},
\]
from which it follows that
$\langle \sigma_\gamma\rangle_+=\omega_{\gamma+}(\sigma_\gamma)=|\gamma|$.
One easily computes the ES-functional 
$$
e(\alpha)=\begin{cases} 
-\alpha |\gamma| & \text{if }\alpha \leq 1/2,\\
-(1-\alpha)|\gamma|&\text{if }\alpha \geq  1/2,
\end{cases}
$$
while the GC-functional is $e_+(\alpha)= -\alpha|\gamma|$. Hence, 
$e(\alpha)\not=e_+(\alpha)$ for $\alpha >1/2$,  the ES-symmetry holds, 
but GC does not.

Consider $\gamma$ as a control parameter. The associated flux observable
$\Phi=(x^2-1)/(x^2+1)$ does not depend on $\gamma$, $\omega(\Phi)=0$ and 
$\omega_{\gamma +}(\Phi)=\mathrm{sign}\gamma$. The kinetic transport coefficient
$$
L=\partial_{\gamma}\left.\omega_{\gamma+}(\Phi)\right|_{\gamma=0},
$$
is  not well defined.  On the other hand, $ \omega(\Phi_t)=\mathrm{th}(\gamma t /2)$  
is a real analytic function of $\gamma$, the finite time kinetic transport coefficient is  
well defined, the finite time Green-Kubo formula holds, and 
\[
L_t =\partial_{\gamma}\left.\left(\frac{1}{t}\int_0^t 
\omega(\Phi_s)\d s\right)\right|_{\gamma=0} =
\frac{1}{2}\int_{-t}^t\omega(\Phi \Phi_s)\left( 1 -\frac{|s|}{t}\right) \d s
= \frac{t}{2}.
\]
Note that, as in the  example of Section \ref{gasprom}, $L_t$ diverges as $t\to\infty$.
The reader may have noticed the similarity between this toy example and the
microcanonical ideal gas of Section \ref{mucangasfinitet} (see also \cite{CG} for a 
related example).

By slightly modifying this example we can illustrate another point. Consider 
$(M, \phi_{-\gamma^2}, \omega)$ and let again $\gamma$ be the control parameter.
Then 
\[
\sigma_\gamma(x)=-\gamma^2 \frac{1-x^2}{1+x^2},\qquad 
\Phi_\gamma(x)=-\gamma \frac{1-x^2}{1+x^2},
\]
$\omega_{\gamma+}=\delta_0$ for all $\gamma\not=0$, and the functions  
$\langle \sigma_\gamma\rangle_+=\gamma^2$, 
$\langle \Phi_\gamma\rangle_+=\gamma$ are 
entire analytic. The kinetic transport coefficient is equal to $1$ and, 
since $\Phi_{0}\equiv 0$, the Green-Kubo formula fails. In this example the finite time 
linear response theory holds with 
$L_t=0$, 
$\gamma \mapsto \langle \Phi_\gamma\rangle_t$ is real analytic, 
$\gamma\mapsto \langle \Phi_\gamma\rangle_+$ is entire analytic, but the limit and 
derivative in the expression 
$\partial_\gamma (\lim_{t\rightarrow \infty}\langle \Phi_\gamma\rangle_t)|_{\gamma=0}$
{\em cannot} be interchanged.

\section{Gaussian dynamical systems}
\label{subsection-grf}  

This class of dynamical systems is treated in detail in the forthcoming article \cite{JLTP}
and for reason of space we shall be brief (in particular we will  omit all the proofs). 
The thermally driven harmonic chain of Section \ref{harmonicbig} is an example of 
a Gaussian dynamical system.

Let $\Gamma$ be a countably infinite set and
\[
M=\rr^\Gamma=\{x =(x_n)_{n\in \Gamma}\,|\, x_n\in \rr\}.
\]
A sequence $l=\{l_n\}_{n\in \Gamma}$ of strictly positive real numbers such that 
$\sum_{n \in \Gamma}l_n=1$ defines a metric
\[
d(x, y)=\sum_{n \in \Gamma}l_n \frac{|x_n-y_n|}{1 + |x_n- y_n|},
\]
on $M$. Equipped with $d$, $M$ is a complete separable metric space. Its Borel 
$\sigma$-algebra ${\cal F}$ is generated by the cylinders 
\begin{equation}
C(B;n_1, \ldots, n_k)= \{ x\in M \,|\, (x_{n_1},\ldots,x_{n_k}) \in B \},
\label{cyl}
\end{equation}
for $k\ge0$, $n_1,\ldots,n_k\in\Gamma$ and Borel sets $B\subset\rr^k$. 

We denote by $\ell^2_\rr(\Gamma)\subset M$ (respectively $M_l\subset M$) the real 
Hilbert space  with the inner product $(x,y)=\sum_{n\in\Gamma}x_ny_n$ (respectively 
$(x, y)_l =\sum_{n\in\Gamma} l_n x_n y_n$). $\ell^2_\rr(\Gamma)$ is dense in $M_l$
and $M_l$ is dense in $M$.  
All the measures on $(M,\mathcal F)$ we will consider  in this example will be supported
on $M_l$. We denote  by  $A_{nm}=(\delta_n, A\delta_m)$ the matrix elements of a linear
operator $A$ on $\ell^2_{\rr}(\Gamma)$ w.r.t. its standard basis 
$\{\delta_n\}_{n\in\Gamma}$.

Let ${\cal L}$ be  a bounded linear operator on $\ell^2_{\rr}(\Gamma)$ which has a 
continuous extension to $M_l$. For $x\in M$ and $t\in\rr$ we set 
\begin{equation}
\phi^t(x)=
\begin{cases}
\e^{t{\cal L}}x &\mbox{if  } x\in M_l, \\[3mm]
 x &\mbox{if  } x\not\in M_l.
\end{cases}
\label{dyn-gauss}
\end{equation}
$\phi^t$ is a group of automorphisms of $(M,\mathcal F)$ describing the time evolution.  
Note that the  map $(t, x)\mapsto \phi^t(x)$ is measurable and so Assumptions 
\Fone{}-\Ftwo{} of Section \ref{subsec-flows} hold for $(M, \phi)$.

Let $D$ be a bounded, strictly positive  operator on $\ell^2_{\rr}(\Gamma)$. The 
centered Gaussian measure on $(M, {\cal F})$ of covariance $D$ is the unique measure 
$\omega$ specified by its value on cylinders
\[
\omega(C(B;n_1, \ldots, n_k)) =\frac1{\sqrt{(2\pi)^k{\rm det} D_c}}
\int_{B}\e^{-\frac{1}{2}(x, D_c^{-1}x)}\d x,
\]
where $D_c= [D_{n_{i}n_j}]_{1\leq i, j\leq k}$. For any finite subset
$\Lambda\subset\Gamma$ one has
$$
\int_M\sum_{n\in\Lambda}l_nx_n^2\,\d\omega(x)=\sum_{n\in\Lambda}l_n D_{nn}
\le\|D\|,
$$
which shows that $\omega(M_l)=1$.

Our starting point is the dynamical system $(M, \phi, \omega)$. 
$\omega_t = \omega \circ \phi^{-t}$ is a Gaussian measure of covariance 
\[
D_t = \e^{t {\cal L}}D\e^{t {\cal L}^\ast}.
\]
$D_t$ is a bounded strictly positive operator on $\ell_\rr^2(\Gamma)$ and 
$\omega_t (M_l)=1$ for all $t$. Denote by ${\cal T}$ the real vector space of all trace 
class operators on $\ell_\rr^2(\Gamma)$. The trace norm 
$\|T\|_1={\rm tr}((T^\ast T)^{1/2})$ turns ${\cal T}$ into a Banach space.
By the Feldman-Hajek-Shale theorem $\omega_t$ and $\omega$ are equivalent
iff $D_t^{-1} - D^{-1}\in {\cal T}$. We shall assume more: 
\begin{quote}
\label{G1-def}
{\bf (G1)}  The map $\rr \ni t \mapsto D_t^{-1}-D^{-1}\in {\cal T}$ is differentiable.
\end{quote}

It follows that
$$
\varsigma =-\frac{1}{2}\frac{\d}{\d t} (D_t^{-1} - D^{-1})|_{t=0}
= \frac{1}{2}({\cal L^\ast} D^{-1} + D^{-1}{\cal L}),
$$
is trace class. Set
\begin{equation}
\sigma(x)=(x,\varsigma x) -{\rm tr}(D\varsigma).
\label{ent-gaussian}
\end{equation}

\bep  Suppose that {\rm (G1)} holds. Then:
\begin{enumerate}[{\rm (1)}]
\item $\sigma \in L^1(M, \d\omega)$ and $t\mapsto\sigma_t$ is
strongly continuous in $L^1(M, \d\omega)$.
\item $\ell_{\omega_t|\omega}=\int_0^{t}\sigma_{-s}\d s$
and $t\mapsto\e^{\ell_{\omega_t|\omega}}$ is strongly $C^1$ in $L^1(M, \d\omega)$.
\item $\omega_t(\sigma)={\rm tr}(\varsigma (D_t- D))$ and in particular
$\omega(\sigma)=0$.
\item ${\rm Ent}(\omega_t|\omega)=-\int_0^t{\rm tr}(\varsigma (D_{s}-D))\d s.$
\end{enumerate}
\label{gauss-prop-1}
\eep
This proposition implies that Assumption \Eone{} holds and that $\sigma$ is the entropy 
production observable of 
$(M, \phi, \omega)$. In some examples only finitely many matrix elements 
$\varsigma_{nm}$ are non-zero and in this case $\sigma$ is continuous.  
$\sigma$ is bounded only in the trivial case  $\varsigma=0$ and so Assumption 
\Etwo{} is not satisfied.  

Our next assumptions are:
\begin{quote}
\label{G2-def}
{\bf (G2)} For some constants $m_\pm$ and  all $t\in \rr$, $0<m_-\leq D_t\leq m_+<\infty$. 

\label{G3-def}
{\bf (G3)} The strong limit 
\[
\slim_{t \rightarrow \infty}D_t = D_+
\]
exists.
\end{quote}
Clearly, $m_-\leq D_+\leq m_+$. Let $\omega_+$ be the Gaussian measure on 
$(M, {\cal F})$ with covariance $D_+$.  
\bep   Suppose that {\rm (G1)-(G3)} hold. Then: 
\begin{enumerate}[{\rm (1)}]
\item For all $f\in C_\rr(M)$, 
\[
\lim_{t\rightarrow \infty}\omega_t(f)=\omega_+(f).
\]
\item $\sigma \in L^1(M, \d\omega_+)$ and 
\[
\langle \sigma\rangle_+ =\lim_{t\rightarrow \infty}\omega_t(\sigma) 
= {\rm tr}(\varsigma (D_+-D)) = \omega_+(\sigma).
\]
\end{enumerate}
\label{gauss-prop-2}
\eep
We shall call $\omega_+$ the NESS of $(M, \phi, \omega)$. Note that assumptions  (NESS2) and (NESS3) of Section 
\ref{subsec-bn-steady}
do not hold for $(M, \phi, \omega)$. 

Finally, we assume the existence of time reversal in the following form:
\begin{quote}
\label{G4-def}
{\bf (G4)} The exists a linear involution $\vartheta$ on $\ell_{\rr}^2(\Gamma)$ such that 
$\vartheta{\cal L}= -{\cal L}\vartheta$ and $\vartheta D=D \vartheta$. 
\end{quote}
This assumption implies that 
$D_{-t}= \vartheta D_t \vartheta$ also satisfies (G2) and (G3) and  
\[
D_-=\slim_{t\rightarrow -\infty} D_{t}= \vartheta D_+\vartheta.
\]
Moreover, $\vartheta\varsigma = -\varsigma\vartheta$ and ${\rm tr}(D\varsigma)=0$.

Since $\sigma$ is unbounded, 
\[
e_{t}(\alpha)=\log \omega(\e^{\alpha \int_0^t \sigma_{-s}\d s}),
\]
is a priori finite only for $\alpha \in [0,1]$.  Note that $e_t(\alpha)$ is real analytic on 
$]0,1[$. (G4) implies that 
\[
e_t(\alpha)=\log \omega(\e^{-\alpha \int_0^t \sigma_{s}\d s}),
\]
and that $e_t(\alpha) =e_t(1-\alpha)$. Set $\delta=m_-/(m_+-m_-)$. 
\bep  Suppose that {\rm (G1)-(G4)} hold. Then:
\begin{enumerate}
\item $e_t(\alpha)$ is finite and  real analytic on the interval $]-\delta, 1+\delta[$.
\item $D_\alpha=\left((1-\alpha)D_+^{-1} + \alpha D_-^{-1}\right)^{-1}$
is a real analytic, bounded  operator valued function of 
$\alpha$  on this interval.
\end{enumerate}
\eep
 
With these preliminaries, the Evans-Searles Fluctuation Theorem holds in the following 
form.
\bet Suppose that {\rm (G1)-(G4)} hold. Then:
\begin{enumerate}[{\rm (1)}]
\item For  $\alpha \in ]-\delta, 1+\delta[$, 
\begin{equation}
e(\alpha)=\lim_{t\rightarrow \infty}\frac{1}{t} e_t(\alpha)= 
- \int_0^\alpha {\rm tr}(\varsigma D_\gamma)\d \gamma.
\label{grf-es}
\end{equation}
The ES-functional $e(\alpha)$ is real analytic and convex on the interval 
$]-\delta, 1+\delta[$, satisfies the ES-symmetry and
$e^\prime(0)=-\langle \sigma \rangle_+$.
\item If $t_n$ is a regular sequence, then 
\[
\lim_{n\rightarrow \infty}\frac{1}{t_n}\int_0^{t_n}\sigma_s(x)\d s =\langle \sigma \rangle_+,
\]
for $\omega$-a.e. $x$.
\item The Large Deviation Principle holds in the following form. The function
\[
I(s)= \inf _{\alpha \in]-\delta, 1+\delta[}(\alpha s + e(\alpha)).
\]
is a concave with values in $[-\infty,0]$, $I(s)=0$ iff $s=\langle \sigma \rangle_+$ and 
$I(s)= s + I(-s)$. Moreover, there is $\epsilon>0$ such that for any open interval 
$J \subset ]-\langle \sigma \rangle_+-\epsilon, \langle \sigma \rangle_+ + \epsilon[$, 
\[
\lim_{t \rightarrow \infty} 
\frac{1}{t} \log \omega\left(\left\{
 x\,\, \bigg |\,\, \frac{1}{t}\int_0^t \sigma_s(x) \d s\in J\right\}\right) = 
\sup_{s\in J} I(s).
\]
\end{enumerate}
\label{prop-ga-1}
\eet

We now turn to the Gallavotti-Cohen Fluctuation Theorem. Let 
\[
e_{t+}(\alpha)=\log \omega_+(\e^{-\alpha \int_0^t \sigma_s\d s}).
\]
A priori  $e_{t+}(\alpha)$ might not be finite for any $\alpha$.

\bep Suppose that {\rm (G1)-(G4)} hold. Then: 
\begin{enumerate}[{\rm (1)}]
\item $e_{t+}(\alpha)$ is real analytic on the interval $]-\delta, 1+\delta[$ and for any 
$\alpha$ in this interval, 
\begin{equation}
e_+(\alpha)=\lim_{t\rightarrow \infty}\frac{1}{t} e_{t+}(\alpha)= 
- \int_0^\alpha {\rm tr}(\varsigma D_\gamma)\d \gamma = e(\alpha).
\label{grf-gc}
\end{equation}
In particular, $(M, \phi, \omega, \omega_+)$ has regular  entropic fluctuations.

\item If $t_n$ is a regular sequence, then 
\[
\lim_{n\rightarrow \infty}\frac{1}{t_n}\int_0^{t_n}\sigma_s(x)\d s 
=\langle \sigma \rangle_+,
\]
for $\omega_+$-a.e. $x$.

\item The Large Deviation Principle holds for $\sigma$ and $(M, \phi, \omega_+)$, i.e.,
for some $\epsilon>0$  and for any open interval
$J \subset ] -\langle \sigma \rangle_+-\epsilon, \langle \sigma \rangle_+ + \epsilon[$, 
\[
\lim_{t \rightarrow \infty} 
\frac{1}{t} \log \omega_+\left(\left\{ x\,\, \bigg |\,\,
 \frac{1}{t}\int_0^t \sigma_s(x) \d s\in J\right\}\right) = 
\sup_{s\in J} I(s).
\]
\end{enumerate}
\label{prop-ga-2}
\eep

The proofs of all the results described in this section can be found in \cite{JLTP} and
here we will only sketch the computations 
leading to  the Formulas (\ref{grf-es}) and (\ref{grf-gc}).

Using
\[
\partial_\alpha e_t(\alpha) =-\frac{ \omega\left(\left[ \int_0^t \sigma_s \d s\right]\e^{-\alpha \int_0^t \sigma_s \d s}\right)}
{\omega(\e^{-\alpha \int_0^t \sigma_s \d s})},
\]
and the fact that 
$\e^{-\alpha \int_0^t \sigma_s}\d\omega\big/\omega(\e^{-\alpha \int_0^t \sigma_s \d s})$
is a Gaussian measure of covariance 
$\left[(1-\alpha)D^{-1} + \alpha D_{-t}^{-1}\right]^{-1}$,
we get 
\[
\begin{split}
e_t(\alpha) &= -\int_0^{\alpha}\int_0^t {\rm tr}\left(
\e^{s {\cal L}^\ast }\varsigma \e^{s {\cal L}}
\left[ (1-\gamma)D^{-1} + \gamma D_{-t}^{-1}\right]^{-1}\right)\d s \d\gamma\\[3mm]
&=-\int_0^{\alpha}\int_0^t {\rm tr}\left(\varsigma\left[
(1-\gamma)D_s^{-1} + \gamma D_{-t+s}^{-1}\right]^{-1}\right) \d s \d\gamma\\[3mm]
&=-t\int_0^{\alpha}\int_0^1 {\rm tr}\left(\varsigma \left[ 
(1-\gamma)D_{ts}^{-1} + \gamma D_{-t(1-s)}^{-1}\right]^{-1}\right) \d s\d\gamma,
\end{split}
\]
and so 
\[
\begin{split}
\lim_{t \rightarrow \infty} \frac{1}{t}e_t(\alpha) =
-\lim_{t\rightarrow \infty}\int_0^{\alpha}\int_0^1 
{\rm tr}\left(\varsigma \left[ (1-\gamma)D_{ts}^{-1} 
+ \gamma D_{-t(1-s)}^{-1}\right]^{-1}\right) \d s\d\gamma
=-\int_0^\alpha {\rm tr}(\varsigma D_\gamma)\d\gamma.
\end{split}
\]
Regarding (\ref{grf-gc}),  we have
\[
\partial_\alpha e_{t+}(\alpha) =-\frac{ \omega_+\left(\left[ \int_0^t
 \sigma_s \d s\right]\e^{-\alpha \int_0^t \sigma_s \d s}\right)}
{\omega_+(\e^{-\alpha \int_0^t \sigma_s \d s})},
\]
where $\e^{-\alpha \int_0^t \sigma_s}\d\omega_+\big/
\omega(\e^{-\alpha \int_0^t \sigma_s \d s})$ is a Gaussian measure with covariance 
$\left[ D_+^{-1} + \alpha D_{-t}^{-1} - \alpha D^{-1}\right]^{-1}$. 
Proceeding as before, we get
\[
e_{t+}(\alpha)= -t\int_0^{\alpha}\int_0^1 
{\rm tr}\left(\varsigma \left[ D_+^{-1} +\gamma D_{-t(1-s)}^{-1} -
 \gamma D_{ts}^{-1}\right]^{-1}\right) \d s\d\gamma,
\]
and hence, 
\[
\lim_{t \rightarrow \infty} \frac{1}{t}e_{t+}(\alpha) = 
-\lim_{t\rightarrow \infty}\int_0^{\alpha}\int_0^1 
{\rm tr}\left(\varsigma \left[ D_+^{-1} +\gamma D_{-t(1-s)}^{-1} 
- \gamma D_{ts}^{-1}\right]^{-1}\right) \d s\d\gamma
=-\int_0^\alpha {\rm tr}(\varsigma D_\gamma)\d\gamma.
\]

We finish with several remarks.

Regarding the resonance 
interpretation of $e(\alpha)$ and $e_+(\alpha)$, since $\sigma$ is unbounded the study of  
Liouvilleans and their resolvents requires some care. Regarding generalized functionals 
and symmetries, if  ${\cal L}_X$ and $D_X$ depend on  control parameters $X$,  then 
under mild additional regularity assumptions one can compute  $g(X, Y)$, $g_+(X,Y)$ and
prove the Fluctuation Dissipation Theorem. Again, $g(X, Y)= g_+(X,Y)$ and the principle
of  regular fluctuations holds.

\section{Homeomorphisms  of compact metric spaces}
\label{sec-homeo}

Let $(M, d)$ be a compact metric space and $\phi: M \rightarrow M$ a 
homeomorphism. In this section we consider  the discrete time  dynamical system 
on $M$ generated by $\phi$. We shall show, using the thermodynamic formalism, 
that if $\phi$ is sufficiently "chaotic", then the ES and GC Fluctuation Theorems,
the Fluctuation-Dissipation theorem and the principle of regular entropic fluctuations
hold for a large class of reference states $\omega$.
Our  treatment generalizes \cite{MV,Ma1}. Some of the results presented
here extend also to flows (see Remark 4 on Anosov flows at the end Section 
\ref{sec-anosov}).

The material in this section is organized as follows.
Subsection \ref{TopoDynRev} is a brief review of some basic aspects of 
topological dynamics (see \cite{Wa1} for a more detailed introduction).
Subsection \ref{TopoChaos} deals more specifically with two classes of "chaotic"
topological dynamics: expansive homeomorphisms with specification (see \cite{KH})
and Smale spaces (see \cite{Ru1}). The reader familiar with these topics can skip 
Subsections \ref{TopoDynRev},\ref{TopoChaos} and proceed directly to Subsection 
\ref{TopoEP} where, adopting the point of view of \cite{MV,Ma1}, we discuss entropy
production and fluctuation theorems for chaotic homeomorphisms.  We show in
Subsection \ref{ContactFinaly} how these results relate with the general approach,
based on the notion of reference state, advocated in this work. Finally, we
discuss, as an example, the simple case of a topological Markov chain in Subsection 
\ref{TopoMark}.

\subsection{Topological dynamics}
\label{TopoDynRev}
\subsubsection{Entropy and pressure}

Let $\nu \in {\cal S}_I$ and let $\xi=(C_1, C_2, \ldots, C_r)$, $C_j\in {\cal F}$,
be a finite measurable partition of $M$. By a standard subadditivity argument, the limit
\[
h_\nu(\phi,\xi) \,=\,- \lim_{k \to \infty}\frac1k \sum_{j_1,\ldots,j_{k}\in\{1,\ldots,r\}} 
\nu( \phi^{-1}(C_{j_1})\cap\cdots\cap\phi^{-k}(C_{j_k}) )
\log\nu( \phi^{-1}(C_{j_1})\cap\cdots\cap\phi^{-k}(C_{j_k})),
\]
exists.  The Kolmogorov-Sinai entropy of $\phi$ w.r.t. $\nu\in {\cal S}_I$
is given by 
\begin{equation}
h_\nu(\phi) \,=\, \sup_{\xi} h_\nu(\phi, \xi).
\label{KSEnt}
\end{equation}
The map
$\nu\mapsto h_\nu(\phi)$ is affine, that is,
$$
h_{\lambda\nu+(1-\lambda)\mu}(\phi)=\lambda h_\nu(\phi)+(1-\lambda)h_\mu(\phi),
$$
for any $\nu,\mu\in{\cal S}_I$ and $\lambda\in[0,1]$.

For $x\in M$  and $\epsilon >0$ we denote by 
\[
B_{\pm n}(x, \epsilon)=\{y\in M\,|\, \max _{0\leq j\leq n-1}
d(\phi^{\pm j}(x), \phi^{\pm j}(y))<\epsilon\},
\]
the Bowen ball of order $\pm n$ .
The following  result is known as the Brin-Katok local entropy formula \cite{BK} 
and is a  topological version of the Shannon-McMillan-Breiman theorem.
\bet Suppose that  $\nu\in {\cal S}_I$ is non-atomic and that $h_\nu(\phi)$ is finite. 
Then for $\nu$-a.e. $x\in M$,
\[
\lim_{\epsilon \rightarrow  0}\limsup_{n\rightarrow \infty}\,
-\frac1n \log\nu(B_n(x, \epsilon))
=\lim_{\epsilon \rightarrow  0}\liminf_{n\rightarrow \infty}\,
-\frac1n \log\nu(B_n(x, \epsilon))= h_\nu(x).
\]
The function $h_\nu(x)$  is $\phi$-invariant and $\nu(h_\nu)=h_\nu(\phi)$. If $\nu$
is ergodic, then $h_\nu(x)=h_\nu(\phi)$ for $\nu$-a.e. $x$.
\label{KatBrin}
\eet

For any function $\varphi\in C(M)$ we set 
\[
S_n\varphi(x) \,=\, \sum_{j=0}^{n-1}\varphi\circ\phi^j(x).
\]
A set $E\subset M$ is said  to be $(n, \epsilon)$-separated 
if for every $x, y \in E$, $x\not=y$, we have $y \not \in B_n(x, \epsilon)$. 
For $\varphi\in C_\rr(M)$  let
\[
Z_n(\varphi,\epsilon)=
\sup_E\sum_{x \in E}\e^{ S_n\varphi(x)}, 
\]
where the supremum is taken over all $(n, \epsilon)$-separated subsets of $M$.
The limit
\begin{equation}
P(\varphi)=\lim _{\epsilon \rightarrow 0}\limsup_{n \rightarrow \infty}\frac{1}{n}
\log Z_{n}(\varphi,\epsilon),
\label{tpres}
\end{equation}
exists and $P(\varphi)$ is called the {\em topological pressure} of $\phi$ with respect to
the potential $\varphi$. Alternative representations of the pressure are
\begin{equation}
P(\varphi)=\lim _{\epsilon \rightarrow 0}
\limsup_{n \rightarrow \infty}\frac{1}{n}\log 
\sum_{x \in E_{n, \epsilon}}\e^{S_n\varphi(x)}
=\lim _{\epsilon \rightarrow 0}\liminf_{n \rightarrow \infty}
\frac{1}{n}\log \sum_{x \in E_{n, \epsilon}}\e^{S_n\varphi(x)},
\label{PressUre}
\end{equation}
where the $E_{n, \epsilon}$  are arbitrary maximal $(n, \epsilon)$-separated sets.
The pressure also satisfies the variational principle
\begin{equation}
P(\varphi)=\sup_{\nu \in {\cal S}_I}
\left(\nu(\varphi) + h_\nu(\phi)\right).
\label{defpre}
\end{equation}
The special case $P(0)= \sup_{\nu \in {\cal S}_I}h_\nu(\phi)$ is the {\em topological
entropy} of $\phi$. An immediate consequence of the variational principle is  that  
$P(\varphi)$ depends only on the topology of $M$ and not on the choice of metric $d$. 
Another consequence is that the map $C_\rr(M)\ni\varphi\mapsto  P(\varphi)$ is convex and that 
either $P(\varphi)=+\infty$ for all $\varphi$, or $P(\varphi)$ is finite for all $\varphi$.
In what follows we assume that $P(\varphi)$ is finite for all $\varphi$. Occasionally we
shall use an additive normalization $\widehat\varphi=\varphi-P(\varphi)$ which ensures
that $P(\widehat\varphi)=0$.

\subsubsection{Potentials and equilibrium states}
An invariant measure $\nu$ is called an {\em equilibrium state} for the potential $\varphi$
if the supremum in (\ref{defpre}) is realized at $\nu$. In the case
$\varphi=0$ the equilibrium states are called measures of maximal entropy. 
In general, equilibrium states do not necessarily exist, and if they
exist they are not necessarily unique. The set ${\cal S}_\eq(\varphi)$ of all equilibrium
states for $\varphi$ is obviously convex. If ${\cal S}_{\eq}(\varphi)$ is singleton, we 
denote by $\nu_\varphi$ the unique equilibrium state for $\varphi$. 
\bet Suppose that the entropy map 
\begin{equation}
{\cal S}_I \ni \nu\mapsto h_\nu(\phi),
\label{ent-map}
\end{equation}
is upper-semicontinuous. Then:
\begin{enumerate}[{\rm (1)}]
\item For all $\varphi\in C_\rr(M)$ the set ${\cal S}_\eq(\varphi)$ is non-empty and compact.
A measure $\nu$ is an extreme point of ${\cal S}_\eq(\varphi)$ iff $\nu$ is $\phi$-ergodic.
\item For a dense set of $\varphi$ in $C_\rr(M)$ the set ${\cal S}_\eq(\varphi)$ is a singleton.
\item For $\varphi\in C_\rr(M)$, the map $\rr^N\ni Y \mapsto P(\varphi+Y\cdot{\mathbf f})$
is differentiable at $0$ for all ${\mathbf f}\in C_\rr(M)^N$ iff 
${\cal S}_\eq(\varphi)$ is a singleton and in this case 
\[
\nabla_Y P(\varphi+Y\cdot{\mathbf f})\biggr|_{Y=0}=\nu_\varphi({\mathbf f}).
\]
\end{enumerate}
\label{c-song}
\eet

The proofs of (1), (2) and (3) for $N=1$ can be found in \cite{Wa1}.
For a proof of (3) for $N>1$ see \cite{Je}. 

\bigskip
For a given $\varphi \in C_\rr(M)$ we denote by ${\cal S}_\varphi$ the collection of all 
$\nu \in {\cal S}$  such that  for all $x\in M$, $n>0$ and sufficiently small $\epsilon>0$, 
\begin{equation}
C_n(\epsilon)^{-1}\leq 
\nu(B_n(x, \epsilon)) \e^{- S_n\varphi(x)}\leq 
C_n(\epsilon),
\label{kiff1}
\end{equation}
where $C_n(\epsilon)>0$ satisfies
\begin{equation}
\lim_{n \rightarrow \infty}\frac{1}{n}\log C_n(\epsilon)=0.
\label{kiff2}
\end{equation}

The main property of the class $S_\varphi$ is the following basic result of Kifer \cite{Ki}
which relates it to the large deviation formalism trough the topological pressure.

\begin{proposition}\label{kif} For  $\nu \in {\cal S}_{\varphi}$ and $\psi\in C_\rr(M)$, 
\begin{equation}
\lim_{n \rightarrow \infty}\frac{1}{n}\log\nu (\e^{S_n\psi})=P(\varphi+\psi).
\label{kif-form}
\end{equation}
\end{proposition}

\proof Let $E_{n, \epsilon}$ be arbitrary maximal $(n, \epsilon)$-separated sets. Note that
for two different $x, y\in E_{n, \epsilon}$ one has
$B_n(x, \epsilon/2)\cap B_n(y, \epsilon/2)=\emptyset$ and by maximality 
$\cup _{x\in E_{n, \epsilon}}B_n(x, \epsilon)= M$.
It follows that
$$
 \sum_{x\in E_{n,\epsilon}}\nu(1_{B_n(x, \epsilon/2)}\e^{S_n\psi})
\leq 
\nu(\e^{S_n\psi})
 \leq
 \sum_{x\in E_{n,\epsilon}}\nu(1_{B_n(x, \epsilon)}\e^{S_n\psi}),
$$
where $1_B$ denotes the indicator function of the set $B$.
Setting $ \delta_\epsilon=\sup_{x, y \in M, d(x,y)<\epsilon}|\psi(x)-\psi(y)|$, we
obtain
 \[
 \sum_{x\in E_{n,\epsilon}}\nu(B_n(x, \epsilon/2))\e^{S_n\psi(x)- n \delta_\epsilon}
\leq 
\nu(\e^{S_n\psi})
 \leq
 \sum_{x\in E_{n,\epsilon}}\nu(B_n(x, \epsilon))\e^{S_n\psi(x)+n \delta_\epsilon}.
 \]
Combining these estimates with (\ref{kiff1}) we get 
\[
 C_n(\epsilon/2)^{-1}\sum_{x\in E_{n, \epsilon}}\e^{S_n(\varphi+\psi)(x)-n \delta_\epsilon}
\leq \nu(\e^{S_n\psi})\leq
 C_n(\epsilon)\sum_{x\in E_{n,\epsilon}}\e^{S_n(\varphi+\psi)(x)+n \delta_\epsilon},
\]
and so 
\begin{equation*}
\begin{split}
\limsup_{n\rightarrow \infty}\frac{1}{n}\log
\sum_{x\in E_{n, \epsilon}}\e^{S_n(\varphi+\psi)(x)}&\geq 
\limsup_{n\rightarrow \infty}\frac{1}{n}\log \nu(\e^{S_n\psi}) -\delta_{\epsilon},\\[3mm]
\liminf_{n\rightarrow \infty}\frac{1}{n}\log
\sum_{x\in E_{n, \epsilon}}\e^{S_n(\varphi+\psi)(x)}&\leq
\liminf_{n\rightarrow \infty}\frac{1}{n}\log \nu(\e^{S_n\psi}) +\delta_{\epsilon}.
\end{split}
\end{equation*}
Since $\delta_\epsilon \downarrow 0$ as $\epsilon \downarrow 0$, the statement now
follows from Equ. (\ref{PressUre}). \qed

\bec
Suppose that the entropy map (\ref{ent-map}) is upper-semicontinuous and that
 ${\cal S}_\eq(\varphi)$ is a singleton. Then, for all $\nu\in {\cal S}_{\varphi}$, 
\[
\lim_{n\rightarrow \infty}\frac{1}{n}S_nf(x)
=\nu_\varphi(f),
\]
for all $f \in C_\rr(M)$ and $\nu$-a.e. $x\in M$. In particular, $\nu_\varphi$ is the unique 
NESS of the system $(M,\phi,\nu)$.
\label{PhysicalCriter}
\eec

\proof By Proposition \ref{kif}, the generating function
$$
\lim_{n \rightarrow \infty}\frac{1}{n}\log\nu (\e^{-\alpha S_nf})=P(\varphi-\alpha f),
$$
exists for all $\alpha\in\rr$. By Theorem \ref{c-song}, it is differentiable at $\alpha=0$
so the statement follows from the G\"artner-Ellis Theorem
(Proposition \ref{gartnerellis} (1)).
\qed

As an immediate consequence of the Katok-Brin local entropy 
formula (Theorem \ref{KatBrin}) we have:

\bep Suppose that $\nu \in {\cal S}_I\cap{\cal S}_\varphi$ is non-atomic. 
Then for small enough $\epsilon$ and $\nu$-a.e. $x\in M$, 
\[
h_\nu(x)=
-\lim_{n\rightarrow \infty}\frac{1}{n}\log \nu(B_n(x, \epsilon))
= -\lim_{n\rightarrow \infty}\frac{1}{n}S_n\varphi(x),
\]
and in particular $\nu(\varphi) + h_\nu(\phi)=0$.
\label{rain-v}
\eep

\begin{corollary}
If ${\cal S}_\varphi$ is non-empty,  then   $P(\varphi)=0$. Moreover, if 
$\nu\in {\cal S}_I \cap {\cal S}_{\varphi}$ is non-atomic, then 
$\nu\in {\cal S}_\eq(\varphi)$.
\label{cor-kif}
\end{corollary}
\proof Setting $\psi=0$ in (\ref{kif-form}) we get $P(\varphi)=0$. This fact and Proposition 
\ref{rain-v} imply the second statement. \qed

We shall say that the potentials $\varphi$ and $\psi$ are {\em physically equivalent}, 
denoted $\varphi \sim \psi$, if
\[
\lim_{n\rightarrow \infty}\frac{1}{n}\sup_{x\in M}\big|S_n\varphi(x) - S_n\psi(x)\big|=0.
\]
This clearly defines an equivalence relation on $C_\rr(M)$. The following facts are easy to 
prove: 
\bep\begin{enumerate}[{\rm (1)}]
\item If $\varphi_{1}\sim \psi_{1}$ and $\varphi_{2}\sim \psi_{2}$, then
$a\varphi_1+b\varphi_2\sim a\psi_1+b\psi_2$ for all $a,b\in\rr$.
\item $\varphi\sim \psi$ implies $P(\varphi)=P(\psi)$ and $\nu(\varphi)=\nu(\psi)$
for all $\nu\in {\cal S}_I$. In particular, ${\cal S}_\eq(\varphi)={\cal S}_\eq(\psi)$.
\item ${\cal S}_\varphi= {\cal S}_\psi$ iff $\varphi\sim \psi$.
\item Either ${\cal S}_\varphi\cap {\cal S}_\psi=\emptyset$ or 
${\cal S}_\varphi= {\cal S}_\psi$.
\item Note that ${\cal S}_\varphi$ depends on the choice of metric $d$. If $\widetilde d$
is a metric  equivalent to $d$ ({\sl i.e.,} $C^{-1}d \leq \widetilde d \leq C d$), 
then ${\cal S}_{\varphi, d}={\cal S}_{\varphi,\widetilde d}$.
\end{enumerate}
\label{lotofrain-v}
\eep

\subsection{Chaotic homeomorphisms}
\label{TopoChaos}

\subsubsection{Expansiveness and specification}
The non-triviality of ${\cal S}_\varphi$ can be deduced from  suitable "chaoticity"
assumptions on $\phi$. A homeomorphism $\phi$ is called {\em expansive} if
\begin{quote}
{\bf (ES1)} There exists $r>0$ such that if $d(\phi^n(x), \phi^n(y))\leq r$
for all $n\in \zz$ then $x=y$. 
\end{quote}

$r$ is called expansive constant of $\phi$.  If $\phi$ is expansive then
$h_\nu(\phi)=h_\nu(\phi,\xi)$ for any measurable partition $\xi$ such that
${\rm diam}(\xi)=\sup\{d(x,y)\,|\,x,y\in C, C\in\xi\}\le r$ and the entropy
map ${\cal S}_I\ni\nu\mapsto h_\nu(\phi)$ is upper-semicontinuous
(see Proposition 6.5 in \cite{Ru1}).

$\phi$ is called an {\em expansive homeomorphism with specification} if in addition to
(ES1) 
\begin{quote}
{\bf (ES2)} For each $\delta>0$, there exists an integer $p(\delta)>0$ such that the 
following holds: if $a<b$ are integers, $I_1,\ldots, I_n$ finite intervals of $\zz$ 
contained in $\{a,\ldots,b\}$ with ${\rm dist}(I_j,I_k)>p(\delta)$ for $j\not=k$, and 
$x_1,\ldots,x_n\in M$, then there is $x\in M$ such that $\phi^{b-a+p(\delta)}(x)=x$, and
$$
d(\phi^k(x), \phi^k(x_i))<\delta,
$$
for $k\in I_i$, $i=1,\ldots,n$.
\end{quote}

A potential $\varphi$ is called {\em regular} if for all sufficiently small $\epsilon >0$ there 
exists $C_\epsilon>0$ such that for 
all $x\in M$, all $n>0$ and all $y\in B_n(x, \epsilon)$, 
\[\big|S_n\varphi(x) -S_n\varphi(y) \big|<C_\epsilon.\]
If  $\varphi$ is regular so is  $\widehat \varphi =\varphi-P(\varphi)$.

Expansive homeomorphisms with specification were introduced by Bowen in \cite{Bo1} 
and have been much studied since then. We recall the following classical result of Bowen
 \cite{Bo1} (see also \cite{KH}).
\bet Suppose that $\phi$ is an expansive homeomorphism with specification
and $\varphi$ a regular potential. Then ${\cal S}_\eq(\varphi)$ is a singleton  and 
$\nu_\varphi\in{\cal S}_{\widehat \varphi}$. 
 \label{bow}
\eet

An expansive homeomorphism with specification $\phi$ has a rich set of periodic points
which completely determine the equilibrium state $\nu_\varphi$ of a regular potential 
$\varphi$. Indeed,
\begin{equation}
\nu_\varphi(f)=\lim_{n\to\infty}\frac1{Z_n(\varphi)}
\sum_{x\in{\rm Fix}(\phi^n)}\e^{S_n\varphi(x)}f(x),
\label{ExpanEquiState}
\end{equation}
for all $f\in C(M)$, where
$$
Z_n(\varphi)=\sum_{x\in{\rm Fix}(\phi^n)}\e^{S_n\varphi(x)},
$$
and ${\rm Fix}(\phi^n)=\{x\in M\,|\, \phi^n(x)=x\}$, the set of periodic points of $\phi$
of period $n$ (see \cite{Bo1}). Moreover, the pressure of $\varphi$ is given by
\begin{equation}
P(\varphi)=\lim_{n\to\infty}\frac1n\log Z_n(\varphi),
\label{ExpanPress}
\end{equation}
(see {\sl e.g.} \cite{KH}, Proposition 20.3.3). These two approximation results
lead to the following characterization of physical equivalence.

\bep Suppose that $\phi$ is an expansive homeomorphism with specification and that 
$\varphi, \psi$ are regular potentials. Then the following statements are equivalent:
\begin{enumerate}[{\rm (1)}]
\item$\varphi\sim\psi$.
\item For all $n$ and all $x\in{\rm Fix}(\phi^n)$, 
$
S_n\varphi(x) = S_n\psi(x).
$
\end{enumerate}
\label{last-v}
\eep
\proof For $x\in{\rm Fix}(\phi^n)$ and $k\in\nn$, one has 
$S_{kn}(\varphi-\psi)(x)=kS_n(\varphi-\psi)(x)$ and hence
$$
|S_n(\varphi-\psi)(x)|\le n\,\frac1{kn}\sup_{x\in M}|S_{kn}(\varphi-\psi)(x)|.
$$
Letting $k\to\infty$ shows that (1) $\Rightarrow$ (2). Suppose (2) holds, then 
Equ. (\ref{ExpanPress}) implies $P(\varphi)=P(\psi)$ and Equ. (\ref{ExpanEquiState})
$\nu_\varphi=\nu_\psi$. Theorem \ref{bow} now implies that
$\nu_\varphi\in{\cal S}_{\widehat\varphi}\cap{\cal S}_{\widehat\psi}$
and Part (5) of Proposition \ref{lotofrain-v} yields 
${\cal S}_{\widehat\varphi}={\cal S}_{\widehat\psi}$. By Part (4) of the same
proposition we have $\widehat\varphi\sim\widehat\psi$ and, since $P(\varphi)=P(\psi)$,
we conclude that $\varphi\sim\psi$. \qed

\subsubsection{Smale spaces}
One can say more under a stronger "chaoticity" assumption. $(M, \phi)$ is called
{\em Smale space} if the following holds:
\begin{quote}
{\bf (S1)} For some $\epsilon>0$, there exists a continuous map 
\[ 
[\,\cdot\, , \,\cdot\,]: \{ (x,y)\in M\times M\,|\, d(x, y)<\epsilon\} \rightarrow M,
\]
such that $[x,x]=x$, $[[x,y], z]=[x,z]$, $[x, [y,z]]=[x,z]$ and $\phi([x,y])=[\phi (x), \phi(y)]$ whenever both sides
of these identities are defined.

{\bf (S2)} For some $\delta >0$, $0 <\lambda <1$ and all $n\in\nn$ one has
\[
\begin{split}
d(\phi^n(y), \phi^n(z))&\leq \lambda^n d(y, z) \qquad \hbox{if}
\qquad y,z \in \bigcap_{m=1}^\infty B_m(x, \delta),\\[3mm]
d(\phi^{-n}(y), \phi^{-n}(z))&\leq \lambda^n d(y, z) \qquad \hbox{if}
\qquad y,z \in \bigcap_{m=1}^\infty B_{-m}(x, \delta).
\end{split}
\]
\end{quote}
A Smale space is called {\em regular} if there exists $C>0$ such that
$d(x, [x,y])\leq Cd(x,y)$. 
A Smale space is {\em topologically + transitive} if there exists $x\in M$ such that the set 
$\{\phi^n (x) | n \geq 0\}$ is dense in $M$. $(M,\phi)$ is  a topologically + transitive
iff for any open sets  $U, V$ and any $N\geq 0$ there exists $n\geq N$ such that 
$\phi^n(U)\cap V\not= \emptyset$.
A Smale space is {\em topologically mixing} if, for any open sets  $U, V$ there exists 
$N\geq 0$ such that, for all $n\geq N$, $\phi^n(U)\cap V\not= \emptyset$.
If $(M,\phi)$ is a topologically + transitive Smale space then $\phi$ is an expansive 
homeomorphism with specification.  Note that topologically mixing $\Rightarrow$ 
topologically + transitive. 

Smale spaces can be studied using powerful tools of symbolic dynamics and are very 
well understood.  We will recall some classical result (see Chapter 7 in \cite{Ru1} and 
in particular Corollaries 7.10 and 7.12). For $\alpha\in]0,1[$ we denote by 
$C_{\rr/\cc}^{\alpha}(M)$ the real/complex vector space of all H\"older continuous 
functions with exponent $\alpha$, {\sl i.e.,} all $f\in C_{\rr/\cc}(M)$ such that, for some
$C>0$ and  all $x, y\in M$,  $|f(x)-f(y)|\leq C\,d(x,y)^\alpha$. The norm 
\[
\|f\|_\alpha =\sup_{x\not=y}\frac{|f(x)-f(y)|}{d(x,y)^\alpha} + \sup_x|f(x)|,
\]
turns $C_{\rr/\cc}^\alpha(M)$ into a real/complex  Banach space. 
\bet Let $(M, \phi)$ be a topologically + transitive Smale space and let $\alpha \in]0,1[$ 
be given.
\begin{enumerate}[{\rm (1)}]
\item The map $C_\rr^\alpha(M)\ni \varphi \mapsto P(\varphi)$ is real analytic.
\item For any  $\varphi\in C_\rr^\alpha(M)$, ${\cal S}_\eq(\varphi)$ is singleton and
$\nu_\varphi\in {\cal S}_{\widehat \varphi}$.
\item If $\varphi,\psi\in C_\rr^\alpha(M)$, then $\nu_\varphi=\nu_\psi$ iff $\varphi$ and 
$\psi$ are homologous, {\sl i.e.,} $\varphi = \psi +c + f\circ \phi -f$, 
where $c= P(\varphi)-P(\psi)$  and $f\in C_\rr(M)$  is unique up to an additive constant. 
If $(M, \phi)$ is regular, then $f \in C_\rr^\alpha(M)$.
\item If $\varphi, \psi \in C_\rr^{\alpha}(M)$, then  $\varphi \sim \psi$ iff  $\varphi$ and 
$\psi$ are homologous and $P(\varphi)=P(\psi)$.

\bigskip
Suppose that $(M, \phi)$ is topologically mixing. Then 
\bigskip

\item If $\varphi, f, g \in C_\rr^\alpha(M)$, then for some $A, B>0$ and all $n\in \zz$, 
\[
|\nu_\varphi(gf_n)-\nu_\varphi(g)\nu_\varphi(f)|\leq A\e^{-B|n|}.
\]
\item Suppose that $\varphi, f^{(1)}, \cdots, f^{(N)}\in C_\rr^\alpha(M)$. Then the
Central Limit Theorem holds  for ${\mathbf f}=(f^{(1)}, \cdots, f^{(N)})$ w.r.t. 
$(M, \phi, \nu_\varphi)$ with covariance matrix 
\[
D_{jk}=\sum_{n\in \zz}\left[\nu_\varphi(f^{(j)}f^{(k)}_{n})
-\nu_\varphi(f^{(j)})\nu_\varphi(f^{(k)})\right].
\]
Moreover, if $(M, \phi)$ is regular, then  $D_{kk}>0$ unless $f^{(k)}$ is homologous to 
$0$.
\item Let  $\varphi, \psi  \in C_\rr^\alpha(M)$. Consider  the transfer operator 
\[
U_\psi f  = \e^{\psi} f\circ \phi,
\]
on $C_\rr(M)$ and let 
\[
{\cal R}(z)= \sum_{n=0}^\infty \e^{-n z}\nu_\varphi(U_\psi^n 1), 
\qquad \Re z > \sup_x|\psi(x)|.
\]
Then for some $\epsilon >0$ the  function ${\cal R}(z)$ has a meromorphic continuation 
to the half-plane  
\[
\Re z >P(\psi + \varphi)-P(\varphi)-\epsilon,
\]
and its only singularity is a simple pole at $P(\psi +\varphi)-P(\varphi)$.
\item Let $\varphi \in C_\rr^\alpha(M)$. Then there exists $\epsilon >0$  and 
$C_\epsilon >0$ such that, for all $\psi \in C_\cc^\alpha(M)$ with 
$\|\psi\|_\alpha <\epsilon$, 
\[ 
\sup _{n >0} \frac{1}{n}|\log \nu_\varphi (\e^{S_n\psi})| \leq C_\epsilon.
\]
\end{enumerate}
\label{rue-van}
\eet

For (1)-(6)  see  \cite{Ru1}. (7) and (8) are implicit in \cite{Ru1, Ba1} and are easily 
established using the well-known spectral properties of Ruelle transfer operators. Note that
(8) and Proposition \ref{bryc-prop} yield  CLT for  
$f^{(1)},\ldots, f^{(N)}\in C_\rr^\alpha(M)$ w.r.t. $(M,\phi,\nu_\varphi)$.

\subsection{Entropy production}
\label{TopoEP}

In this section we suppose that $\phi$ is \TRI{}  with a continuous time reversal $\vartheta$.
Note that the map $\nu\mapsto\nu\circ\vartheta$ preserves ${\cal S}_I$.
One easily checks, using the definition (\ref{KSEnt}), that
\begin{equation}
h_{\nu\circ\vartheta}(\phi)=h_\nu(\phi^{-1})=h_\nu(\phi).
\label{KSTR}
\end{equation}
It follows that, for any $\varphi\in C_\rr(M)$,
\begin{align}
P(\varphi)
=\sup_{\nu\in{\cal S}_I}\left(\nu(\varphi)+h_\nu(\phi)\right)
&=\sup_{\nu\in{\cal S}_I}\left(\nu\circ\vartheta(\varphi)
+h_{\nu\circ\vartheta}(\phi)\right)\nonumber\\
&=\sup_{\nu\in {\cal S}_I}\left(\nu(\varphi\circ\vartheta)+h_\nu(\phi)\right)
=P(\varphi\circ\vartheta).
\label{PresTRI}
\end{align}

To each potential $\varphi \in C_\rr(M)$ we associate the function 
\begin{equation}
\widetilde\sigma_\varphi = \varphi-\varphi\circ\vartheta.
\label{def-ent-dyn}
\end{equation}
As we shall see in the next section, $\widetilde\sigma_\varphi$ is closely related to
the entropy production observable $\sigma$ of the dynamical system $(M,\phi,\omega)$
(as defined in Equ. (\ref{sigmadefdiscrete}) ) for $\omega\in{\cal S}_\varphi$.
In this section, we investigate the intrinsic properties of $\widetilde\sigma_\varphi$
and its fluctuations.

\bep
\begin{enumerate}[{\rm (1)}]
\item$\widetilde\sigma_{\varphi}\circ\vartheta=-\widetilde\sigma_{\varphi}$
(compare with Proposition \ref{bas-ep}).
\item If $\varphi\sim\psi$ (i.e., ${\cal S}_\varphi={\cal S}_\psi$),
then $\widetilde\sigma_\varphi\sim\widetilde\sigma_\psi$.
\item Suppose that ${\cal S}_\eq(\varphi)$ is a singleton. Then 
$\nu_\varphi(\widetilde\sigma_\varphi)=0$ iff $\nu_\varphi\circ\vartheta =\nu_\varphi$.

\bigskip
In the remaining statements we assume that $\nu\in{\cal S}_I\cap{\cal S}_\varphi$ 
is non-atomic.
\bigskip
\item For $\epsilon$ small enough 
\[
\lim_{n\rightarrow \infty}\frac{1}{n}
\log \frac{\nu(B_n(x,\epsilon))}{\nu(B_n(\vartheta\circ\phi^{n-1}(x),\epsilon))}
=\lim_{n\rightarrow \infty}\frac{1}{n} S_n\widetilde\sigma_\varphi(x),
\]
for $\nu$-a.e. $x\in M$. If  $\nu$ is ergodic, then
\begin{equation}
\lim_{n\rightarrow \infty}\frac{1}{n}
\log\frac{\nu(B_n(x,\epsilon))}{\nu(B_n(\vartheta\circ\phi^{n-1}(x),\epsilon))}
=\nu(\widetilde\sigma_{\varphi}),
\label{hola}
\end{equation}
for $\nu$-a.e. $x\in M$.
\item $\nu(\widetilde\sigma_\varphi)\geq 0$.
\item Suppose that  $\nu$ is ergodic and that $\nu$ and $\nu\circ\vartheta$ are 
equivalent measures. Then $\nu(\widetilde\sigma_{\varphi})=0$.
\end{enumerate}
\label{theend}
\eep

{\bf Remark 1.}  Apart from (1), all statements of the previous proposition hold with the
same proofs if one replaces $\vartheta$ by $\vartheta\circ\phi^k$ with an arbitrary
$k\in\zz$. Our choice of $k=0$ differs from the one in \cite{MV}. A different choice
would not 	affect any result in this and the next section.

{\bf Remark 2.} By part (4), the observable $\widetilde\sigma_{\varphi}$ associated to a potential $\varphi \in C_\rr(M)$
quantifies the Brin-Katok local entropy produced by changing the reference  point  from 
$x$ to $\vartheta\circ\phi^{n-1}(x)$, {\sl i.e.,} by reversing the orbit of $x$.  

\proof (1)-(2) are obvious. 

(3) $\nu_\varphi(\widetilde\sigma_\varphi)=0$ is equivalent to
$\nu_\varphi(\varphi)=\nu_\varphi(\varphi \circ \vartheta)$ which, by
the variational principle and Equ. (\ref{KSTR}), is  equivalent to
\[
P(\varphi)=\nu_\varphi(\varphi)+h_{\nu_\varphi}(\phi)
=\nu_\varphi\circ\vartheta(\varphi)+h_{\nu_\varphi\circ\vartheta}(\phi).
\]
Hence $\nu_\varphi\circ\vartheta$ is also an equilibrium state for $\varphi$ 
and the uniqueness implies $\nu_\varphi=\nu_\varphi\circ\vartheta$.

(4) follows easily from the conditions (\ref{kiff1})-(\ref{kiff2})
and Birkhoff ergodic theorem. If $\nu\in{\cal S}_I\cap{\cal S}_\varphi$ is non-atomic, 
then by Corollary \ref{cor-kif}, $\nu\in{\cal S}_{\rm eq}(\varphi)$. Since 
$\nu\circ\vartheta\in{\cal S}_I$, the variational principle and Equ. (\ref{KSTR}) lead to
\[
\nu(\varphi)+h_\nu(\phi)=P(\varphi)
\geq\nu\circ \vartheta(\varphi)+h_{\nu\circ\vartheta}(\phi)
=\nu(\varphi\circ\vartheta)+h_\nu(\phi),
\]
and (5) follows.

(6) The Brin-Katok formula and (\ref{hola}) imply that 
$\nu(\widetilde\sigma_\varphi)=h_\nu(\phi)-h_\nu(\phi)=0$.
\qed

The  fluctuations of the observable $\widetilde\sigma_\varphi$ in the states
$\nu\in{\cal S}_\varphi$ are described in our next result.

\bep
\begin{enumerate}[{\rm (1)}]
\item For all $\nu \in{\cal S}_\varphi$ the functional
\[
\rr\ni\alpha\mapsto
e_\varphi(\alpha)=
\lim_{n \rightarrow \infty}\frac{1}{n}
\log\nu(\e^{-\alpha S_n\widetilde\sigma_{\varphi}}),
\]
exists and is given by $e_\varphi(\alpha)=P(\varphi-\alpha\widetilde\sigma_\varphi)$.
\item The symmetry $e_\varphi(\alpha) = e_\varphi(1-\alpha)$ holds.
\item If $\varphi\sim\psi$, then $e_{\varphi}(\alpha)=e_\psi(\alpha)$.

\item Suppose that the entropy map ${\cal S}_I \ni \nu\mapsto h_\nu(\phi)$ is 
upper-semicontinuous and that ${\cal S}_\eq (\varphi-\alpha\widetilde\sigma_\varphi)$ 
is singleton for all $\alpha\in\rr$ and let $\nu\in{\cal S}_\varphi$. Then the Large Deviation 
Principle holds for $\widetilde\sigma_\varphi$ w.r.t. $(M, \phi, \nu)$ with the concave rate
function $I_\varphi(s)= \sup_{\alpha \in {\mathbb R}}(\alpha s + e_\varphi(\alpha))$ 
which satisfies the relation $I_\varphi(s)= s + I_\varphi(-s)$.
\end{enumerate}
\label{believe}
\eep
\proof
(1) follows from Proposition \ref{kif}. Writing
$\varphi-\alpha\widetilde\sigma_\varphi=(1-\alpha)\varphi+\alpha\varphi\circ\vartheta$,
(2) follows immediately from Equ. (\ref{PresTRI}).
(3) is a direct consequence of (1) and Proposition \ref{lotofrain-v}.
(4) follows from Theorem \ref{c-song}, Proposition \ref{kif}, 
 and  the G\"artner-Ellis Theorem
(Proposition \ref{gartnerellis}).\qed

One can introduce control parameters in the above framework and discuss  
the generalized symmetry and linear response theory. Consider a map 
$\rr^N \ni X \mapsto \phi_X$, where each $\phi_X$ is a homeomorphism of $M$ with
continuous time reversal $\vartheta_X$, and a map 
$\rr^N \ni X \mapsto\varphi_X \in C_\rr(M)$. Let 
$\widetilde\sigma_X=\widetilde\sigma_{\varphi_X}$. We shall assume that
there exists ${\bf\Phi}_X=(\Phi_X^{(1)}, \cdots,\Phi_X^{(N)})$, with
$\Phi_X^{(j)}\in C_\rr(M)$, such that
\begin{equation}
\widetilde\sigma_X\sim X\cdot{\mathbf\Phi}_X, 
\label{WellBehaved}
\end{equation}
and ${\mathbf\Phi}_X\circ\vartheta_X\sim-{\mathbf\Phi}_X$. 
Finally, denote by ${\cal S}_{IX}$ the set of $\phi_X$-invariant states and by $P_X$ 
the pressure functional for the map $\phi_X$.

\bep
\begin{enumerate}[{\rm (1)}]
\item For all $\nu \in {\cal S}_{\varphi_X}$, the functional
\[
\rr^N\ni Y\mapsto{\mathfrak g}(X, Y)=
\lim_{n \rightarrow \infty}\frac{1}{n}
\log\nu (\e^{- S_n(Y\cdot{\bf\Phi}_{X})}),
\]
exists and is given by ${\mathfrak g}(X, Y) = P_X(\varphi_X-  Y \cdot {\bf\Phi}_X)$.
\item The symmetry ${\mathfrak g}(X, Y)={\mathfrak g}(X, X-Y)$ holds.

\bigskip
In the remaining statements we assume  that the entropy maps 
${\cal S}_{IX} \ni \nu\mapsto h_\nu(\phi_X)$ are upper-semicontinuous.
\bigskip

\item Suppose that ${\cal S}_\eq(\varphi_X)$ is a singleton and denote 
$\nu_X=\nu_{\varphi_X}$. Then for all $f \in C(M)$ and $\nu\in{\cal S}_{\varphi_X}$, 
\[
\lim_{n\rightarrow \infty} 
\frac{1}{n}S_nf(x) = \nu_{X}(f),
\]
for $\nu$-a.e. $x\in M$.
\item Suppose that ${\cal S}_\eq(\varphi_X)$ is a singleton for $X$ small enough 
and that  ${\mathfrak g}(X, Y)$ is $C^{1,2}$ in a neighborhood of $(0,0)$. Then the 
transport coefficients are defined and satisfy the Onsager reciprocity relations.
\item Suppose that the assumptions of {\rm (4)} hold and write $\nu=\nu_0$ and 
$\Phi^{(j)}=\Phi^{(j)}_0$. Then $\nu(\Phi^{(j)})=0$. 
Suppose in addition that $\nu(\Phi^{(k)}\Phi^{(j)}_n)=O(n^{-1})$ for $n\to\infty$
and that for some $\epsilon>0$, 
\[
\sup_{Y\in D_\epsilon,n>0}\frac1n|\log\nu(\e^{-S_n(Y\cdot {\bf \Phi})})|<\infty.
\]
Then the Fluctuation-Dissipation Theorem holds. 
\item Suppose that $Y\mapsto {\mathfrak g}(X,Y)$ is differentiable for all $Y$ and let 
$\nu \in {\cal S}_{\varphi_X}$. Then the Large Deviation Principle holds for ${\bf \Phi}_X$
w.r.t. $(M,\phi_X, \nu)$ with the concave rate function  
$ I_X(s)= \sup_{Y \in {\mathbb R^N}}(Y\cdot s + {\mathfrak g}(X, Y))$ 
which  satisfies the GGC-symmetry $I_X(s)= X \cdot s + I_X(-s)$. 
\end{enumerate}
\label{believe1}
\eep
The proof of Proposition \ref{believe1} (1) is the same as the proof of Proposition 
\ref{believe} (1). The  proof of the remaining  statements is the same as the proof of 
the corresponding statements in Proposition \ref{whistler-2}.

Assuming "chaoticity" one can say more. For example if  $\phi_X$ is an expansive
homeomorphism with specification and $\varphi_X$ and ${\bf \Phi}_X$ are regular, 
then the entropy maps are upper-semicontinuous,
${\cal S}_\eq(\varphi_X- Y\cdot {\bf \Phi}_X)$ is a singleton for all $Y$ (Theorem
\ref{bow}) and  the map $Y \mapsto {\mathfrak g}(X, Y)$ is everywhere differentiable
(Theorem \ref{c-song}). If $(M, \phi_X)$ is a topological + transitive Smale space  and 
$\varphi_X$, ${\bf \Phi}_X$ are H\"older  continuous, then the map 
$Y\mapsto {\mathfrak g}(X, Y)$ is real analytic.  
In addition, Theorem \ref{rue-van} yields: 

\bep  Suppose that $(M, \phi_X)$ is a topologically mixing Smale space and that 
 $\varphi_X$, 
${\bf \Phi}_X$,  are H\"older  continuous for $X$ in a neighborhood of $0$.   
Suppose also that 
${\mathfrak g}(X,Y)$ is $C^{1,2}$ in a neighborhood of $(0,0)$.  Then the 
Fluctuation-Dissipation Theorem holds. 
\label{sunny-van}
\eep


\subsection{Reference measure and physical equivalence} 
\label{ContactFinaly}

In this section we investigate the relation between the observable 
$\widetilde\sigma_\varphi$ introduced in the previous section and the entropy production
observable $\sigma$ defined by equ. (\ref{sigmadefdiscrete}). 
Throughout the section we make the following assumptions:  

The homeomorphism $\phi$ admit a time reversal $\vartheta$ and $\varphi\in C_\rr(M)$
is a potential.

The metric $d_\vartheta(x,y)=d(\vartheta(x),\vartheta(y))$ 
is equivalent to $d$, {\sl i.e.,} that there exists a constant $C>0$ such that 
$C^{-1}\,d(x,y)\le d_\vartheta(x,y)\le C\,d(x,y)$ for all $x,y\in M$. This requirement is
a mild regularity assumption on $\vartheta$. It follows that $d$ is also equivalent to 
$d+d_\vartheta$, so that we can as well assume that $\vartheta$ is isometric.
 
$\omega\in{\cal S}_\varphi$ is a \TRI{} reference state and the dynamical system 
$(M,\phi,\omega)$ satisfies Assumption \C{} with an entropy production observable
$\sigma=\ell_{\omega_1|\omega}\circ\phi\in C_\rr(M)$. 

The following key proposition relates $\sigma$ to $\widetilde\sigma_\varphi$.

\bep Under the above assumptions one has $\sigma\sim\widetilde\sigma_\varphi$.
\label{SimSigma}
\eep

\proof Using the elementary identity
$B_n(\vartheta\circ\phi^{n-1}(x),\epsilon)=\vartheta\circ\phi^{n-1}(B_{n}(x,\epsilon))$
and the fact that $\omega$ is \TRI{} we can write
$$
\omega(B_n(\vartheta\circ\phi^{n-1}(x),\epsilon))
=\omega_{-n+1}(B_{n}(x,\epsilon))
=\omega\left(\e^{-S_{n-1}\sigma}\,1_{B_{n}(x,\epsilon)}\right).
$$
We derive the inequalities
$$
\e^{-S_{n-1}\sigma(x)-(n-1)\delta_\epsilon}\omega(B_{n}(x,\epsilon))
\le\omega(B_n(\vartheta\circ\phi^{n-1}(x),\epsilon))
\le\e^{-S_{n-1}\sigma(x)+(n-1)\delta_\epsilon}\omega(B_{n}(x,\epsilon)),
$$
where $\delta_\epsilon=\sup_{x,y\in M,d(x,y)<\epsilon}|\sigma(x)-\sigma(y)|$.
With $c=\max_{x\in M}|\sigma(x)|$, we thus obtain, for arbitrary $n\in\nn$, $x\in M$
and $\epsilon>0$,
\begin{equation}
\e^{-n\delta_\epsilon-c}
\le\frac{\omega(B_n(\vartheta\circ\phi^{n-1}(x),\epsilon))}{\omega(B_{n}(x,\epsilon))}
\,\e^{S_n\sigma(x)}\le\e^{n\delta_\epsilon+c}.
\label{SunnyButCold}
\end{equation}
Since $\omega\in{\cal S}_\varphi$ and
$(S_n\varphi)(\vartheta\circ\phi^{n-1}(x))=S_n(\varphi\circ\vartheta)(x)$, the
estimate  (\ref{kiff1}) leads, for small enough $\epsilon>0$, to
$$
C_n(\epsilon)^{-1}
\le\omega(B_n(\vartheta\circ\phi^{n-1}(x),\epsilon))\,\e^{-S_n(\varphi\circ\vartheta)(x)}
\le C_n(\epsilon).
$$
Using again (\ref{kiff1}), we obtain 
$$
C_n(\epsilon)^{-2}
\le\frac{\omega(B_{n}(x,\epsilon))}{\omega(B_n(\vartheta\circ\phi^{n-1}(x),\epsilon))}
\,\e^{-S_n\widetilde\sigma_\varphi(x)}\le C_n(\epsilon)^2,
$$
which, combined with (\ref{SunnyButCold}), yields
$$
\frac1n\left|S_n(\sigma-\widetilde\sigma_\varphi)(x)\right|\le
\delta_\epsilon+\frac{c}n+\frac2n\log C_n(\epsilon).
$$
It follows that
$$
\limsup_{n\to\infty}\frac1n\sup_{x\in M}
\left|S_n(\sigma-\widetilde\sigma_\varphi)(x)\right|\le
\delta_\epsilon,
$$
and the proof is completed by noticing that $\delta_\epsilon\downarrow0$ as
$\epsilon\downarrow0$.
\qed

\bec 
\begin{enumerate}[{\rm (1)}]
\item The ES-functional of the dynamical system $(M,\phi,\omega)$
$$
\rr\ni\alpha\mapsto e(\alpha)=\lim_{n\to\infty}\frac1n\log\omega(\e^{-\alpha S_n\sigma}),
$$
exists and satisfies the ES-symmetry $e(1- \alpha)=e(\alpha)$.

\bigskip
In the remaining statements, we assume that $\phi$ is expansive with specification and
that the potential $\varphi$ is regular.
\bigskip

\item The ES-functional $e(\alpha)$ is everywhere differentiable.
\item The ES Fluctuation Theorem hold.
\item The system $(M,\phi,\omega)$ has a unique NESS $\omega_+$ and for
any $f\in C(M)$,
$$\lim_{n\to\infty}\frac1nS_nf(x)=\omega_+(f),
$$
holds for $\omega$-a.e. $x\in M$.
\item The GC Fluctuation Theorem hold.
\item The principle of regular entropic fluctuations hold.
\item If $\omega_+$ in non-atomic, the system is entropy producing iff 
$\omega_+\not=\omega_+\circ\vartheta$.
\end{enumerate}
\label{SoCold}
\eec

\proof (1) Proposition \ref{kif} yields the existence and the relation
$e(\alpha)=P(\varphi-\alpha\sigma)$. Proposition  \ref{SimSigma}
implies $\varphi-\alpha\sigma\sim\varphi-\alpha\widetilde\sigma_\varphi$
from which Proposition \ref{lotofrain-v} (2) allow us to
conclude that $e(\alpha)=e_\varphi(\alpha)$. Thus, the ES-symmetry follows
either from Proposition \ref{believe} (2) or directly from 
Proposition \ref{hang-over} (2).

Invoking Theorem \ref{bow}, (2)-(6) are direct consequences 
of Proposition \ref{believe}. In particular $\omega_+=\nu_\varphi\in{\cal S}_\varphi$ so
that (7) is a direct application of Proposition \ref{theend} (6).
\qed

{\bf Remark.} Suppose that $(M, \phi)$ is a topologically + transitive Smale space and
that $\widetilde\sigma_\varphi$ and $\sigma$ are H\"older continuous. 
Applying Propositions \ref{rue-van}, we conclude that the ES-functional $e(\alpha)$
is real analytic. Moreover, $\widetilde\sigma_\varphi$ and $\sigma$ are homologous, 
{\sl i.e.,} 
\begin{equation*}
 \widetilde\sigma_\varphi=\sigma+h\circ\phi-h,
 \end{equation*}
for some $h\in C_\rr(M)$ which is unique up to an additive constant and 
also H\"older continuous if $(M, \phi)$ is regular. 

Invoking Proposition \ref{believe1} similar results can be obtained for systems
$(M,\phi_X,\omega_X)$, depending on control parameters, with
\TRI{} reference states $\omega_X\in{\cal S}_{\varphi_X}$.
Note that if ${\bf\Phi}_X$ is a continuous flux relation for the corresponding
entropy production observable $\sigma_X$, then (\ref{WellBehaved}) holds.
We leave the details to the reader.


\subsection{Markov chains} 
\label{TopoMark}

We shall illustrate the results of this section on  the simple example of a Markov chain 
with finitely many states. The set of states is $\Omega=\{1, \cdots, l\}$,
\[
M=\Omega^\zz=\{x=(x_j)_{j\in \zz}\,|\, x_j \in\Omega\},
\]
with the usual product topology and $\phi: M\rightarrow  M$ is the left shift, 
$\phi(x)_j=x_{j+1}$. $M$ is metrizable and a convenient metric for our purposes is 
\[
d(x,y)=\lambda^{k(x,y)},
\]
where $\lambda \in ]0,1[$ is fixed and $k(x,y)=\inf\{|j|\,|\, x_j\not=y_j\}$.
If $d(x,y)<1$, then $x_0=y_0$. It follows that $\phi$ is expansive and that any
$r\in]0,1[$ is an expansive constant. Setting
\[
[x, y]= (\ldots, y_{-2}, y_{-1}, x_0, x_1,\ldots).
\]
one easily shows that $(M,\phi)$ is a regular, topologically mixing Smale space.

Any  function $f: M\rightarrow \rr$ which depends only on finitely many $x_j$'s
is H\"older continuous.  The map $\vartheta(x)_j=x_{-j}$ is an isometric time-reversal.

Let ${\mathbb P}=[p_{ij}]_{i,j\in\Omega}$ with $p_{ij} > 0$,  $\sum_{j}p_{ij}=1$, be 
a transition matrix. By the Perron-Frobenius theorem there is a unique probability vector
$\bar p = [\bar p_i]_{i\in\Omega}$ with $\bar p_i>0$ and $\sum_i \bar p_i=1$
such that $\bar p\,{\mathbb P}=\bar p$.

The (two-sided) Markov chain with transition matrix ${\mathbb P}$ is the invariant 
Borel probability measure $\nu\in {\cal S}_I$ such that, for any cylinder, 
\[
C=\{x\in M\,|\, x_{k}=j_1, x_{k+1}=j_2,  \ldots, x_{k+n-1}=j_n\},
\] 
we have 
\[
\nu(C)=\bar p_{j_1}p_{j_1 j_2}\cdots p_{j_{n-1}j_n}.
\]
The assumption that $p_{ij}>0$ implies that $\nu$ is mixing w.r.t. $\phi$. 

For the potential
\begin{equation}
\varphi(x) = \log p_{x_0 x_1},
\label{MarkoPot}
\end{equation}
and the corresponding observable
\[
\widetilde\sigma_\varphi(x)=\varphi(x)-\varphi\circ\vartheta(x)
=\log\frac{p_{x_0x_1}}{p_{x_{0} x_{-1}}}.
\]
one computes
\[
\nu(\varphi)=\sum_{i, j\in\Omega}\bar p_i p_{ij}\log p_{ij},\qquad
\nu(\widetilde\sigma_\varphi)=\sum_{i, j\in\Omega}\bar p_i p_{ij}\log \frac{p_{ij}}{p_{ji}}.
\]
Since $B_n(x,\lambda^k)=\{y\in M\,|\,y_i=x_i\text{ for }-k\le i\le k+n-1\}$ for $k\ge0$,
it easily follows that for any $0<\epsilon<1$ there is a constant $C_\epsilon$
such that, for all $n>0$,
\[
C_\epsilon^{-1}\leq\nu(B_n(x, \epsilon))\,\e^{-S_n\varphi(x)}\leq C_\epsilon.
\]
We conclude that $\nu\in{\cal S}_\varphi$.

The partition $\xi=(\{x\in M\,|\,x_0=i\})_{i\in\Omega}$ has ${\rm diam}(\xi)=\lambda<1$.
Hence, the Kolmogorov-Sinai entropy can be computed from $h_\nu(\phi)=h_\nu(\phi,\xi)$
and a simple calculation leads to
\[
h_\nu(\phi)=-\sum_{i,j\in\Omega}\bar p_i p_{ij}\log p_{ij}.
\]

The pressure $e_\varphi(\alpha)=P(\varphi-\alpha\widetilde\sigma_\varphi)$, a 
real analytic function of $\alpha$, is most easily computed from Equ. (\ref{ExpanPress}),
where we have
$$
Z_n(\varphi-\alpha\widetilde\sigma_\varphi)
=Z_n((1-\alpha)\varphi+\alpha\varphi\circ\vartheta)
=\sum_{x\in\Omega^n}(p_{x_1x_2}^{1-\alpha}p_{x_2x_1}^\alpha)\cdots
(p_{x_nx_1}^{1-\alpha}p_{x_1x_n}^\alpha)=\mathrm{tr}\,{\mathbb P}_\alpha^n,
$$
with the matrix ${\mathbb P}_\alpha=[p_{ij}(\alpha)]$, 
$p_{ij}(\alpha)= p_{ij}^{1-\alpha}p_{ji}^\alpha$.
Since $p_{ij}(\alpha)>0$, the Perron-Frobenius theorem applies to ${\mathbb P}_\alpha$
and consequently $e_\varphi(\alpha)$ is equal to the logarithm of its dominant eigenvalue.
In particular one checks
$$
P(\varphi)=0=\nu(\varphi)+h_\nu(\phi),
$$
so that $\nu$ is the unique equilibrium state for the potential $\varphi$.
Note that ${\mathbb P}_{\alpha}^\ast={\mathbb P}_{1-\alpha}$. The resulting
identity $\mathrm{tr}\,{\mathbb P}_\alpha^n=\mathrm{tr}\,{\mathbb P}_{1-\alpha}^n$
provides an alternative proof of the symmetry $e_\varphi(\alpha)=e_\varphi(1-\alpha)$.

Proposition \ref{theend} implies that $\nu(\widetilde\sigma_\varphi)\geq 0$ and
$\nu(\widetilde\sigma_\varphi)=0$ iff $\nu\circ\vartheta=\nu$.  The latter
condition is easily seen to equivalent to $\bar p_i p_{ij}=\bar p_j p_{ji}$ for all 
$i,j\in\Omega$. In other words, $\nu(\widetilde\sigma_\varphi)=0$ iff the Markov
chain satisfies detailed balance. Note that in this case
$\widetilde\sigma_\varphi=g\circ\phi-g\sim0$, with $g(x)=\log\bar p_{x_0}p_{x_0x_{-1}}$.

Let $q=[q_i]_{i\in\Omega}$ be a probability vector and $\omega$ the state uniquely
determined by
$$
\omega\left(\left\{x\in M\,|\,x_k=j_k, k=-m,\ldots,n\right\}\right)=
q_{j_0}\left(p_{j_0j_1}\cdots p_{j_{n-1}j_n}\right)\,
\left( p_{j_0j_{-1}}\cdots p_{j_{-m+1}j_{-m}}\right).
$$
A simple calculation shows that $\omega$ is \TRI{}. Moreover, $\omega\in{\cal S_\varphi}$
provided $q_i>0$ for all $i\in\Omega$. Thus, Corollary \ref{SoCold} applies to the \TRI{}
system $(M,\phi,\omega)$. Note in particular that its  entropy production observable
$$
\sigma(x)=\log\frac{q_{x_0}p_{x_0x_1}}{q_{x_1}p_{x_1x_0}},
$$
is homologous to $\widetilde\sigma_\varphi$. Explicitly,
$\widetilde\sigma_\varphi-\sigma=h\circ\phi-h$ with 
$h(x)=\log(q_{x_0}p_{x_0x_{-1}})$. The unique NESS of the system is
${\omega}_+=\nu$ and the system is entropy producing iff $\nu$ does not
satisfy detailed balance.

\bigskip
Suppose that the transition matrix ${\mathbb P}_X=[p_{ij}(X)]$ depends  on the control
parameters $X\in\rr^N$. We assume that  the functions $X\mapsto p_{ij}(X)$ are $C^2$
and that $p_{ij}(X)>0$ for all $X$. Denote by $\bar p(X)=[\bar p_i(X)]$ the corresponding
equilibrium vector. Let $\varphi_X$ be the corresponding potential, $\nu_X$ its equilibrium
state and set $\widetilde\sigma_X=\varphi_X-\varphi_X\circ\vartheta$. We assume that 
detailed balance holds for $X=0$ so that $\widetilde\sigma_0\sim0$.

For each $X\in\rr^N$ let $q(X)=[q_i(X)]$ be a probability vector such that 
$q_i(X)>0$ and assume that $q(0)=\bar p(0)$. Construct the \TRI{} state $\omega_X$ as
above and denote by $\sigma_X$ the corresponding entropy production observable.
The detailed balance condition at $X=0$ implies that $\omega_0=\nu_0$ and hence
$\sigma_0=0$.

Setting
$$
{\mathbf F}_X(i,j)=\int_0^1\frac{({\mathbf\nabla}p_{ij})(uX)}{p_{ij}(uX)}\,\d u,
$$
and ${\bf \Phi}_X(x)={\mathbf F}_X(x_0,x_1)-{\mathbf F}_X(x_1,x_0)$ we obtain
a flux relation,
\[
\sigma_X\sim\widetilde\sigma_X \sim X\cdot{\mathbf\Phi}_X,
\]
such that the map $X\mapsto{\mathbf\Phi}_X\in C(M)^N$ is differentiable and 
${\mathbf\Phi}_X\circ \vartheta\sim-{\mathbf\Phi}_X$. Arguing as before, the assumption
$p_{ij}(X)>0$ and the relation
$$
Z_n(\varphi_X- Y \cdot {\bf \Phi}_X)={\rm tr}\,{\mathbb P}(X,Y)^n,
$$
where ${\mathbb P}(X,Y)
=[p_{ij}(X)\e^{-Y\cdot({\mathbf F}_X(i,j)-{\mathbf F}_X(j,i))}]$ imply that
${\mathfrak g}(X, Y)=P_X(\varphi_X- Y \cdot {\bf \Phi}_X)$
is the logarithm of the dominant eigenvalue of ${\mathbb P}(X,Y)$.
The perturbation theory of isolated simple eigenvalue further implies that $\mathfrak g$
is $C^{1,2}$ in a neighborhood of $(0, 0)$ and all the conclusions of Propositions
\ref{believe1} and \ref{sunny-van} hold. Finally, for the family
$(M,\phi_X,\omega_X,{\mathbf\Phi}_X)$, one shows that 
\[
g(X,Y)=g_+(X,Y)={\mathfrak g}(X, Y).
\]

{\bf Remark.}  The above approach can be used to discuss entropic fluctuations of a wide 
range of stochastic processes with suitable modifications to accommodate
general (non-compact) state space, continuous-time,  as well as Gibbs measures
rather than 
Markov measures.  The entropic fluctuations of Markov chains were 
first discussed by Kurchan \cite{Ku1},   Lebowitz and Spohn  \cite{LS2}, and 
Maes et al. \cite{Ma1,MRV,MN}  who used the path measure approach  and the 
Gibbsian formalism.

\section{ Anosov diffeomorphisms} 
\label{sec-anosov}
Let $M$ be a  compact connected smooth Riemannian manifold with a given Riemannian
metric and let $\omega$ be the induced volume measure on $M$. We denote by 
${\rm Diff}^k(M)$ the set of all $C^k$ diffeomorphisms of $M$ equipped with the 
usual $C^k$-topology.

$\phi\in{\rm Diff}^1(M)$ is called  {\em Anosov}  if $M$ is a hyperbolic set,
{\sl i.e.,} if there exist constants 
$0 < \lambda < 1$, $K>0$, and a decomposition of the tangent bundle
\begin{equation}\label{sp1}
TM \,=\, E^u \oplus E^s,
\end{equation}
into $D\phi$-invariant unstable and stable subbundles, such that 
for each $x \in M$ and every $n\in \nn$  
\begin{equation}\label{sp2}
\| \left.D_x \phi^n\right|_{E_x^s} \| \le K \lambda^n \,, \quad   
\|\left.D_x \phi^{-n}\right|_{E_x^u} \| \le K \lambda^n \,. 
\end{equation}
The above bounds should hold for some norm equivalent to the
Riemannian metric of $M$ and there always exists such a norm for which 
$K=1$.

The set ${\cal A}_k(M)$ of all $C^k$-Anosov diffeomorphisms of $M$ is an open subset of 
${\rm Diff}^k(M)$ (which can be empty). An Anosov diffeomorphism is called transitive 
if for any two non-empty open sets $U$ and $V$ and any $N\ge0$ there exists $n>N$ 
such that $\phi^{-n}(U)\cap V\not=\emptyset$.  Any Anosov diffeomorphism on a torus
${\mathbb T}^n$ is transitive. More generally, it is conjectured that all  Anosov 
diffeomorphisms  are  transitive (see \cite{KH} for various partial results). 

In this section we consider dynamical systems $(M, \phi, \omega)$ where
$\phi\in{\cal A}_2(M)$. We use freely notations and results from Section
\ref{sec-homeo}.

Let $D_x \phi:T_x M \rightarrow T_{\phi(x)} M $ be the derivative
map of $\phi$ at $x$. The entropy production observable of $(M, \phi, \omega)$ is
\[
\sigma(x)=-\log D(x),
\] 
where $D(x)=|{\rm det}D_x\phi|$ is the Jacobian of $\phi$ at $x$. Setting
$D^u(x)=|\left.{\rm det}D_x\phi\right|_{E_x^u}|$,  
$D^s(x)=|\left.{\rm det}D_x\phi\right|_{E_x^s}|$, we shall consider the potential
\[
\varphi(x) =-\log D^u(x),
\]
which is known to be H\"older continuous for $\phi\in{\cal A}_2(M)$
(see \cite{Bo2,PS1}).

We shall say that the system $(M, \phi, \omega)$ is \TRI{} if it satisfies the conditions
of Section \ref{TRIDefSect} with a continuous time reversal $\vartheta$.

{\bf Remark 1.} If $\vartheta\in{\rm Diff}^1(M)$ then $D_x\vartheta$ provides an
isomorphism between $E_x^{s/u}$ and $E_{\vartheta(x)}E_{\vartheta(x)}^{u/s}$, 
in particular the stable and unstable subbundles have the same dimension. Moreover,
\begin{equation}
\log D^u \circ \vartheta =-\log D^s \circ \phi^{-1},
\label{beautiful-day}
\end{equation}
so that $\sigma$ is homologous to $\widetilde\sigma_\varphi=\varphi-\varphi\circ\vartheta$,
$$
\widetilde\sigma_\varphi-\sigma=\log D^s-\log D^s\circ\phi^{-1}.
$$
We stress however that we shall not assume $\vartheta$ to be of class $C^1$
in the following.

{\bf Remark 2.} One  can always construct \TRI{} Anosov systems starting with 
an Anosov system 
$(M, \phi,\omega)$ and applying the construction described at the end  of
Section \ref{TRIDefSect}. The time reversal obtained in this way is $C^\infty$. 

\bigskip
The following classical result is known as the Volume Lemma (\cite{Bo2,KH}):
\bet If $\phi \in {\cal A}_2(M)$ then $\omega\in{\cal S}_\varphi$.
More precisely, for sufficiently small $\epsilon>0$ there exists  $C_\epsilon>0$
such that, for all $x\in M$ and $n>0$,
\[
C_\epsilon^{-1}\leq 
\omega(B_n(x, \epsilon))\, \e^{- S_n\varphi(x)}\leq 
C_\epsilon.
\]
\label{volume}
\eet

Note that, by Corollary \ref{cor-kif}, this implies $P(\varphi)=0$. Set 
\[
e_n(\alpha)= \omega(\e^{- \alpha S_n\sigma}).
\]
If $(M, \phi, \omega)$ is \TRI{}, the finite time ES-theorem (Proposition \ref{es-sym1})
yields that 
\begin{equation}
e_n(\alpha)=e_n(1-\alpha).
\label{sim-dif0}
\end{equation}
Proposition  \ref{kif} immediately imply the existence of the ES-functional.

\bep Suppose that $\phi\in {\cal A}_2(M)$. Then for all $\alpha \in \rr$, 
\[
e(\alpha)=\lim_{n\rightarrow \infty}\frac{1}{n}\log e_n(\alpha)= P(\varphi-\alpha\sigma).
\]
If $(M, \phi, \omega)$ is {\rm\TRI{}} then 
\begin{equation}
e(\alpha)=e(1-\alpha).
\label{sim-dif}
\end{equation}
\label{an1}
\eep
{\bf Remark.}  The symmetry (\ref{sim-dif}) is forced by the finite time symmetry 
(\ref{sim-dif0}) and so  Relation (\ref{beautiful-day}) (and the fact that $\vartheta$ is
$C^1$) is not used. One gets a direct proof of (\ref{sim-dif}) based on  (\ref{beautiful-day})
using the variational principle for the pressure in the same way as in the proof of  
Proposition \ref{believe} (2).

To improve  Proposition \ref{an1} we need to assume more. We recall another classical
result in the theory of Anosov diffeomorphisms:

\bet Suppose that $\phi\in{\cal A}_2(M)$ is transitive. Then $(M, \phi)$ is a  topologically
mixing Smale space.
\eet

In particular $\phi$ is expansive with specification and Theorem \ref{bow}, together with
the fact that $P(\varphi)=0$, yield that $\phi$ has a unique equilibrium state 
$\nu_\varphi\in{\cal S}_\varphi$. Furthermore, by Corollary \ref{PhysicalCriter},
\[
\lim_{n\rightarrow \infty}\frac1n S_nf(x)=\nu_\varphi(f),
\]
for all $f\in C(M)$ and $\omega$-a.e. $x$. Thus, $\omega_+=\nu_\varphi$ is the unique
NESS of $(M,\phi,\omega)$.
Proposition \ref{kif} yields that GC-functional exists and is equal to 
the  ES-functional.

\bep Suppose that $\phi \in {\cal A}_2(M)$ is transitive. Then for all $\alpha \in \rr$, 
\[
e_+(\alpha)
=\lim_{n\rightarrow \infty}\frac{1}{n}\log\omega_+(\e^{-\alpha S_n\sigma})
=P(\varphi-\alpha \sigma).
\]
In particular, $e(\alpha)=e_+(\alpha)$ and $(M, \phi, \omega, \omega_+)$ has regular entropic fluctuations.
\label{an2}
\eep

The analyticity of $e(\alpha)$ and $e_+(\alpha)$ (Theorem \ref{rue-van}) yields the 
respective Large Deviation Principles for the entropy production  observable. 

Regarding the strict positivity of entropy production we have the following result.
\bep
\begin{enumerate}[{\rm (1)}]  
\item Suppose that $\phi \in {\cal A}_2(M)$ is transitive. Then $\omega_+(\sigma)=0$ 
iff $\omega_+\ll \omega$.

\item Suppose that all Anosov diffeomorphisms of $M$ are  transitive (for example, 
$M={\mathbb T}^n$). Then there is an open dense set
$\tilde {\cal A}\subset {\cal A}_2(M)$
such that for all $\phi \in \tilde {\cal A}$, $\omega_+(\sigma)>0$.
\end{enumerate}
\eep
\proof (1) If $\omega_+\ll\omega$, then Corollary \ref{no-no-end} implies that $\omega_+(\sigma)=0$. The other direction follows from 
the result of Ruelle \cite{Ru6}. (2) follows from (1) and the stability result of Sinai \cite{Si}
which states that for an open dense set of $\phi$'s in ${\cal A}_2(M)$ 
the NESS is singular w.r.t. $\omega$. \qed

The resonance interpretation of $e_+(\alpha)$  follows from Theorem \ref{rue-van} (6).  
The resonances interpretation of  $e(\alpha)$  follows from recent results of Baladi and 
Tsujii \cite{Ba2,Ba3} and Gouezel and Liverani \cite{BKL,GL1,GL2,Li2,LT} on the spectrum 
of transfer operators in anisotropic Banach spaces and on the zeta function for Anosov maps.

We now turn to the discussion of linear response theory for Anosov diffeomorphisms.  
Let $k\geq4$ and let $X\mapsto\phi_X$ be a $C^k$ map from some 
neighborhood of the origin in  $\rr^N$ into ${\cal A}_k(M)$  such that $\omega$ is an 
invariant state for 
$\phi=\phi_0$ (${\cal A}_k(M)$ is a Banach manifold so the notion  differentiability 
makes sense). The map $X\mapsto \sigma_X=-\log |{\rm det}D\phi_X|\in C_\rr(M)$ is 
$C^{k-1}$. We shall consider only the  flux relation
\[
{\bf \Phi}_X=\int_0^1 \nabla \sigma_Y|_{Y=uX}\d u.
\]
Clearly, $X \mapsto \Phi_X^{(j)}\in C_\rr(M)$ is $C^{k-2}$.  
\bet Suppose that all Anosov diffeomorphisms of $M$ are transitive.  Let $k\geq 4$ and let 
$X \mapsto \phi_X\in {\cal A}_k(M)$ be a $C^k$ map from some neighborhood of the origin in 
$\rr^N$ such that $\omega$ is an invariant state for $\phi=\phi_0$. 
Suppose that $(M, \phi_X, \omega)$ is {\rm\TRI{}} with a time-reversal independent of $X$. 
Then the  Fluctuation-Dissipation Theorem  holds:
the transport coefficients 
\[
L_{jk}=\partial_{X_k}\left.\omega_{X+}(\Phi_X^{(j)})\right|_{X=0},
\]
are defined and satisfy the Onsager reciprocity relations 
\[
L_{jk}=L_{kj}.
\]
For some $A, B>0$ and all $n$, $|\omega(\Phi^{(k)}\Phi^{(j)}_n)|\leq A\e^{- B |n|}$ and the Green-Kubo formula 
\[
L_{jk} =\frac{1}{2}\sum_{n\in \zz} \omega(\Phi^{(k)}\Phi^{(j)}_n),
\]
holds. The Central Limit Theorem 
holds for $\bf\Phi$ with covariance matrix $[D_{jk}]=2[L_{jk}]$.
\label{sunny-final-van}
\eet
\proof $(M, \phi_X)$ is a topologically mixing Smale space for $X$ small enough. Arguing as in Propositions 
\ref{an1} and \ref{an2} we deduce that 
\[
g(X, Y)=g_+(X, Y)= P_X\left(\varphi_X-Y\cdot{\bf\Phi}_X\right).
\]
Ruelle \cite{Ru3} has proven that the map $X \mapsto \varphi_X\in C^\alpha(M)$ is $C^{k-2}$.
Combining this result  with Theorem 7 in \cite{KKPW} one deduces that $(X, Y)\mapsto g(X,Y)$ 
is $C^{1,2}$ in the neighborhood of the origin and  the result follows from Proposition 
\ref{sunny-van}. \qed

We finish with some  remarks.

{\bf Remark 1.} The Green-Kubo formula and Onsager reciprocity relations 
for Anosov diffeomorphisms were first proven in \cite{GR}.  This proof was based on 
explicit computations and the differentiation formula established in \cite{Ru3}. 

{\bf Remark 2.} The proof of Theorem \ref {sunny-final-van}  looks deceptively simple.
It stands on the shoulders of deep results established in \cite{Ru3, KKPW}.

{\bf Remark 3.} The linear response theory for Anosov diffeomorphisms can be also 
established starting with  the finite time Green-Kubo formula and following the  strategy 
outlined in Section \ref{sec-no-end}.  Obviously, the assumptions of the finite time 
linear response theory discussed in Section  \ref{sec-finite-time} hold under  the 
conditions of Theorem \ref{sunny-final-van}.  Under the same conditions Ruelle \cite{Ru3} 
has proven that the functions  $X\mapsto \omega_{X+}(\Phi_X^{(j)})$  are differentiable 
(see also \cite{KKPW} and  \cite{GL2}). That the limit and derivative in 
the expression (\ref{lin3}) can be interchanged follows  from the  results of 
Gouezel  and Liverani \cite{GL2,Li3} (see also \cite{Po}).  

{\bf Remark 4.} Most of the results in this section extend to a certain class of Anosov flows
for which sufficiently fast mixing has been proved, such as contact Anosov flows 
and flows  with smooth stable and unstable foliations, see \cite{BGM,Ge,Do1,Do2,Li1,BL}.

\section*{Table of Abbreviations and Symbols}
\label{AbbTable}
\addcontentsline{toc}{section}{Table of Abbreviations and Symbols}
\begin{center}
\begin{tabular}{lll}
(C)&Equivalence of $\omega$ and $\omega_t$&page \pageref{C-def}\\
(E1)&Regularity of $t\mapsto\Delta_{\omega_t|\omega}$&page \pageref{E1-def}\\
(E2)&Boundedness of $\sigma$&page \pageref{E2-def}\\
(E3)&Existence of $\langle\sigma\rangle_+$&page \pageref{E3-def}\\
(F1)&Group property of $\phi^t$&page \pageref{F1-def}\\
(F2)&Measurability of $(t,x)\mapsto\phi^t(x)$&page \pageref{F2-def}\\
(NESS1)&$M$ is a complete separable metric space&page \pageref{NESS1-def}\\
(NESS2)&Continuity of $(t,x)\mapsto\phi^t(x)$&page \pageref{NESS2-def}\\
(NESS3)&Continuity of $\sigma$&page \pageref{NESS3-def}\\
(NESS4)&Precompactness of $\{t^{-1}\int_0^t\omega_s\d s\,|\,t\ge1\}$&page \pageref{NESS4-def}\\
(T1)&$\omega_0$ is $\phi_0^t$-invariant&page \pageref{T1-def}\\
(T2)&${\bf\Phi}_X\circ\vartheta_X=-{\bf\Phi}_X$&page \pageref{T2-def}\\
(T3)&Differentiability of $\langle{\bf\Phi}_X\rangle_t$ at $X=0$&page \pageref{T3-def}\\
(T4)&Existence of $\langle{\bf\Phi}_X\rangle_+$&page \pageref{T4-def}\\
(T5)&$\omega(\Phi^{(k)}\Phi^{(j)}_t)=O(t^{-1})$&page \pageref{T5-def}\\
&&\\
ES-functional&Evans-Searles functional $e(\alpha)$&page \pageref{ES-functional-def}\\
GC-functional&Gallavoti-Cohen functional $e_+(\alpha)$&page \pageref{GC-functional-def}\\
GES-functional&Generalized Evans-Searles functional $g(X,Y)$&page \pageref{GES-functional-def}\\
GGC-functional&Generalized Gallavoti-Cohen functional $g_+(X,Y)$&page \pageref{GGC-functional-def}\\
NESS&Non-Equilibrium Steady State(s)&page \pageref{NESS-def} \\
TRI&Time-Reversal Invariance&page \pageref{TRI-def}\\
&&\\
$\langle f\rangle_t$&$t^{-1}\int_0^t\omega(f_s)\d s$&page \pageref{T-Av-def}\\
$\langle f\rangle_+$&$\lim_{t\to\infty}\langle f\rangle_t$&page \pageref{T-As-def}\\
$c^t$&Entropy cocycle&page \pageref{Ecocycle-def}\\
$\ell_{\nu|\omega}$&$\log\Delta_{\nu|\omega}$&page \pageref{RD-def}\\
${\bf L}_t=[L_{jkt}]$&Finite time Onsager matrix, transport coefficients&page \pageref{FTL-def}\\
${\bf L}=[L_{jk}]$&Onsager matrix, transport coefficients&page \pageref{lin2}\\
${\cal N}_\omega$&Normal states w.r.t. $\omega$&page \pageref{omega-N-def}\\
${\cal S}_I$&Invariant states&page \pageref{SI-def}\\
$\Delta_{\nu|\omega}$&Radon-Nikodym derivative&page \pageref{RD-def}\\
${\bf\Phi}_X$, $\Phi^{(j)}_X$&Flux observables&page \pageref{ent-flux}\\
$\sigma$&Entropy production observable&pages \pageref{sigmadefdiscrete} and \pageref{r-sigma-def}\\
$\Sigma^t$&Mean entropy production rate&page \pageref{SigmaDef}\\
${\bf\Sigma}_X^t$&Mean fluxes&page \pageref{BSigma-def}\\

\end{tabular}
\end{center}

\end{document}